\documentclass[aps,prl,floatfix,twocolumn,superscriptaddress,twoside]{revtex4-1}
\usepackage{amsmath,amssymb,amsfonts}
\usepackage{graphicx}
\usepackage[english]{babel}
\usepackage{psfrag}
\usepackage{color}
\usepackage{tabularx}
\usepackage{times}
\usepackage{hyperref}
\usepackage{float}

\begin{document}

\title{Multicritical deconfined quantum-criticality and Lifshitz point of a helical valence-bond phase}

\author{Bowen Zhao}
\email{bwzhao@bu.edu}
\affiliation{Department of Physics, Boston University, 590 Commonwealth Avenue, Boston, Massachusetts 02215, USA}

\author{Jun Takahashi}
\email{jt@iphy.ac.cn}
\affiliation{Beijing National Laboratory for Condensed Matter Physics and Institute of Physics, Chinese Academy of Sciences, Beijing 100190, China}

\author{Anders W. Sandvik}
\email{sandvik@bu.edu}
\affiliation{Department of Physics, Boston University, 590 Commonwealth Avenue, Boston, Massachusetts 02215, USA}
\affiliation{Beijing National Laboratory for Condensed Matter Physics and Institute of Physics, Chinese Academy of Sciences, Beijing 100190, China}

\date{\today}

\begin{abstract}
The $S=1/2$ square-lattice $J$-$Q$ model hosts a deconfined quantum phase transition between antiferromagnetic and dimerized (valence-bond
solid) ground states. We here study two deformations of this model---a term projecting staggered singlets as well as a modulation of the
$J$ terms forming alternating ``staircases'' of strong and weak couplings. The first deformation preserves all lattice symmetries. Using 
quantum Monte Carlo simulations, we show that it nevertheless introduces a second relevant field, likely by producing topological defects.
The second deformation induces helical valence-bond order. Thus, we identify the deconfined quantum critical point as a multicritical Lifshitz 
point---the end point of the helical phase and also the end point of a line of first-order transitions. The helical--antiferromagnetic transitions 
form a line of generic deconfined quantum-critical points. These findings extend the scope of deconfined quantum criticality and resolve a 
previously inconsistent critical-exponent bound from the conformal-bootstrap method.
\end{abstract}  

\maketitle

The deconfined quantum critical point (DQCP) is a paradigmatic ``beyond Landau'' quantum phase transition in two-dimensions \cite{Senthil04a}.
Building on field theories for quantum magnets \cite{Haldane88,Chakravarty89,Read89,Read90,Murthy90} and stimulated by intriguing numerical
simulations \cite{Sandvik02,Motrunich04}, the DQCP proposal posits that the transition between an antiferromagnetic (AFM) ground state and
a valence-bond solid (VBS, where singlets condense on groups of two or more spins) is continuous and described by spinons coupled to a
U(1) gauge field without topological defects. With the symmetry of the spinons extended from SU($2$) to SU($N$), the proposed CP$^{N-1}$ field
theory can be solved for $N \to \infty$. In violation of the Landau rules, which prescribe a first-order transition, the critical exponents
including $1/N$ corrections agree remarkably well \cite{Dyer15} with simulations \cite{Kaul12,Block13} of lattice models with AFM-VBS transitions
for moderately large $N$.

\begin{figure}[b]
\centering
\includegraphics[width=70mm]{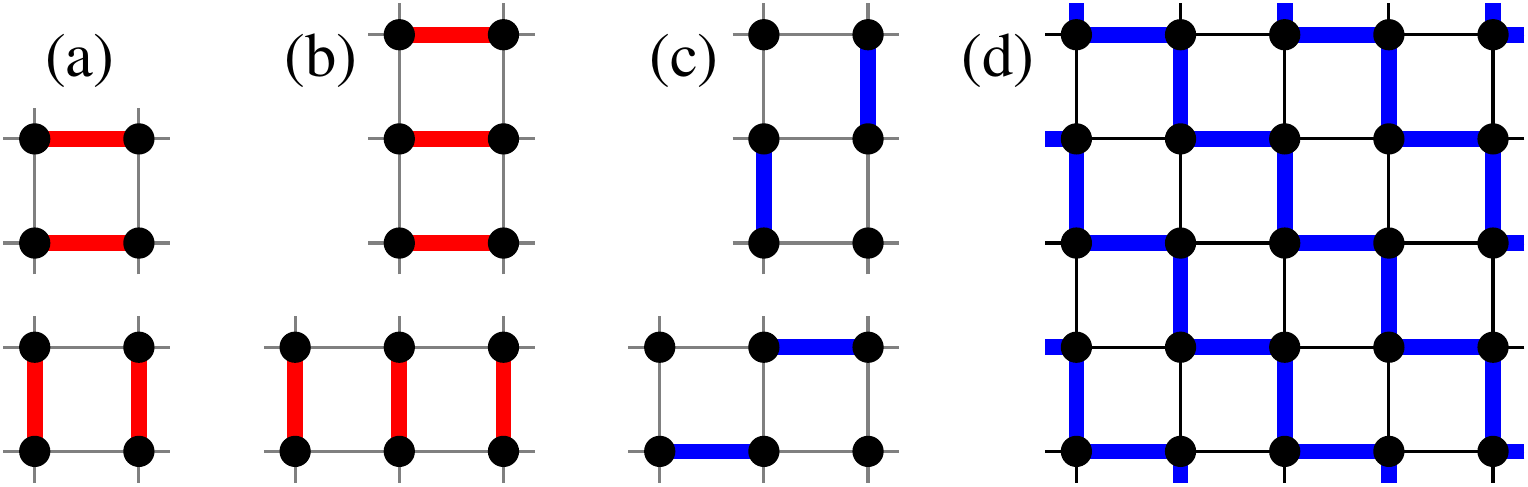}
\caption{The multi-spin columnar $Q$ interactions are products of two ($Q_2$) in (a) or three ($Q_3$) in (b) singlet projectors.
(c) The $Z$ perturbation consists of all four-spin interactions $(\mathbf{S}_i\cdot \mathbf{S}_j)(\mathbf{S}_k\cdot \mathbf{S}_l)$
with the site pairs $ij$ and $kl$ forming two staggered bonds, as shown as well as the $\pi/2$ rotated cases. (d) Staircase exchange
pattern $W$, with thick blue and thin black links representing $J(1\pm h)\mathbf{S}_i\cdot \mathbf{S}_j$.}
\label{fig:model}
\end{figure}

A contentious aspect of the DQCP scenario is the suggestion that the continuous transition persists down to $N=2$. This conjecture \cite{Senthil04a,Nogueira07}
found early support in quantum Monte Carlo (QMC) simulations of the $J$-$Q$ model, in which the $S=1/2$ Heisenberg model with exchange $J$ on the square lattice
is supplemented by four-spin \cite{Sandvik07} or six-spin \cite{Lou09} terms $Q$, illustrated in Figs.~\ref{fig:model}(a) and \ref{fig:model}(b), 
that induce correlated singlets and lead
to VBS order for large $Q/J$. Many QMC studies of these and other variants of the $J$-$Q$ model
\cite{Melko08,Jiang08,Sandvik10a,Kaul11,Sandvik12,Harada13,Chen13,Pujari15,Suwa16,Shao16,Ma18,Zhao20}, as well as
related 3D classical loop models \cite{Nahum15a,Nahum15b}, have characterized the signatures of the DQCP, including an emergent U($1$) symmetry of
the VBS fluctuations \cite{Sandvik07,Jiang08,Sandvik12,Nahum15a}. However, anomalous scaling behaviors have been interpreted by some as precursors to
a first-order transition \cite{Jiang08,Chen13,Kuklov08}. Attempts to explain the observations as a weakly first-order ``walking'' transition invoke
a non-unitary conformal field theory (CFT) with a DQCP slightly outside the accessible model space, e.g., in dimensionality different from two
\cite{Wang17,Gorbenko18a,Gorbenko18b,Ma19,Nahum19,He20}. In this scenario, the transition reflects the properties of the inaccessible fixed point but eventually,
for large lattices, flows away from it. No concrete predictions have been put forward, however, and concurrently further QMC studies have
provided compelling evidence of a continuous transition \cite{Sandvik20}.

A puzzling issue is that the critical correlation-length exponent $\nu \approx 0.45$ \cite{Shao16,Nahum15b,Sandvik20} violates
a bound $\nu > 0.51$ from the CFT bootstrap \cite{Nakayama16}. We here identify a loophole in this bound and also discover a previously
unknown helical valence-bond (HVB) phase. We consider two deformations of the $J$-$Q$ model and demonstrate that they are renormalization-group (RG)
relevant at the DQCP. First, we study a four-spin term $Z$ of staggered bond operators; Fig.~\ref{fig:model}(c). We recently showed that strong
staggered interactions lead to a first-order transition \cite{Zhao20}, likely by suppressing the emergent U($1$) symmetry associated with the 
DQCP \cite{Levin04}. Using QMC simulations, we here show that an infinitesimal $Z$ perturbation is relevant and invalidates the bootstrap 
$\nu$-bound, which is conditional on a single symmetry-preserving relevant field. The second deformation is a staircase $J$-modulation,
Fig.~\ref{fig:model}(d), which is also relevant and evolves the DQCP into a HVB phase. 

{\it Model.}---We consider the $J$-$Q_2$ and $J$-$Q_3$ models with exchange $J_b$ on links $b$ connecting nearest-neighbor sites
$i_b,j_b$. Using singlet projectors $P_b=P_{ij}=1/4-\mathbf{S}_j\cdot \mathbf{S}_j$, we write the Hamiltonian on periodic lattices with $N=L^2$ spins as
\begin{equation}
H = - \sum_{b=1}^{2N} J_{b}P_b - Q  \sum_{p=1}^{2N} \prod_{\{b_p\}} P_{b_p},
\label{hdef}
\end{equation}
where the products have either two or three singlet projectors in the sets $\{b_p\}$, arranged as in Figs.~\ref{fig:model}(a) and \ref{fig:model}(b).

Defining $g=J/(J+Q)$,
the $J$-$Q_2$ and $J$-$Q_3$ models with uniform $J_b=J$ have AFM--VBS transitions at $g_c \approx 0.0432$ \cite{Sandvik20} and $g_c \approx 0.400$ \cite{Lou09},
respectively. The DQCP has been better characterized in the $J$-$Q_2$ model \cite{Shao16,Sandvik20}, and we use it to study the relevance of the
infinitesimal staggered bond interactions, Fig.~\ref{fig:model}(c), and staircase $J$ modulation, Fig.~\ref{fig:model}(d). The $J$-$Q_3$ model is a more robust
VBS for large $g$ \cite{Sandvik12} and we use it to study finite staircase modulation. By universality, our results should apply also to other DQCP
systems.

{\it Scaling dimensions.}---To characterize the $Z$ and $W$ deformations, we compute corresponding correlation functions in the critical $J$-$Q_2$ model.
With $H_c=H(g=g_c)$ in Eq.~(\ref{hdef}), we write the perturbed Hamiltonian as
\begin{equation}
H = H_c + \delta V,~~~~ V=\sum_{a}V(\mathbf{r}_a), 
\label{hhdef}
\end{equation}
where $V(\mathbf{r}_a)$ is a subset of terms of $V$ in a suitable lattice cell. Following standard quantum-criticality and RG
notation, the correlation function $C_V(\mathbf{r})= \langle V(\mathbf{r})V(0)\rangle-\langle V(0)\rangle^2$ at $\delta=0$ should decay as
$C_V(\mathbf{r})\propto r^{-2\Delta_V}$, where $\Delta_V$ is the scaling dimension of $V$. We have used a projector QMC method in
the valance-bond basis \cite{Sandvik10b} to calculate $C_V(\mathbf{r})$, using operator cells that will be described below for the two different perturbations.
Technical details and additional results are presented in the Supplemental Material \cite{sm}.

\begin{figure}[t]
\includegraphics[width=70mm]{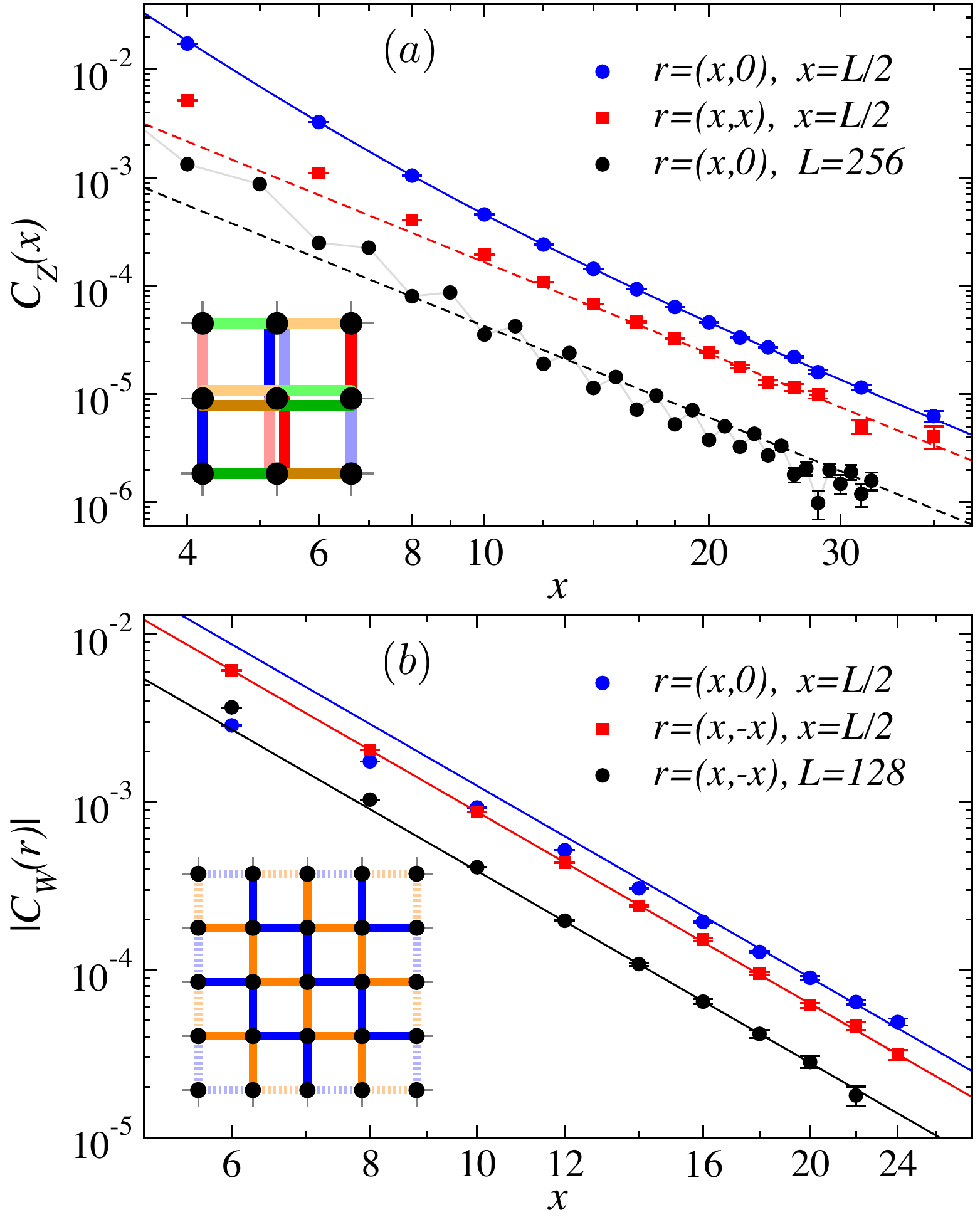}
\caption{Correlation functions at $r\propto L/2$ and $r \ll L/2$  of the operators illustrated in the insets.
(a) Staggered product operators [Fig.~\ref{fig:model}(c)], where $Z(\mathbf{r})$ is a sum of eight terms 
(indicated with different colors). The blue curve is a fit to the $r=(x,0)$ data for $x=L/2 \ge 6$ of the form $ax^{-2\Delta'_0}(1 + cx^{-\omega})$
giving $\Delta_0' = 1.40(2)$ and $\omega \approx 2.0$. The $L=256$ data (black symbols) have been divided by $4$ for visibility. The dashed lines
show the leading power law $x^{-2\Delta'_0}$. (b) Staircase $J$ modulation [Fig.~\ref{fig:model}(d)] with
$W(\mathbf{r})$ defined on a $5\times 5$ site cell with$+\mathbf{S}_i\cdot \mathbf{S}_j$ and $-\mathbf{S}_i\cdot \mathbf{S}_j$ on the blue
and orange links, respectively. The dashed edge links indicate prefactors $1/2$ needed for the cell summation in Eq.~(\ref{hhdef}).  The correlations being
negative, absolute values are shown. A fit (red line) of the form $ax^{-2\Delta_W}$ to the $r=(x,-x)$ data for $x=L/2 \ge 8$ gives $\Delta_W = 1.90(2)$.
The other lines have the same slope.}

\label{fig:scaledim}
\end{figure}

Results for the staggered bonds, $V=Z$, are shown in Fig.~\ref{fig:scaledim}(a). Here a sum of eight local terms defines the symmetric operator $Z(\mathbf{r})$. The observed power-law decay corresponds to the scaling dimension $\Delta_Z \approx 1.40(2)$,
considerably larger than the dimension $\Delta_0 \approx 0.800(4)$ of the previously known primary symmetric scalar operator $O_0$ \cite{Sandvik20}. All
correlations are positive and clearly represent the spatially uniform perturbation in Eq.~(\ref{hhdef}). While we can not rigorously prove that
$Z$ contains a second primary operator $O'_0$, its scaling dimension matches neither the dimensions $\Delta_0 + n$ ($n=1,2,\ldots$) of the
decendants of $O_0$ nor those of the order parameters $O_{\rm VBS}$ and $O_{\rm AFM}$, both of which have scaling dimensions of approximately $0.63$
\cite{Sandvik12,Nahum15b} (see also the Supplemental Material \cite{sm}). Thus, we conjecture that a second symmetric primary operator exists.
In the Supplemental Material \cite{sm} we provide further results supporting this conclusion and show examples of other bond-products
that exhibit the conventional scaling dimension $\Delta_0$.

It is surprising that an interaction with the symmetries of the unperturbed Hamiltonian can introduce a primary operator not already present
in the $J$-$Q$ model. The most likely scenario is that $Z$ generates topological defects (monopoles). The $Q$ terms in Figs.~\ref{fig:model}(a,b) are
conducive to the emergent U($1$) symmetry that is required within the DQCP scenario and which can be traced to the irrelevance of the quadrupled monopoles
associated with the $\mathbb{Z}_4$ symmetric VBS order parameter. Staggered singlets induced by the $Z$ interaction may counteract the emergent
symmetry, as we recently showed with similar terms that, when strong enough, render the AFM--VBS transition clearly first-order \cite{Zhao20}. Our present
results suggest that already an infinitesimal $Z$ causes a first-order transition.

Next, we consider the staircase $J$ modulation, $V=W$, which breaks lattice symmetries. The four-fold degenerate columnar VBS state of the 
model still retains its $\mathbb{Z}_4$ symmetry, with clock-like angular fluctuations between neighboring states characterized by a complex order
parameter $D = |D|{\rm e}^{i\phi}$ (as we demonstrate in Supplemental Material \cite{sm}). The unit cell is doubled when $h>0$ in
Fig.~\ref{fig:model}(d), but there is no symmetry implying destruction of the DQCP due to Berry phase cancelations, unlike systems such as the bilayer
SU($N$) model \cite{Kaul12b}. The $W$ perturbation being irrelevant in the VBS and AFM phases, it could a priori also be RG irrelevant at the DQCP, even
though it breaks lattice symmetries; in the Supplemental Material \cite{sm} we show an example of an irrelevant perturbation breaking the $\pi/2$
lattice rotation symmetry.

Figure \ref{fig:scaledim}(b) shows that $C_W$ defined with a $5\times 5$-site cell operator gives $\Delta_W=1.90(2)$, i.e., the staircase perturbation
is also relevant. Thus, the DQCP is unstable, but from the scaling dimension alone we do not know what fixed point the system flows to for a finite
$W$ perturbation. We will next show that a new phase opens between the VBS and AFM phases.

{\it HVB phase.}---To characterize bond order beyond regular patterns with small unit cells, we here first define a local order parameter
coarse-grained on a cell of $3\times 3$ spins,
\begin{eqnarray}
T_{x}(\mathbf{r}) &= & (-1)^{r_x} \left(S^z_{\mathbf{r}}  S^z_{\mathbf{r} + \hat{x}} 
+ S^z_{\mathbf{r} + \hat{y}}  S^z_{\mathbf{r} + \hat{x} +\hat{y}} 
+ S^z_{\mathbf{r} - \hat{y}}  S^z_{\mathbf{r} + \hat{x} -\hat{y}} \right. \nonumber \\
&-&  S^z_{\mathbf{r}}  S^z_{\mathbf{r} - \hat{x}} 
- S^z_{\mathbf{r} + \hat{y}}  S^z_{\mathbf{r} - \hat{x} +\hat{y}} 
- S^z_{\mathbf{r} - \hat{y}}  S^z_{\mathbf{r} - \hat{x} -\hat{y}})/6,
\label{txydef}
\end{eqnarray}
and $T_{y}(\mathbf{r}) = T_{x}(\mathbf{r}) (\hat{x} \leftrightarrow \hat{y})$.
We will demonstrate that the $J$-$Q_3$ model with finite staircase modulation $h>0$ hosts a phase with helical order parameter
\begin{equation}
m\left( \mathbf{k}\left( w_x, w_y \right) \right) = \sum_{\mathbf{r}} [T_{\hat{x}}(\mathbf{r}) + iT_{\hat{y}}(\mathbf{r}) ]e^{-i \mathbf{r} \cdot {\bf k} (w_x, w_y)},
\label{mkdef}
\end{equation}
where $(w_x,w_y)$ are positive integer winding numbers and $\mathbf{k}(w_x,w_y) ={2\pi}(w_x,-w_y)/L$. Here the minus sign on
$w_y$ applies to the choice of $J$ pattern in Fig.~\ref{fig:model}(d), where the stairs are directed along the $(1,-1)$ diagonal. The conventional
columnar VBS order parameter has ${\bf w}=(0,0)$.

\begin{figure}[t]
\includegraphics[width=80mm,clip]{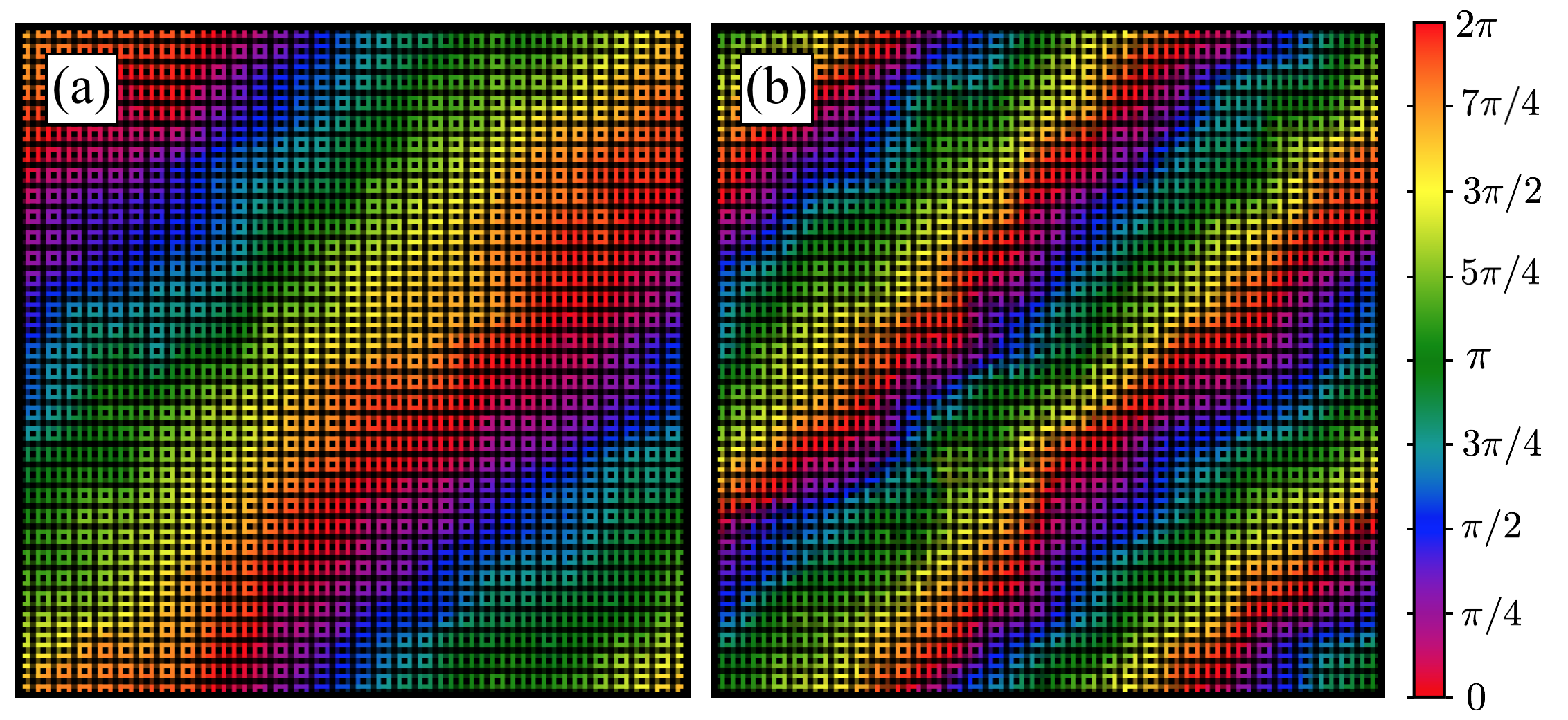}
\caption{Spatial dependence of the local VBS order parameter from short SSE runs of $L=64$ systems
at $h=1$ and $g=0.25$ (a) and $g=0.31$ (b). The bar shows the mapping of the angle extracted from bond-centered combinations of $T_x(\mathbf{r})$ and
$T_y(\mathbf{r})$ defined in Eq.~(\ref{txydef}). The brightness indicates the local bond correlation $|\langle S^z_iS^z_j \rangle|$
by a non-linear map (see Supplemental Material \cite{sm}).}
\label{fig:colorful}
\end{figure}

We used the SSE QMC method \cite{Sandvik10c} at inverse temperature $\beta=L$ to study systems with $0.2 \le h \le 1$. We first visualize the HVB order 
in Fig.~\ref{fig:colorful}, where a bond-centered local angle was extracted from $T_x$ and $T_y$ in short $h=1$ simulations during which symmetries can be 
broken. We observe what appears to be long-range order along the diagonal $(1,1)$ direction and a modulation in the $(1,-1)$
direction, corresponding to winding numbers ${\bf w}=(1,1)$ in Fig.~\ref{fig:colorful}(a) and $(2,2)$ in Fig.~\ref{fig:colorful}(b). We find similar
behaviors also at smaller $h$ values, and below we will present quantitative results showing how the winding increases versus $g$ at fixed $h$. We will also
demonstrate transitions of the HVB phase into a conventional VBS phase at $g_1(h)$ and an AFM phase at $g_2(h) > g_1(h)$.

\begin{figure}[t]
\includegraphics[width=75mm]{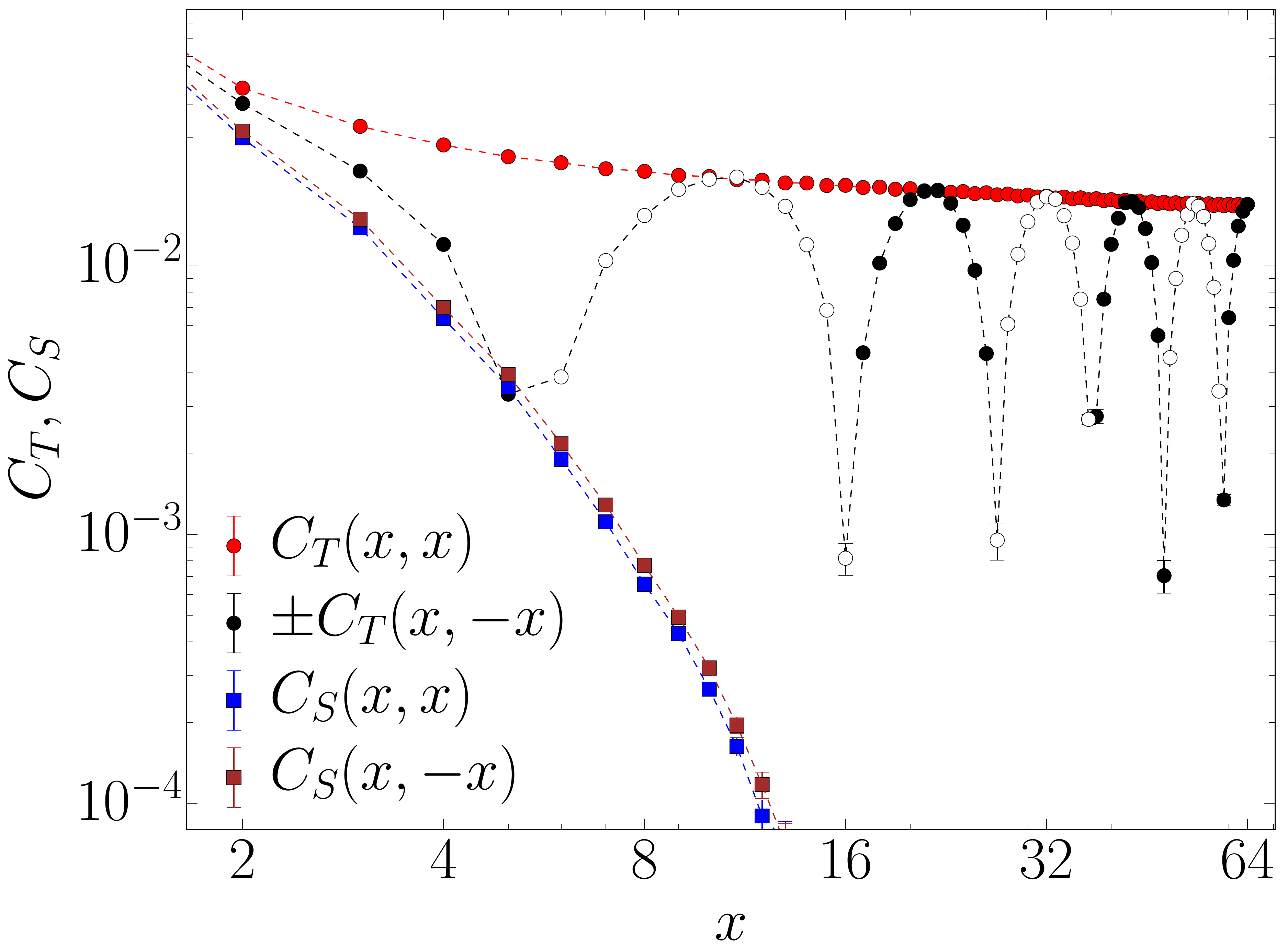}
\caption{Bond $C_{T}(\mathbf{r})$ and spin $C_{S}(\mathbf{r})$ correlations at $h=1$ in the diagonal directions of an $L=128$ system at $g=0.25$,
where $w_x=w_y=3$. Negative $C_{T}(x,-x)$ values have been multiplied by $-1$ and are shown with open circles.}
\label{fig:Corr}
\end{figure}

The following results were obtained by long SSE runs with bona fide quantum mechanical expectation values averaged over the
lattice. We define a correlation function $C_T(\mathbf{r})=\langle T_x(\mathbf{r})T_x(0)\rangle$ and also study the conventional spin correlation
function $C_S(\mathbf{r})=\langle S^z(\mathbf{r})S^z(0)\rangle$. Examples of both are shown in Fig.~\ref{fig:Corr}. As expected from Fig.~\ref{fig:colorful},
$C_T$ in the HVB phase is modulated in the $(x,-x)$ direction, while the correlations along $(x,x)$ are always positive and flatten out when $x \to L/2$.
In contrast, the spin correlations decay monotonically, faster than a power law in both directions.

\begin{figure}[t]
\includegraphics[width=75mm]{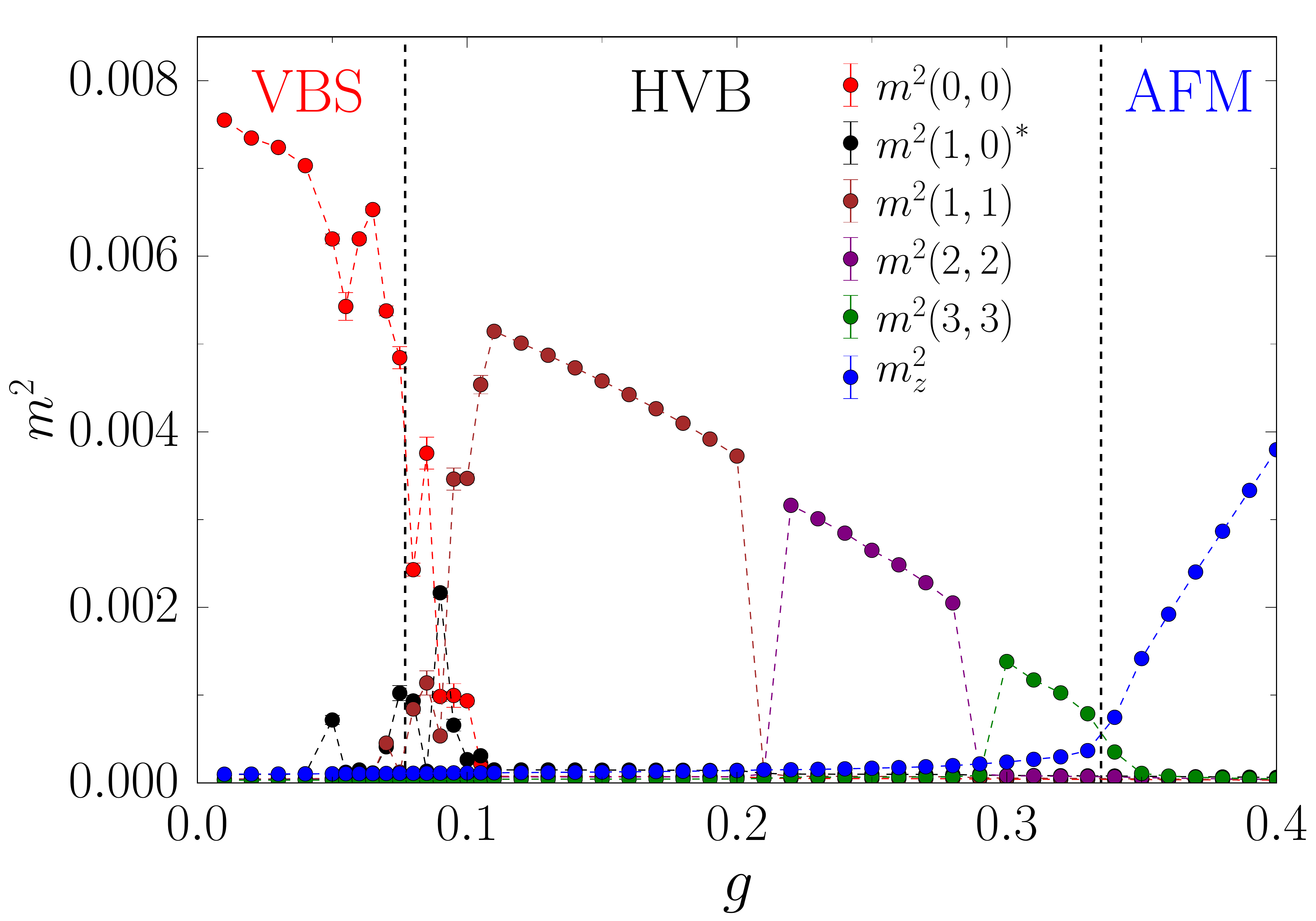}
\caption{Helical order parameters in several $(w_x,w_y)$ sectors vs $g$ for an $L=96$ system at $h=1$. We have defined
$m^2(1,0)^* = m^2(1,0) + m^2(0,1)$, reflecting two degenerate sectors for $g \sim 0.08$-$0.09$. The staggered AFM order parameter
$m^2_z$ is also shown.}
\label{fig:OrderPara}
\end{figure}

Next we consider the squared magnitude of the order parameter, Eq.~(\ref{mkdef}), for different winding numbers. Fig.~\ref{fig:OrderPara}
shows a scan over $g$ for $L=96$. The conventional ${\bf w}=(0,0)$ VBS order successively gives way to HVB order with higher
winding, until ${\bf k^*}$ reaches a maximum ${\bf k}_{\max}(g)$ and AFM order sets in. The finite-size rounded decays of both the HVB and
AFM order parameters suggest a continuous transition. In contrast, the ``micro transitions'' between different winding numbers exhibit 
metastability similar to first-order transitions. In the transition regions where all the displayed $m^2(w_x,w_y)$ values are close to zero 
in Fig.~\ref{fig:OrderPara}, the system fluctuates into winding sectors with $w_x \not = w_y$. Such winding sectors are seen explicitly for 
$g \sim 0.08$-$0.09$, where we observe degenerate ${\bf w}=(0,1)$ and $(1,0)$ helical order adjacent to the conventional ${\bf w}=(0,0)$ VBS 
phase. In the Supplemental Material \cite{sm} we further discuss the metastability and signatures of avoided level crossings in the ground state 
energy in the neighborhood of winding-number transitions. We also demonstrate that these micro transitions are mediated by the creation of spinon 
pairs and their subsequent (after winding) destruction. In the thermodynamic limit, we expect the $w_x=w_y$ states to completely dominate the HVB phase.
	
\begin{figure}[t]
\includegraphics[width=70mm]{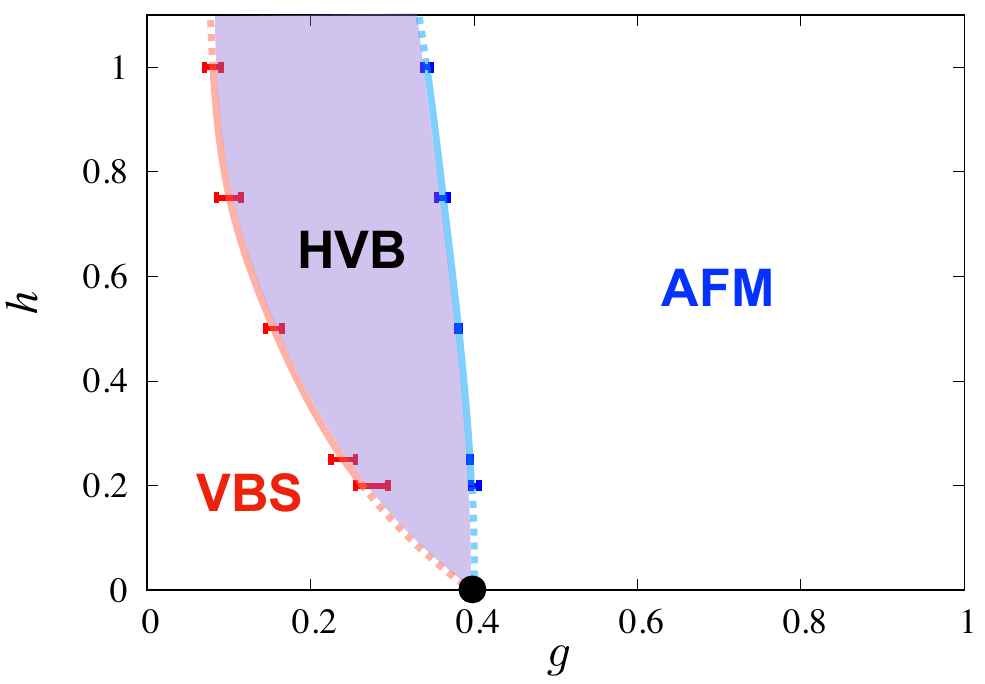}
\caption{Phase diagram of the staircase $J$-$Q_3$ model. The points with error bars are based on $L=96$ results (see Supplemental Material \cite{sm}) and the lines
are guides to the eye. Dotted lines emphasize the unknown shape of the tip of the HVB phase at the multi-critical DQCP (circle) and for $h>1$ (where there
is a QMC sign problem).}
\label{fig:PhaseDiagram}
\end{figure}

{\it Phase Diagram.}---Fig.~\ref{fig:PhaseDiagram} shows the phase diagram constructed from $L=96$ data such as those in Fig.~\ref{fig:OrderPara}.
Results for smaller sizes indicate only minor remaining finite-size effects (see Supplemental Material \cite{sm}). The HVB phase narrows
with decreasing $h$ and should extend all the way to $h=0$, on account of the relevance of the infinitesimal staircase parturbation. The DQCP is then
a kind of Lifshitz point, where the modulated HVB phase meets the VBS and AFM phases. In contrast to the classical Lifshitz point \cite{Hornreich80},
all three phases are ordered, however. The HVB--AFM transitions replace the classical modulated--disordered transitions and may form a line of DQCPs, as 
we show in the Supplemental Material \cite{sm} by examining critical correlation functions and signatures of emergent U($1$) symmetry in the HVB phase.
In the thermodynamic limit, the HVB phase at fixed $h$ should contain infinitely many winding sectors. 
The conventional DQCP approached for $h \to 0$ then has infinite winding degeneracy, as was also argued based
on studies of different winding sectors in the standard $J$-$Q$ model \cite{Shao15}. 

{\it Discussion.}---The CFT bootstrap bound $\nu > 0.51$ \cite{Nakayama16} for the DQCP has been regarded as conflicting with the QMC value $\nu \approx 0.45$
\cite{Shao16,Nahum15b,Sandvik20} and supporting the non-unitary CFT scenario \cite{Wang17,Gorbenko18a,Gorbenko18b,Ma19,Nahum19,He20}. However, the bootstrap
argument can also be interpreted differently \cite{Nakayama16} if the significance of the QMC result is properly recognized: if $\nu < 0.51$ there must
be a second relevant field. We have here identified this field as one induced by the staggered bond operators illsutrated in Fig.~\ref{fig:model}(c) and
conjecture that it destabilizes the DQCP by topological defects. We expect this effect also with other correlated-singlet projectors that are incompatible
with columnar or plaquette VBS states. The values of $1/\nu=3-\Delta_0$ and $1/\nu'=3-\Delta'_0$ are consistent with the CFT bootstrap \cite{Nakayama16} 
and it would be interesting to derive bounds for $\Delta_{\rm AFM}$ and $\Delta_{\rm VBS}$ given $\Delta_0$ and $\Delta'_0$, both with and without the
additional assumption of SO($5$) symmetry. If there is SO(5) symmetry, $\Delta_0$ and $\Delta_0'$ may correspond to cross-over and
SO(5)-preserving fields, respectively.

We have further demonstrated that the staircase perturbation in Fig.~\ref{fig:model}(d) is also relevant and opens up a magnetically disordered 
modulated HVB phase between the conventional VBS and AFM phases. The HVB--AFM transition at $g=g_2$ appears to be a line of generic DQCPs. At the 
VBS--HVB transition at $g=g_1$ we always observe the smallest non-zero winding number. Thus, in the thermodynamic limit ${\bf k}^* \to 0$ continuously 
and the HVB wavelength diverges when $g \searrow g_{1}$. When $g \nearrow g_1$, the VBS amplitude does not vanish and its correlation length 
remains finite. This type of transition is similar to predictions for certain classical helimagnets \cite{Dzyaloshinskii65,Nishikawa16}.
Our results extend the known \cite{Senthil04a} DQCP scenario, suggesting a multi-critical point at which the relevance of topological defects can be 
turned on by a second relevant symmetric field, rendering the VBS--AFM transition first-order. Another relevant field induces winding, establishing the 
the conventional DQCP as a Lifshitz type end point of the HVB phase.

The periodic boundary conditions we have used enforce lattice commensurability, and we have not determined whether there can be incommensurate 
HVB order. The dominant winding vector ${\bf k^*}$ may evolve in steps, as known in 
``devil's staircase" phase diagrams, or continuously, as in ``floating" phases \cite{Bak82,Baccaria07}. In an incommensurate floating HVB phase all 
correlations would decay as power laws and the distribution of $\mathbf{k}$ would be broadened instead of a $\delta$-function at $\mathbf{k}^*$ 
(i.e., ${\bf w}$ would then not be an emergent conserved quantum number). In the future, to settle issues such as these, it would also be useful to study 
the HVB state under different boundary conditions. We note that, even in classical models it has been very challenging to draw definite conclusions 
on the properties of modulated and helical phases \cite{Bak82,Baccaria07,vanderMerwe72,Togawa12,Nishikawa16,Bienzobaz17}.

While transitions between different VBSs have been studied previously \cite{Vishwanath04,Balents05,Zhao20}, a HVB phase with varying pitch was not
considered. The HVB phase with Lifshitz-point resembles the tilted phase and ``Cantor deconfinement'' in a quantum dimer model \cite{Fradkin04}. 
However, the spin physics being left out renders two major differences: instead of a DQCP there is Rokhsar-Kivelson point and 
the critical HVB--AFM line is replaced by first-order transitions of the modulated phase into a staggered dimer phase. 

Starting from the 
critical resonating valence-bond wave function \cite{Tang11,Patil14}, it may be possible to construct a wave function with winding and long-range order 
\cite{Lin12} to describe the HVB phase. In our staircase $J$-$Q$ model the winding is induced with a fixed direction and chirality, as we explain further 
in the Supplemental Material \cite{sm}.

\begin{acknowledgments}
{\it Acknowledgments}.---We thank Ribhu Kaul, Maxim Metlitski, Naoki Kawashima, Yoshihiko Nishikawa, Tin Sulejmanpasic, Chong Wang, and Cenke Xu for valuable 
discussions, and Duncan Haldane for pointing out the irrelevant perturbation discussed in the Supplemental Material, Sec.~3-C.  J.T. acknowledges support by the International Young Scientist Fellowship of the Institute of Physics, Chinese Academy of Sciences, under the Grant No.~202001 and by
Boston University's Condensed Matter Theory Visitors Program. A.W.S. was supported by the NSF under Grant No.~DMR-1710170 and by Simons Investigator 
Award No.~511064. Some of the numerical calculations were carried out on the Shared Computing Cluster managed by Boston University's 
Research Computing Services.
\end{acknowledgments}

\begin{widetext}

\newpage
  
\begin{center}  

\section{Supplemental Material}

{\bf\large \noindent Multicritical deconfined quantum-criticality and Lifshitz point of a helical valence-bond phase}
\vskip5mm

{\noindent
Bowen Zhao,$^{1}$ Jun Takahashi,$^{2}$ and Anders W. Sandvik,$^{1,2}$}
\vskip3mm

{\it
{$^1$Department of Physics, Boston University, 590 Commonwealth Avenue, Boston, Massachusetts 02215, USA} \\
{$^2$Beijing National Laboratory for Condensed Matter Physics and Institute of Physics,\\ Chinese Academy of Sciences, Beijing 100190, China} 
}

\end{center}
\vskip3mm

We here report further details on several technical issues as well as additional results complementing those presented in the main paper.
The materials are organized in sections as follows:

\begin{itemize}
\parskip-0.5mm

\item[1.]
Technical details on the projector QMC calculations of bond correlations.

\item[2.] 
Scaling dimensions of various symmetric operators.
  
\item[3.]
Demonstration of the staircase perturbation maintaining the $\mathbb{Z}_4$ symmetry of the VBS order parameter and discussion of other relevant and
irrelevant perturbations of the $J$-$Q$ model.

\item[4.]
Alternative finite-size scaling methods for the staircase perturbation.
  
\item[5.]
Visualizations of the HVB order and additional results for different winding sectors and spinon states.

\item[6.]    
Results for the order parameters on several different system sizes and finite-size drifts of the phase boundaries.

\item[7.]
Additional results for spin and dimer correlation functions and their critical behaviors.
  
\item[8.]    
Demonstration of emergent U(1) symmetry in the HVB phase.

\item[9.]    
Microscopic origin of the winding (tilt field) generated by the staircase perturbation.

\end{itemize}

\end{widetext}

\vskip5mm

\setcounter{page}{1}
\setcounter{equation}{0}
\setcounter{figure}{0}
\renewcommand{\theequation}{S\arabic{equation}}
\renewcommand{\thefigure}{S\arabic{figure}}

\subsection{1. QMC calculations of bond correlations}

In QMC simulations in the valence-bond singlet basis \cite{Sandvik10b}, the ground state is projected out by acting with a high power $(-H)^n$ of the Hamiltonian
on a trial wave function $|\psi_0\rangle$ expressed in terms of bipartite valence bonds. The convergence with $n$ can be accelerated by using an optimized variational
wave function, but even if $|\psi_0\rangle$ is a poor approximation to the ground state the projected state $|\psi_n\rangle \propto (-H)^n|\psi_0\rangle$ approaches
the ground state for $n \propto N^{1+\alpha}$, where the exponent $\alpha>0$ depends on the scaling of the finite-size gaps in the
system. For the calculations reported here, we used an amplitude product state \cite{Liang88} with amplitudes decaying as $l^{-3.5}$ as a function of the bond length
$l$ and carried out calculations for several values of $n$ until the computed properties were converged to within statistical errors.

In a simulation, contributions to the projected state normalization $\langle \psi_n|\psi_n\rangle$ are sampled and expectation values
$\langle A\rangle = \langle \psi_n|A|\psi_n\rangle/\langle \psi_n|\psi_n\rangle$ are evaluated in the generated valence-bond configurations. The
normalization can be expressed as
\begin{equation}
\langle \psi_n|\psi_n\rangle = \sum_{ab} W_{ab} \langle V_a|V_b\rangle,
\label{vab}  
\end{equation}
where the states $\langle V_{\rm a}|$ and $|V_{\rm b}\rangle$ each originate from $n$ operations with terms of the Hamiltonian on a sampled component (a valence
bond configuration) of the trial state. The weights $W_{ab}$ are not explicitly known but are reflected in the probability of generating the statates in the
simulation procedures (which we do not discuss here but refer to Ref.~\cite{Sandvik10b}).

Many expectation values have expressions in terms of the {\it transition graphs} corresponding to $\langle V_a|V_b\rangle$ in Eq.~(\ref{vab}). A transition
graph is simply a superposition of the valence bond configurations from $\langle V_{\rm a}|$ and $|V_{\rm b}\rangle$, in which a set of loops form from alternating
bonds in the bra and ket state \cite{Liang88}. The simplest case is the standard two-spin correlation function, where $A = \mathbf{S}_i \cdot \mathbf{S}_j$. Then
the properly normalized contribution to the expectation value is $\pm 3/4$ if the sites $i$ and $j$ belong to the same loop and $0$ otherwise, where the sign in
the former case is $+$ for sites on the same sublattice and $-$ else.

Multi-spin correlation functions of any even number of spins also have known expressions in terms of the loop structure \cite{Beach06}. The simplest case is the
general four-spin correlation function, with the operator $A = (\mathbf{S}_i \cdot \mathbf{S}_j)(\mathbf{S}_k \cdot \mathbf{S}_l)$, for which non-zero contributions
from a given transition graph arise if either all sites $i,j,k,l$ belong to the same loop or if pairs of sites belong to two different loops. Some of the
contributions have a rather intricate form, depending also on the relative order in which the sites are connected in a loop. Calculations of the four-spin
correlation functions are therefore time-consuming, especially if averaging is carried out over all translations over the lattice.

Sometimes a simpler, approximate estimator of the four-spin correlation functions is used to speed up the calculations. We here discuss tests of the approximation
applied to the staircase cell correlation function $C_W(\mathbf{r})$ displayed in Fig.~\ref{fig:scaledim}. This correlation function can be expressed as a sum over
dimer correlations of the form
\begin{equation}
C_D(\mathbf{r}_{ij})= \langle (\mathbf{S}_{i} \cdot \mathbf{S}_{i+\hat x})(\mathbf{S}_{j} \cdot \mathbf{S}_{j+\hat \alpha})\rangle,
\label{cdrdef}
\end{equation}
where $\mathbf{r}_{ij}$ is the distance between the lattice sites $i$ and $j$, $\hat \alpha = \hat x$ or $\hat \alpha = \hat y$, and the notation
$j+\hat \alpha$ is a short-hand for the site located one lattice spacing away from $j$ in the $\alpha$ direction.

Let us denote the exact expectation value of an operator $A$ evaluated in a specific transition graph as $[A]$, so that the unbiased
estimate of the quantum mechanical expectation value of $A$ is the mean $\langle [A]\rangle$ of this quantity over the sampled configurations.
As mentioned above, in the case of the spin-spin correlation function we have a very simple estimator, $[\mathbf{S}_i \cdot \mathbf{S}_j] \in \{0,\pm 3/4\}$,
where the non-zero values apply to transition graphs in which $i,j$ belong to the same loop. To avoid computing the much more complicated estimator for
the four-spin operator in Eq.~(\ref{cdrdef}), we can approximate it in the following way:
\begin{equation}
[(\mathbf{S}_{i} \cdot \mathbf{S}_{i+\hat x})(\mathbf{S}_{j} \cdot \mathbf{S}_{j+\hat \alpha})] \approx
[\mathbf{S}_{i} \cdot \mathbf{S}_{i+\hat x}][\mathbf{S}_{j} \cdot \mathbf{S}_{j+\hat \alpha}],
\label{corrappr}
\end{equation}
where the two factors are just the estimators of the two-spin correlation function, evaluated independently of each other in a given transition graph.
This approximation still implicitly contains a one-loop contribution, where all the sites involved belong to the same loop, and two-loop contributions where
$i$ and $i+\hat x$ belong to one loop and $j$ and $i+\hat \alpha$ belongs to another loop. However, it does not contain the contributions present in
the exact estimator from transition graphs where $i$ and $j$ belong to the same loop and $i+\hat x$ and $j+\hat \alpha$ belong to a different loop.
Inside the VBS phase the loops are short (size of order one) and the latter contributions become very unlikely when the separation $\mathbf{r}_{ij}$ is large.
The mean value with the approximated estimator therefore approaches the exact result exponentially fast with increasing separation. In previous works it
has also been argued that the approximation of the VBS order parameter exhibits the same scaling as the exact expectation value in quantum
critical systems \cite{Liu18}.

\begin{figure}[t]
\includegraphics[width=75mm, clip]{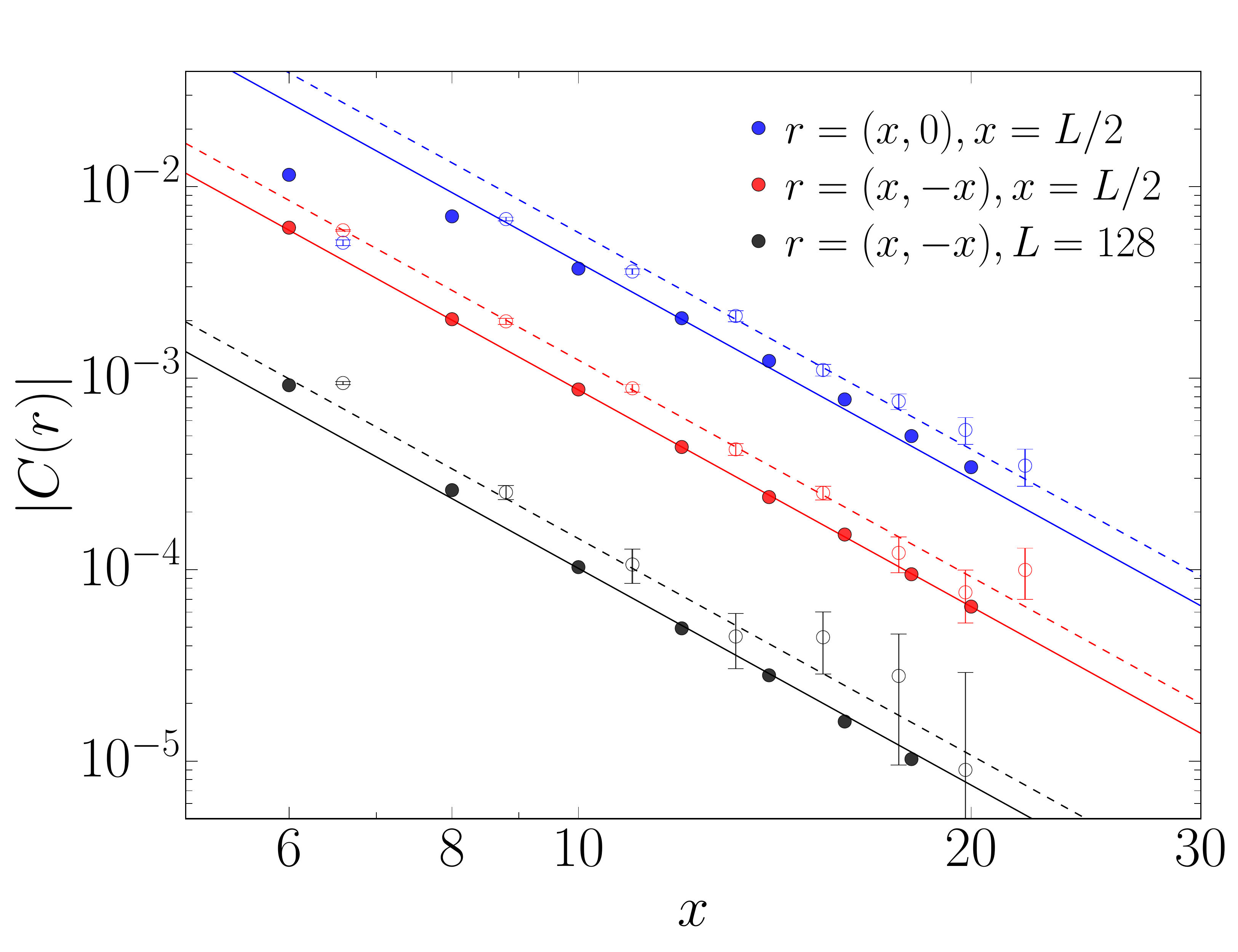}
\caption{The results of Fig.~\ref{fig:scaledim} (solid symbols), which were obtained using the approximate estimator of the dimer correlations, Eq.~(\ref{corrappr}),
compared with results for the same cell correlation functions calculated with the exact estimator (open symbols). The upper (blue) and lower (black) data sets
have been multiplied by $4$ and $1/4$, respectively, in order to make them more clearly distinguishable. Furthermore, the open symbols have been shifted by a
constant factor to the right in order for both sets of points to be more clearly visible. The dashed lines are shifted from the solid ones
by the same factor.}
\label{fig:dcompare}
\end{figure}

The cell correlation function characterizing the relevance of the staircase perturbation, illustrated in Fig.~\ref{fig:model}(a), decays much faster with 
distance than the order parameter correlations, and one may wonder whether the contributions neglected with the simplified estimator could alter the scaling
behavior. To test the quality of the approximation, which was used to generate the results in Fig.~\ref{fig:scaledim}, we have carried out calculations
also with the exact estimator. A comparison of the results is presented in Fig~\ref{fig:dcompare}. Because of the more time consuming calculations, the error
bars of the results obtained with the exact estimator are larger, especially for the larger distances, but it is still clear that the differences between the
two estimators are very small. We can essentially only observe significant differences in the data for $\mathbf{r}=(L/2,0)$ with $L/2=6$. All other points
agree to within the error bars, and there are no indications of differences in the scaling behavior. Thus, we conclude that the approximation,
Eq.~(\ref{corrappr}), is legitimate also for very rapidly decaying correlation functions.

In Sec.~2 below, we will also apply an analogous approximation in our calculations of eight-spin correlation functions of the form $\langle B_iB_jB_kB_l\rangle$,
where $B_i$ is a bond operator $\mathbf{S}_{i} \cdot \mathbf{S}_{i'}$ where the sites $i$ and $i'$ are nearest neighbors. In the cases of interest here, the
locations $i$ and $j$ will furthermore be close to each other, as will $k$ and $l$. Then it may not be a good approximation to replace the estimator $[B_iB_jB_kB_l]$
by $[B_i][B_j][B_k][B_l]$. However, if the two site pairs $ij$ and $kl$ are also far separated from each other (as in the asymptotic form that we are
ultimately interested in) the approximation $[B_iB_jB_kB_l]$ $\approx$ $[B_iB_j][B_kB_l]$ should be good, for the same reasons as discussed above,
and we use it.

\subsection{2. Scaling dimensions of bond-product operators}

Here we will first consider two bond-product operators defined on $2\times 2$ site plaquettes and show that their correlation functions in the critical
$J$-$Q_2$ model are governed by the two known primary scaling fields; the symmetric one with scaling dimension $\Delta_0 \approx 0.80$ \cite{Sandvik20} and
the symmetry-breaking VBS field with scaling dimension $\Delta_{\rm VBS} \approx 0.63$ \cite{Nahum15b}. We then present additional results for the staggered
bond operator $Z$, confirming its scaling dimension $\Delta_Z \approx 1.40$ with a different correlation function than the one used in the main paper.

\begin{figure}[t]
\includegraphics[width=75mm, clip]{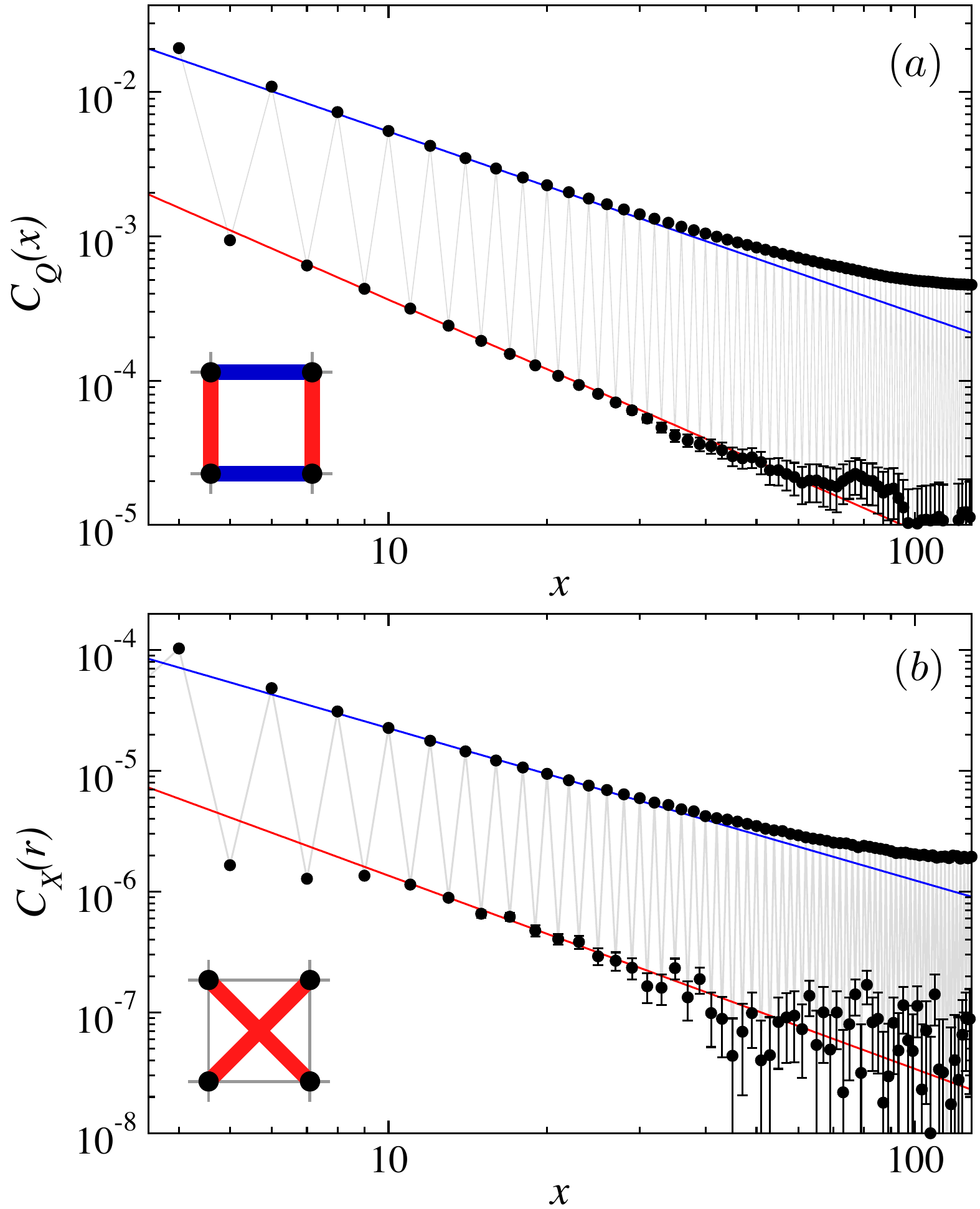}
\caption{Correlation functions of the critical $J$-$Q_2$ model ($g=0.04315$) 
along the $(x,0)$ directions of bond-product operators defined on cells with $2\times 2$ spins as
illustrated in the insets. In $(a)$ the operator $Q$ is a sum of horizontal and vertical bond-products (shown in red and blue in the inset),
while in (b) $X$ is a product of two diagonal bond operators. Both correlation functions are strictly positive in the $(x,0)$ direction and are governed
by the order parameter operator $O_{\rm VBS}$ for even $x$ and by the symmetric primary field $O_0$ for odd $x$. The corresponding power-law
decays $L^{-2\Delta_{\rm VBS}}$ and $L^{-2\Delta_0}$ are shown as blue and red lines, respectively, with the approximately known values
of the scaling dimensions $\Delta_{\rm VBS}=0.63$ \cite{Nahum15b} and $\Delta_0=0.80$ \cite{Sandvik20}.}
\label{fig:bondpair}
\end{figure}

\subsubsection{2-A. Operators defined on $2\times 2$ plaquettes}

In a recent work \cite{Sandvik20}, two of us calculated the critical correlation function of the operator $Q_2$ illustrated in Fig.~\ref{fig:model}(a) and
showed that its behavior along the lattice direction $(x,0)$ is governed by $\Delta_{\rm VBS}$ for even values of $x$, while for odd $x$ the decay
is governed by $\Delta_0$ with very rapidly decaying scaling corrections. Here, instead of using the full singlet projectors defining the $Q_2$
operator, we use the related operator $(\mathbf{S}_{i} \cdot \mathbf{S}_{j})(\mathbf{S}_{k} \cdot \mathbf{S}_{l})$ with the sites arranged exactly
as in Fig.~\ref{fig:model}(a). We combine the cases of horizontal and vertical bond pairs into the operator $Q$ and use the method discussed
in the preceding section to calculate the correlation function $C_Q(x,0)$. Results for system sizes $L=256$ are shown in Fig.~\ref{fig:bondpair}(a).
The correlations take positive values for all $x$ but form clearly distinct even and odd branches. As was previously argued in
Ref.~\onlinecite{Sandvik20}, the decay powers of the two branches are different and give the two scaling dimensions as discussed above. Here
we do not perform fits to obtain independent exponent estimates, but just demonstrate consistency with the expected behavior. Note that there are strong
enhancing effects of the periodic boundaries on the upper (VBS governed) branch, and we should expect clean power-law decays only for $x \ll L/2$ as also
observed. We here do not address the issue of anomalous finite-size scaling for $x \to L/2$ \cite{Nahum15b,Shao16}.

Given that we observe a different scaling dimension in the case of the staggered bond operators, it is interesting to also consider other perturbations
that could possibly couple to the putative second relevant symmetric field. We here investigate the case of a four-spin operator where the two bond
operators connect spins at opposite corners of the $2\times 2$ plaquettes. Such an interaction contains a projection operator of the anti-symmetric plaquette
singlet, in contrast to the symmetric singlet projected by the above operator defined on the edges of the plaquette. 

The diagonal product operator is manifestedly 
strongly frustrating, but, as shown in Fig.~\ref{fig:bondpair}(b), we observed the same behavior as in the case of the bipartite $Q$ operator. While the amplitude 
of the correlations is much smaller, there are still two branches governed by two decay exponents, and those exponents are fully consistent with the two 
known relevant scaling dimensions.

\subsubsection{2-B. Staggered bond operator}

Next we consider the staggered two-bond operators of the $Z$ perturbation, Fig.~\ref{fig:model}(c). In the main paper we combined eight instances of those
operators symmetrically within a $3\times 3$ cell and found the scaling dimension $\Delta_Z \approx 1.40$. To further test the staggered bond operators,
we here consider a smaller cell of $2\times 3$ sites containing two instances of the product operators. We refer to this cell operator as ${Z}_2$.

\begin{figure}[t]
\includegraphics[width=76mm, clip]{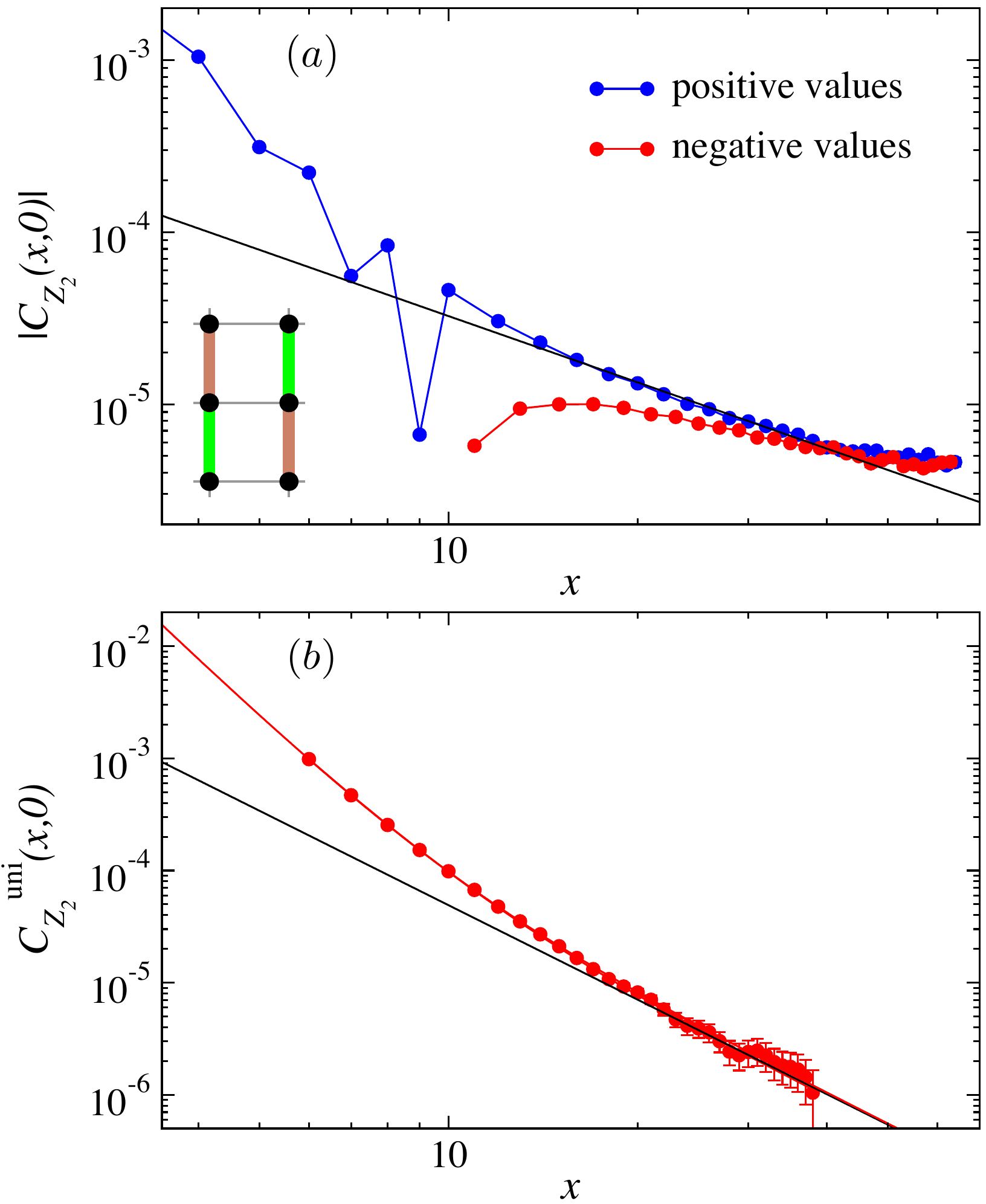}
\caption{(a) Correlation functions along the $(x,0)$ directions of cells containing two $y$-oriented staggered bond-product operators (shown in two
different colors in the inset),
calculated on a system of size $L=128$. The values for odd $x \ge 11$ are negative and are shown as their absolute values (red symbols). All other
values are positive and are shown as the blue symbols. The line shows the asymptotic decay $L^{-2\Delta_{\rm VBS}}$ with $\Delta_{\rm VBS}=0.63$
\cite{Nahum15b} expected due to contributions from the VBS order parameter. (b) The uniform component of the data in (a) extracted using the fourth-derivative
formula, Eq.~(\ref{gx1}). The red curve shows a fit to the form $aL^{-2\Delta'_0} + bL^{-(2\Delta_{\rm VBS}+4)}$ with only $a$ and $b$ adjustable and with
$\Delta'_0=1.40$ determined in the main paper (Fig.~\ref{fig:scaledim}) using a different correlation function and $\Delta_{\rm VBS}=0.63$ \cite{Nahum15b}.
The black line shows the leading term $aL^{-2\Delta'_0}$.}
\label{fig:zuni}
\end{figure}

Since the $Z_2$-cell breaks the $90^\circ$ lattice rotation symmetry, we expect that its correlation function should contain contributions from the VBS order parameter
in addition to the second symmetric field. This is confirmed by our results shown in Fig.~\ref{fig:zuni}(a), where we observe positive values of the correlation
function $C_{Z_2}(x,0)$ for even $x$, while for odd $x$ the values are positive for $x \le 9$ and turn negative for larger $x$. The even-$x$ branch decays
as a power law consistent with the expected value of $\Delta_{\rm VBS}$. We have multiplied the negative values by $-1$ to be able to show the entire odd-$x$
branch on the same log-log plot, and we observe that the two branches approach each other as $x$ increases. Note also again the boundary enhancements of the
correlations, due to which the power-law scaling is violated for large $x$. The fact that the two branches approroach each other asymptotically when graphed
this way shows that the staggered part of the correlation function is governed by the VBS operator while the uniform part may decay with a different exponent.
To test whether the second relevant symmetric field is at play here, we have to extract the uniform part of the correlation function.

Let us write a generic correlation function $C(x)$ as a sum of uniform and staggered components:
\begin{equation}
C(x) = (-1)^x A(x) + B(x),
\end{equation}
where both $A(x)$ and $B(x)$ are positive for all $x$ (and in practice they will also both be monotonically decaying with $x$).
We multiply by $(-1)^x$ to make the staggered part non-oscillating, defining $D(x) = (-1)^xC(x)$ so that
\begin{equation}
D(x) = A(x) + (-1)^x B(x).
\label{dx}
\end{equation}
We want to separate out the uniform part $B(x)$, which in the case here is a subleading contribution.

An often used way to carry out the separation amounts to a numerical second derivative of $(-1)^xC(x)$. Using the form in Eq.~(\ref{dx}) we define
\begin{equation}
f(x) = D(x)-\hbox{$\frac{1}{2}$}D(x-1)-\hbox{$\frac{1}{2}$}D(x+1),
\end{equation}
which with our form of $D(x)$ becomes
\begin{eqnarray}
  f(x) & = & A(x)-\hbox{$\frac{1}{2}$}A(x-1)-\hbox{$\frac{1}{2}$}A(x+1) \\
       & + & (-1)^x[B(x)+\hbox{$\frac{1}{2}$}B(x-1)+\hbox{$\frac{1}{2}$}B(x+1)]. \nonumber
\end{eqnarray}
Asymptotically for large $x$ this becomes
\begin{equation}
f(x) = A''(x) + (-1)^xB(x).
\label{fx}
\end{equation}
Provided that $A''(x)$ decays with $x$ much faster than $B(x)$, we can easily extract $B(x)$ as
$(-1)^xf(x)$ in the limit of large $x$.

In the case at hand, the decay of the staggered part should be governed by $\Delta_{\rm VBS} \approx 0.63$, i.e.,
$A(x) \sim x^{-1.3}$. If the scaling dimension of the uniform part is $\Delta_0' \approx 1.4$, we can see that $A''(x) \sim x^{-3.3}$
does not decay much faster than the uniform part $B(x) \sim x^{-2.8}$, and this method may therefore not work well unless we
have access to very large $x$.

We note that $f(x)$ in Eq.~(\ref{fx}) is exactly of the same form as $D(x)$ in Eq.~(\ref{dx}),
and to better isolate $B(x)$ we can repeat the above procedure with $f(x)$ in place of $D(x)$, defining
\begin{equation}
g(x) = f(x)-\hbox{$\frac{1}{2}$}f(x-1)-\hbox{$\frac{1}{2}$}f(x+1),
\end{equation}
which with the original $D(x)$ function becomes
\begin{eqnarray}
  g(x) & = & \hbox{$\frac{3}{2}$}D(x)-D(x-1)-D(x+1) \nonumber \\
       & + & \hbox{$\frac{1}{4}$}D(x-2)+\hbox{$\frac{1}{4}$}D(x+2),
\label{gx1}
\end{eqnarray}
which of course is the numerical fourth derivative of $D$. In terms of the $A$ and $B$ functions we obtain
the asymptotic form
\begin{equation}
g(x) = A^{(4)}(x) + (-1)^xB(x),
\label{gx2}
\end{equation}
where in the present case we expect $A^{(4)}(x) \sim x^{-5.3}$. Now $B(x)$ should be clearly the dominant contribution
and we can just extract it as $(-1)^xg(x)$ for sufficiently large $x$.

Results for the so extracted uniform component of the $Z_2$ correlations in Fig.~\ref{fig:zuni}(a) are shown in Fig.~\ref{fig:zuni}(b).
We can fit the results to a leading power law with a correction, but instead of doing this we can fix the leading power of $x$ to
$-2\Delta_0'=2.78$ and use a second term with the exponent expected from the fourth derivative of the staggered part, $-5.3$,
according to the arguments above. Remarkably, all data for $x \ge 6$ can be well fitted to this form by only adjusting the
factors in front of the two fixed powers of $x$. A completely independent fit also gives very similar values of the exponents. Thus, we have
shown that the results obtained with the $3\times 3$ symmetric cell in Fig.~\ref{fig:scaledim}(a) in the main paper are fully reproducible with
a different definition of the operator cell, and there should be no doubt that the scaling dimension $\Delta_Z \approx 1.40$
is real and represents a second relevant symmetric operator. The numerical value of $\Delta_Z$ is stable with respect to different
operator cells and distances included in the fit, and we can confidently conclude that it is different from all dimensions of
the decendants of the leading symmetric operator and the order parameter operators.

An interesting aspect of the results presented above, as well as those in Ref.~\onlinecite{Sandvik20}, is that the scaling dimension $\Delta_{\rm VBS}$ is
not manifested just in the staggered component of the plaquette correlation function, but is present also in the uniform component. Moreover, the
correlation function for odd $x$ in Fig.~\ref{fig:bondpair} is not at all contaminated by the VBS correlations. This somewhat counterintuitive aspect of
the plaquette correlations can be traced back to the fact that the VBS order parameter correlations on the line $(x,0)$ are completely canceled out for odd
$x$ when considering a $\mathbb{Z}_4$ or U($1$) symmetric order parameter. However, when using the $2\times 3$ cell of the staggered bond operators, which also detect
VBS order, the averaging argument no longer applies, and the VBS order affects the correlations at both even and odd distances. As we have shown here in
Fig.~\ref{fig:zuni}, the VBS order parameter then is completely contained in the staggered part of the operator and the uniform part is governed by the
new scaling dimension $\Delta'_0$ with very rapidly decaying scaling corrections.

\subsection{3. $\mathbb{Z}_4$ symmetry of the VBS fluctuations}

One of our key observations regarding the staircase perturbation of the $J$-$Q$ model depicted in Fig.~\ref{fig:model}(a) is that
it does not change the symmetry of the VBS order parameter. Without the staircase term, the VBS realized in the $J$-$Q_2$ and
$J$-$Q_3$ models is a columnar one \cite{Lou09,Sandvik12}. It is easy to see that the staircase modulation of the $J$ couplings
has an equivalent effect on all four columnar VBS patterns. Given also that the VBS has a gap above the four degenerate levels,
the four-fold degeneracy must persist for some range of the strength $h$ of the staircase perturbation (until the quantum phase transition
into the HVB phase). However, just having a four-fold degeneracy does not imply that the broken symmetry is $\mathbb{Z}_4$---it
could also in principle be $\mathbb{Z}_2 \times \mathbb{Z}_2$.

\begin{figure}[t]
\includegraphics[width=83mm, clip]{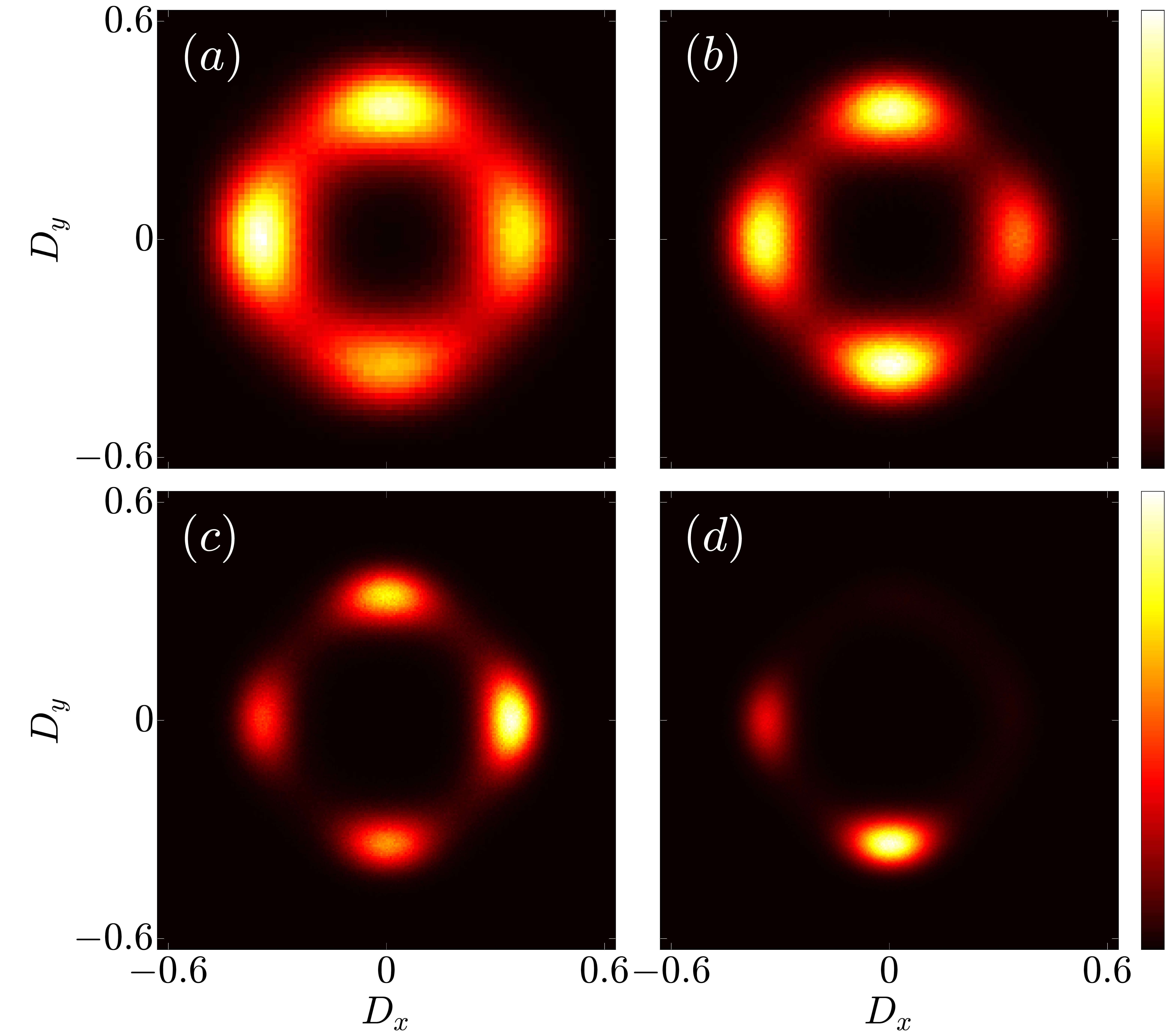}
\caption{Distributions of the VBS order parameter, Eq.~(\ref{dxdydef}) of the staircase $J$-$Q$ model at $g=0.05$, $h=1$, collected in SSE simulations of
four different system sizes: (a) $L=16$, (b) $L=20$, (c) $L=28$, (d) $L=32$. The unequal weights in the different peaks (and the two almost invisible peaks
for $L=32$) is due to the small probability density between the peaks, which implies slow migration of the sampled configurations between the peaks. The
scale used for the horizontal and vertical axes are relative to the maximum value of $D_x$ and $D_y$ in an ideal VBS state.}
\label{fig:dist1}
\end{figure}

What classifies the symmetry as $\mathbb{Z}_4$ is the relationship between the degenerate states, i.e., how they can be physically transformed into each
other. In a finite system, where the symmetries are not broken, fluctuation paths between the states can be investigated and classified in terms
of the order parameter distribution, as recently discussed in detail in Ref.~\onlinecite{Takahashi20}. We will here discuss such distributions
with a conventional definition of the columnar, non-winding VBS order parameter $D=D_x+iD_y=|D|{\rm e}^{i\phi}$, where
\begin{subequations}
\begin{eqnarray}
D_x & = & \frac{1}{N} \sum_{\mathbf{r}} (-1)^{r_x} S^z_{\mathbf{r}}S^z_{\mathbf{r}+\hat x}, \\
D_y & = & \frac{1}{N} \sum_{\mathbf{r}} (-1)^{r_y} S^z_{\mathbf{r}}S^z_{\mathbf{r}+\hat y}.
\end{eqnarray}
\label{dxdydef}
\end{subequations}
In the thermodynamic limit, if there is columnar VBS order, the angle takes the values $\phi=n\pi/2$, $n=0,1,2,3$, and the distribution $P(D_x,D_y)$ comprises
four delta-function peaks at these angles. In a finite system, both the angle and the amplitude fluctuate, and $\mathbb{Z}_4$ symmetry implies that the fluctuations
between the closest angles, $n \to n \pm 1 ~{\rm mod}~4$, are stronger than those between opposite values $n \to n + 2 ~{\rm mod}~4$. The dominant ``clock''
fluctuations correspond to domain walls over which the angle changes by $\pi/2$ (which have also been studied in the $J$-$Q$ models \cite{Shao15}).
In the DQCP scenario, the significance of $\mathbb{Z}_4$ symmetry is that it is enhanced to U($1$) when the AFM phase is approached \cite{Levin04}, through
the mechanism of dangerously irrelevant fields known from classical 3D $q$-state clock models with $q \ge 4$ \cite{Oshikawa00,Leonard15,Okubo15,Shao20}.

\subsubsection{3-A. Staircase perturbation}

The VBS fluctuation pattern is reflected in the order-parameter distribution $P(D_x,D_y)$ accumulated in a QMC simulation. Fig.~\ref{fig:dist1} shows results 
at the largest value, $h=1$, that we can reach with sign-free QMC simulations of the staircase deformed $J$-$Q_3$ model. We set $g=0.05$, which is inside the VBS
phase but close to the HVB boundary according to the phase diagram in Fig.~\ref{fig:PhaseDiagram}. Results are shown for four different system sizes, to
illustrate how the fluctuations gradually diminish with increasing $L$. The distributions clearly show the dominant $\pi/2$ fluctuations between the four
peaks at locations corresponding to the columnar VBS patterns.

The unequal weights in the four peaks, especially for the larger systems, reflect the long
simulation times taken for the system to tunnel between the different symmetry-breaking patterns. The time scale of the angular peak-to-peak fluctuations
diverges exponentially with the system size, and a simulation of a large system may in practice be trapped in one state. Apart from the unequal weight
distribution, the individual peaks show reflection symmetry about the $x$ or $y$ axis on which they are centered, as well $\pi/2$ rotation symmetry. These
distributions are very similar to those for the unperturbed $J$-$Q_3$ model \cite{Lou09} and obey all requirements for the VBS symmetry to be classified as
$\mathbb{Z}_4$. Here it should also be pointed out that the histogram of the diagonal (defined with spin-$z$ operators) reflects the true physical VBS order-parameter
distribution of the model and is not dependent on the simulation algorithm---only the (simulation) time dependence of the fluctuations depend on the
algorithm employed.

The equal and symmetric probability distributions between the dominant clock angles
$\phi=n\pi/2$, $n=0,1,2,3$ can be traced to the staircase perturbation not just maintaining the four-fold degeneracy of the columnar VBS pattern, but
the idealized plaquette VBS states, corresponding to the four angles $\phi=n\pi/2+\pi/4$, are also all identically affected by the perturbation. This fact
already strongly suggests that the fluctuation paths retain their $\mathbb{Z}_4$ clock character, and the observed distributions confirm this.

\begin{figure}[t]
\includegraphics[width=83mm, clip]{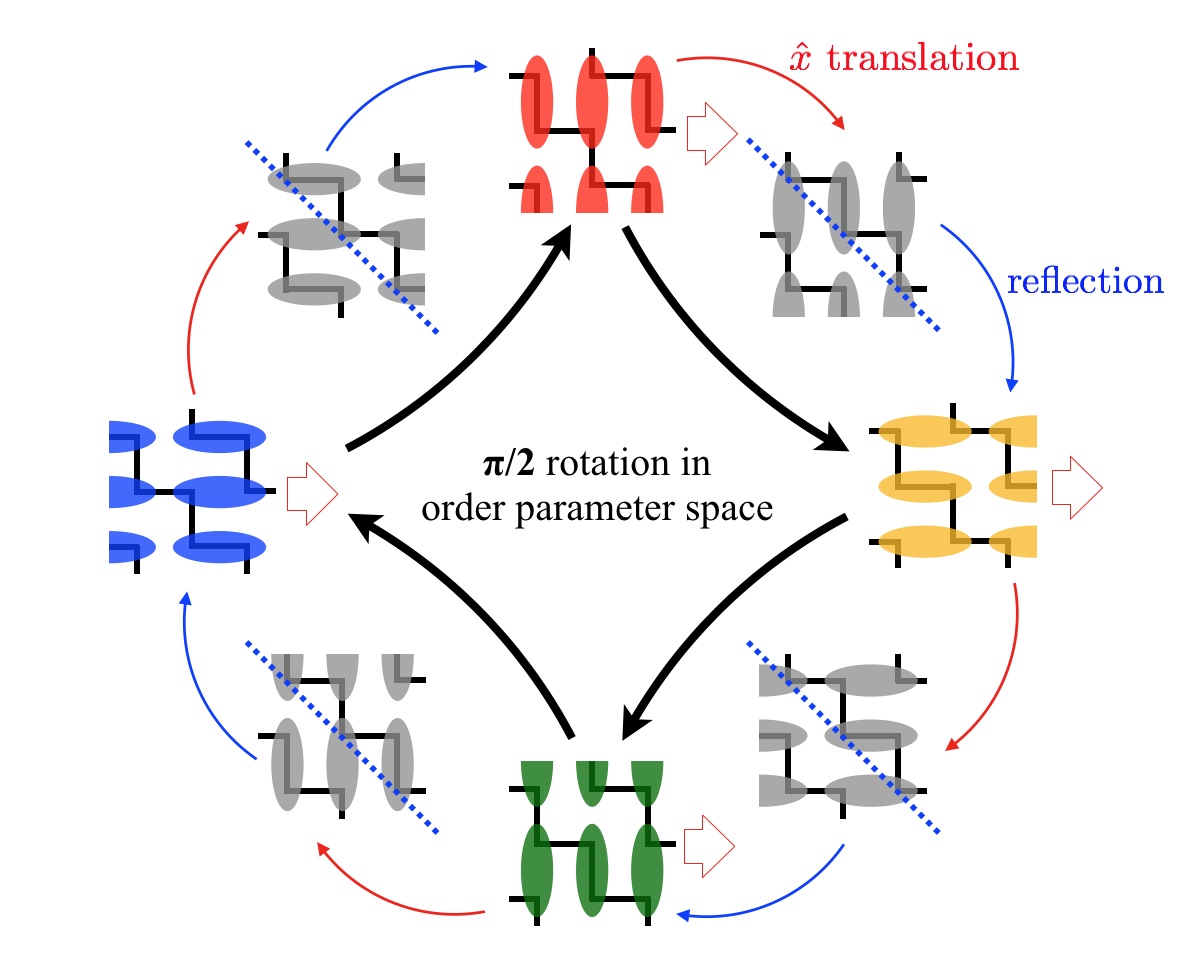}
\caption{Illustration of the $\mathbb{Z}_4$ operator combining a lateral $x$ translation and a diagonal reflection. The combined operation
corresponds to a $\pi/2$ rotation of the VBS order parameter and conserves the staircase pattern.} 
\label{fig:z4op}
\end{figure}

As pointed out above, four-fold degenerate spontaneous symmetry breaking generally has a subtlety in determining if the broken symmetry is
$\mathbb{Z}_4$ or $\mathbb{Z}_2\times \mathbb{Z}_2$, as discussed in detail Ref.~\onlinecite{Takahashi20}.
In our case at hand here, in addition to the empirical evidence from the histogram having a clock-like
fluctuation pattern, we can also formally construct a $\mathbb{Z}_4$ symmetry operation that becomes broken in the VBS phase. It is easy to see that a translation in
the $x$ direction by one lattice spacing followed by a reflection with respect to the diagonal $(1,-1)$  axis will result in a $\pi/2$ rotation of the VBS
pattern, while leaving the staircase pattern invariant. This operation is sometimes called a glide-reflection and is illustrated in Fig.~\ref{fig:z4op}.
The successive four transformations of the VBS pattern can be taken as the representative for the coset between the original and remaining symmetry groups,
forming a $\mathbb{Z}_4$ representation of the symmetry. Alternatively, we can also combine a bond-centered $\pi$ lattice rotation and a reflection along the diagonal
$(1,1)$ axis, which gives the same $\pi/2$ VBS rotation but with operations that are all valid symmetry transformations, i.e., the staircase pattern is
invariant also at the intermediate stage (which is not the case with the glide rotation, as seen in the intermediate configurations shown in gray in
Fig.~\ref{fig:z4op}).

It should be noted here that the lattice transformations that we have discussed above are only intended to show mathematically that an appropriate $\mathbb{Z}_4$ group
can be constructed with the columnar VBS on the staircase deformed square lattice. They have nothing to do with the actual fluctuations of the system (as observed
in Fig.~\ref{fig:dist1}). The underlying mechanism of the physical fluctuations requires the extended spatial structure where domain walls can be formed that
gradually realize the $\pi/2$ rotation through the intermediate VBS angles.

With the $\mathbb{Z}_4$ symmetry of the VBS order parameter maintained, and obviously the O($3$) symmetry of the AFM order parameter also unaffected, the  staircase
perturbation could in principle be RG irrelevant (and example of which is discussed in Sec.~1-C below). As we have shown in the main paper, the perturbation
actually is relevant and leads to the opening up of the HVB phase between the VBS and AFM phases when $h>0$. Thus, there are (at least) two relevant, symmetry
preserving relevant fields at the DQCP separating the VBS and AFM phases at $h=0$, and the bound $\nu > 0.51$ from the CFT bootstrap under the assumption of
a single relevant field \cite{Nakayama16} does not apply to this case.

\begin{figure}[t]
\includegraphics[width=83mm, clip]{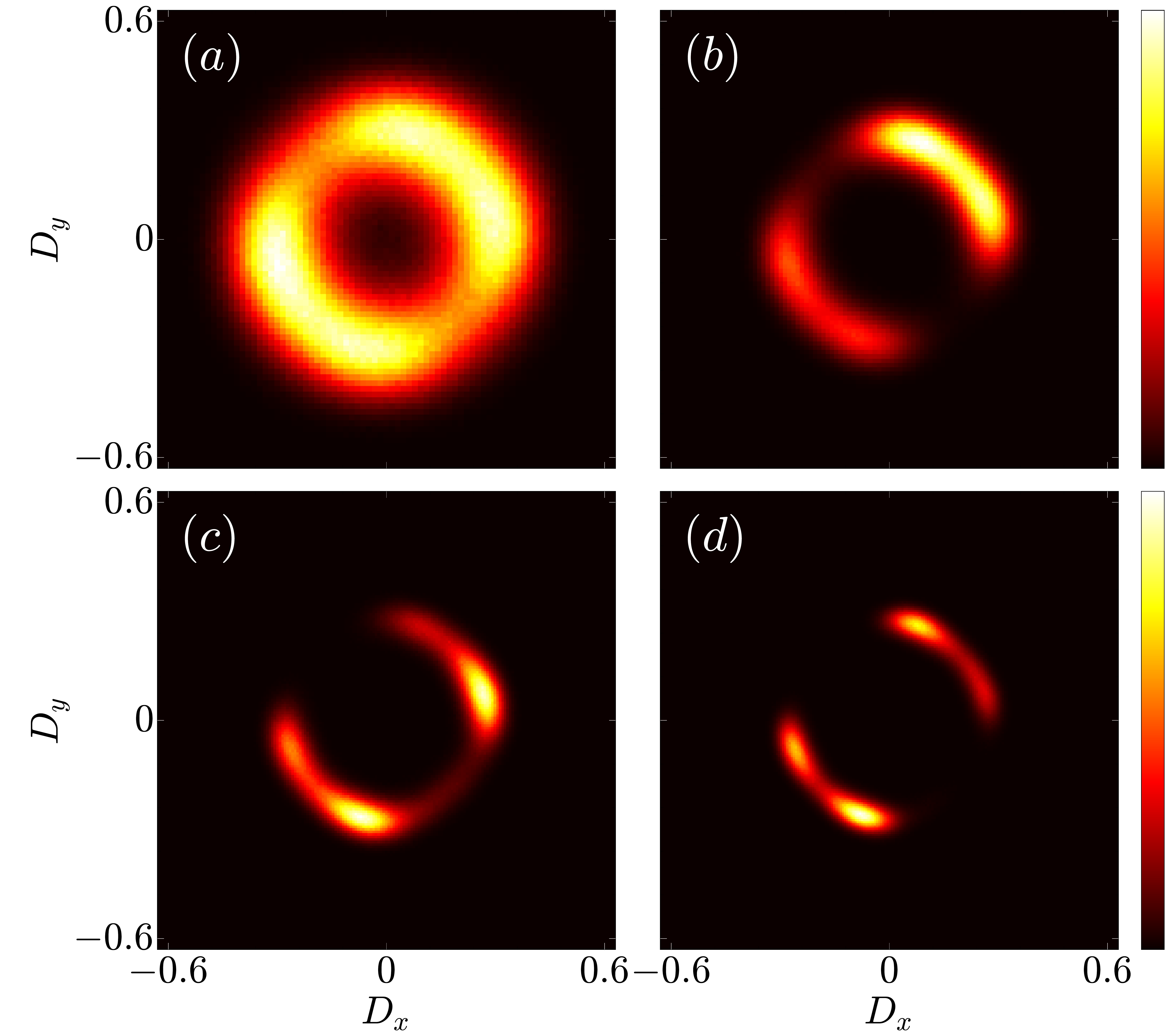}
\caption{VBS order parameter distributions for the model with $Q_3=1$ perturbed by checker-board patterned $Q_2$ interactions of strength
$Q_2(1 \pm q)$ with $Q_2=5/6$ and $q=0.03$.}
\label{fig:dist2}
\end{figure}

\subsubsection{3-B. Checker-board perturbation}

To illustrate that there are also perturbations in the presence of which the $\mathbb{Z}_4$ distribution changes to $Z_\mathbb{Z} \times Z_\mathbb{Z}$,
we consider a checker-board
pattern of $Q_2$ terms, with alternating strengths $Q_{2\pm} = Q_2(1 \pm q)$ (as in the model introduced in Ref.~\cite{Zhao19}). We add this perturbation
to the $Q_3$ model without $J$ terms, so that the ground state is strongly VBS ordered. One can again easily see that the non-uniform interaction has equal
effects on all four columnar VBS states, and, thus, one can expect a four-fold degeneracy, at least for small values of $q$ and $Q_3$ not too small
(noting that the extreme case of $q=1$ and no $Q_3$ interaction has only a two-fold degeneracy \cite{Zhao19}). However, since the $\pi/2$ domain walls of
the columnar VBS are essentially plaquette VBSs, the fluctuations between the four columnar states are no longer clock-like, because the two plaquette
states with singlets on the plaquettes with the $Q_{2+}$ interactions are favored over those with $Q_{2-}$.

Figure \ref{fig:dist2} shows distributions for the model with $Q_3=1$, $Q_2=5/6$, and $q=0.03$, where the distortion of the $\mathbb{Z}_4$ symmetry caused by
the rather weak checker-board distortion $Q_{2+} \not= Q_{2-}$ is clearly visible. Here the $\phi=\pi/4$ and $\phi=5\pi/4$ plaquette states are the ones
most effectively assisting tunneling between the columnar states, and the dominant VBS angles are pulled pairwise toward these angles. If the four unequally
spaced peaks survive when $L \to \infty$, with the weight between them vanishing, as seems plausible based on the results in Fig.~\ref{fig:dist2}, the system
is four-fold degenerate and breaks $\mathbb{Z}_2 \times \mathbb{Z}_2$ symmetry. If the fluctuations between the closely spaced peaks survives, the system
instead breaks $\mathbb{Z}_2$ symmetry, as in the cases studied in Ref.~\onlinecite{Zhao19}.

\subsubsection{3-C. Diagonal perturbation}

\begin{figure}[t]
\includegraphics[width=75mm, clip]{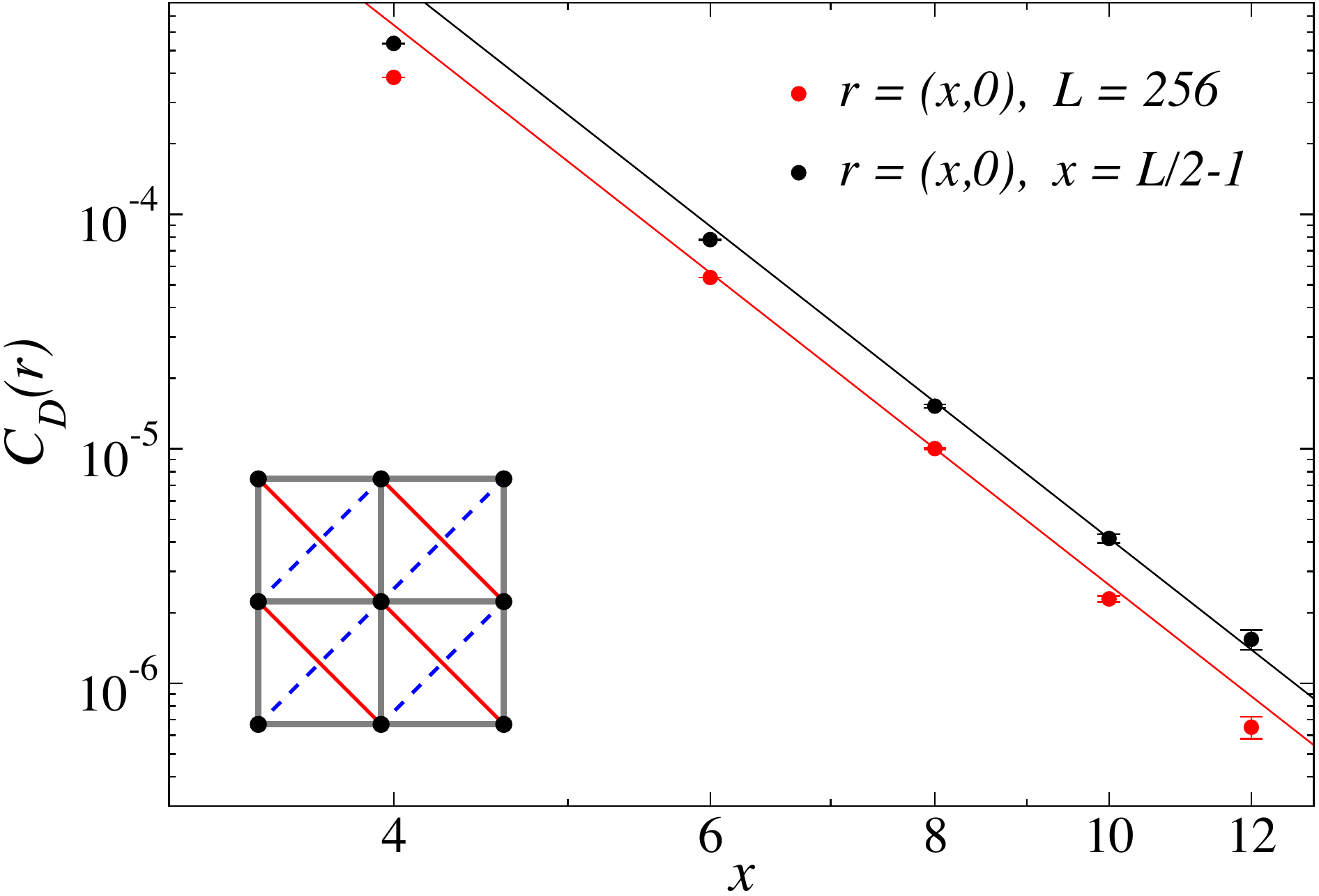}
\caption{Correlation functions in the critical $J$-$Q_2$ model ($g=0.04315$) defined with the $3\times 3$ cell operator $D$ shown as an inset, where the red
solid and blue dashed lines represent second-neighbor operators $\pm \mathbf{S}_i\cdot \mathbf{S}_j$ with positive and negative signs, respectively (and the
whole operator $D$ is a sum of these eight operators). The solid black and red lines drawn close to the $C_D$ data sets both show the limiting form $r^{-6}$
between irrelevant and relevant perturbations. The results were obtained with the projector QMC calculations discussed above in Sec.~1.}
\label{fig:diagpert}
\end{figure}

It is also useful to consider a perturbation that breaks lattice symmetries while preserving the symmetry of the VBS order parameters, but, in contrast to
the staircase perturbation, is RG irrelevant. A candidate for such a perturbation \cite{Haldane} is a next-nearest-neighbor coupling along one of the diagonals.
It can easily be checked that such a coupling in the VBS phase of the $J$-$Q$ model discriminates neither between the four columnar ordered states nor
between the four plaquette states on the clock fluctuation paths, in analogy with subsection 1-A above.

To test whether the perturbation indeed is irrelevant, we consider an operator consisting of all second-neighbor Heisenberg operators
$\pm \mathbf{S}_i\cdot \mathbf{S}_j$ with positive signs in the lattice direction $(1,1)$ and negative signs in the $(1,-1)$ direction. Following the
approach in the main paper, Eq.~(\ref{hhdef}), we divide these interactions into groups that tile the lattice, in this case using the $3\times 3$ spin
cells $D$ depicted in the inset of Fig.~\ref{fig:diagpert}. The correlation function $C_D(\mathbf{r})=\langle D(\mathbf{r})D(0)\rangle$ should decay as
$r^{-2\Delta_D}$ at the critical point of the $J$-$Q_2$ model. As shown in Fig.~\ref{fig:diagpert}, the decay exponent is close to the limiting value $2\Delta_D=6$
separating relevant and irrelevant operators. The error bars are relatively large because of the rapid decay, limiting access to only modest
distances. Nevertheless, the data set with the smaller error bars (where the system size $L=256$ is fixed), shows with reasonable confidence that
the asymptotic decay should be faster than $r^{-6}$. Thus, the diagonal perturbation, which breaks the $\pi/2$ lattice rotation symmetry, is indeed
irrelevant at the DQCP.

\subsection{4. Correlations of the staircase operator}

In the main paper, the scaling dimension of the staircase perturbation was determined by analyzing the decay of a correlation function
defined with cells of $5\times 5$ spins and calculated using the method described in the previous section. The results shown in Fig.~\ref{fig:scaledim}
deliver the scaling dimension $\Delta_W = 3.80(4)$. We here present a consistency test of this value by analyzing the correlation function in the
full $(x,y)$ distances space.

First, the reader may wonder why we chose such a large cell of $5\times 5$ spins (or, equivalently $4\times 4$ plaquettes), Fig.~\ref{fig:model}(c), on which
the staircase operator is defined. In principle any $l\times l$ cell with odd $l \ge 3$ can be used, provided that the $L\times L$ lattice can be completely
tiled by cells with overlapping edges (i.e., $L$ is a multiple of $l$), so that the perturbed Hamiltonian can be expressed in analogy with Eq.~(\ref{hhdef}).
A cell with even $l$ is less useful, because the odd number of rows and columns of lattice links in such a cell implies that the cell-cell correlation
function will contain contributions from the VBS order parameter, and the faster decaying contribution from the staircase field then has to be
extracted as a subleading contribution. With odd $l$, the cells are balanced in positive and negative contributions from the VBS order parameter and the
staircase contribution dominates the correlation function.

\begin{figure}[t]
\includegraphics[width=73mm, clip]{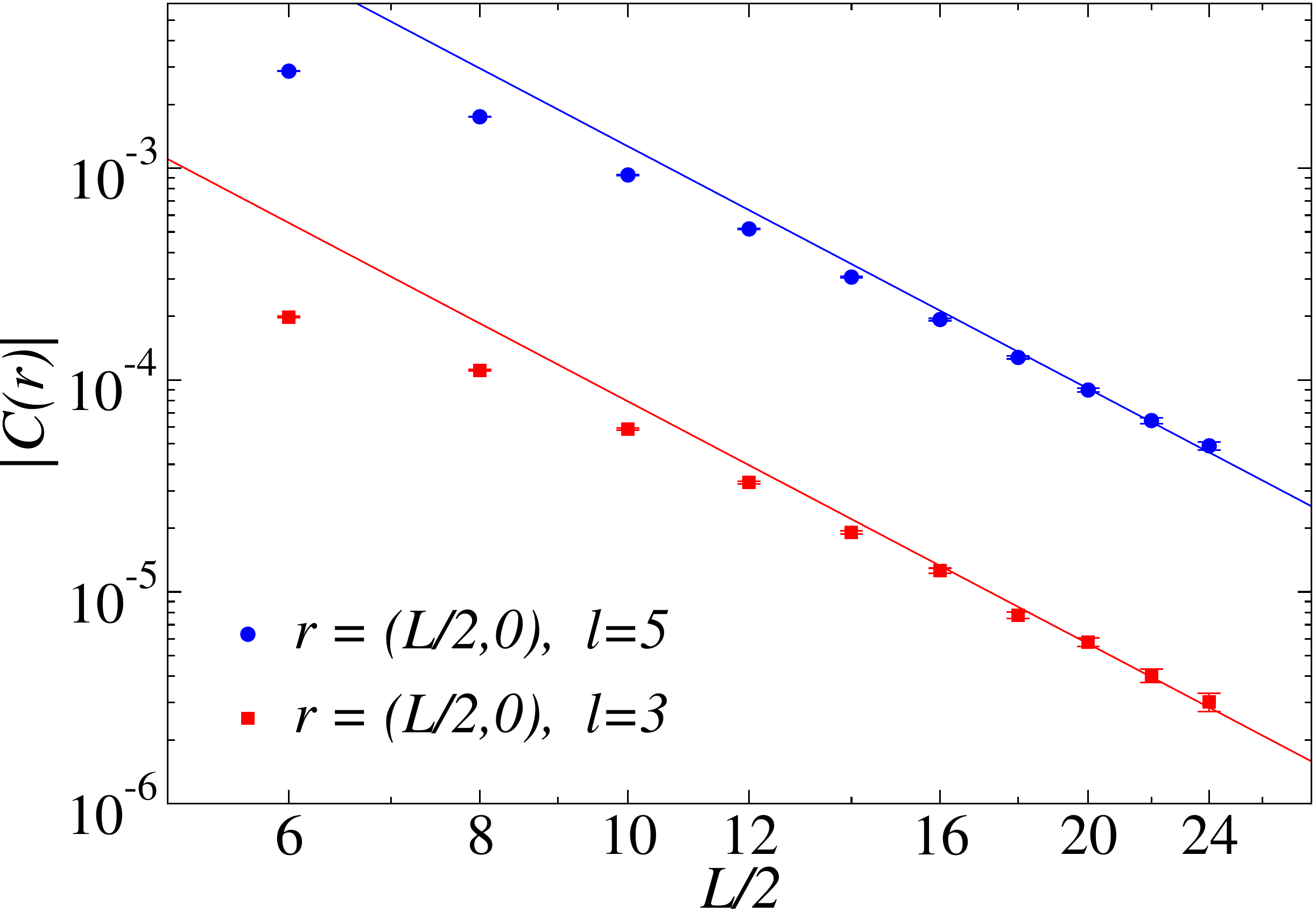}
\caption{Staircase cell-cell correlations at distance $r=L/2$ with the cell operator defined on a $3\times 3$ lattice sites,
compared with results for the $5\times 5$ cell (same data as in Fig.~\ref{fig:cell3}). The lines have the same slope; $-2\Delta_W=-3.80$.}
\label{fig:cell3}
\end{figure}

To demonstrate consistency of results obtained with different $l\times l$ cells, we present results for $l=3$, with the cell operator defined as in
Fig.~\ref{fig:model}(c) but with two rows and columns removed. Results are shown in Fig.~\ref{fig:cell3} along with the previous $5\times 5$
cell data from Fig.~\ref{fig:scaledim}. Here we can see that the asymptotic decay form indeed is the same in both cases, but the overall amplitude
of the correlation function for the $3\times 3$ cell is smaller, and the relative errors are larger. The $5\times 5$ cell is therefore in practice
a better choice.

\begin{figure}[t]
\includegraphics[width=80mm, clip]{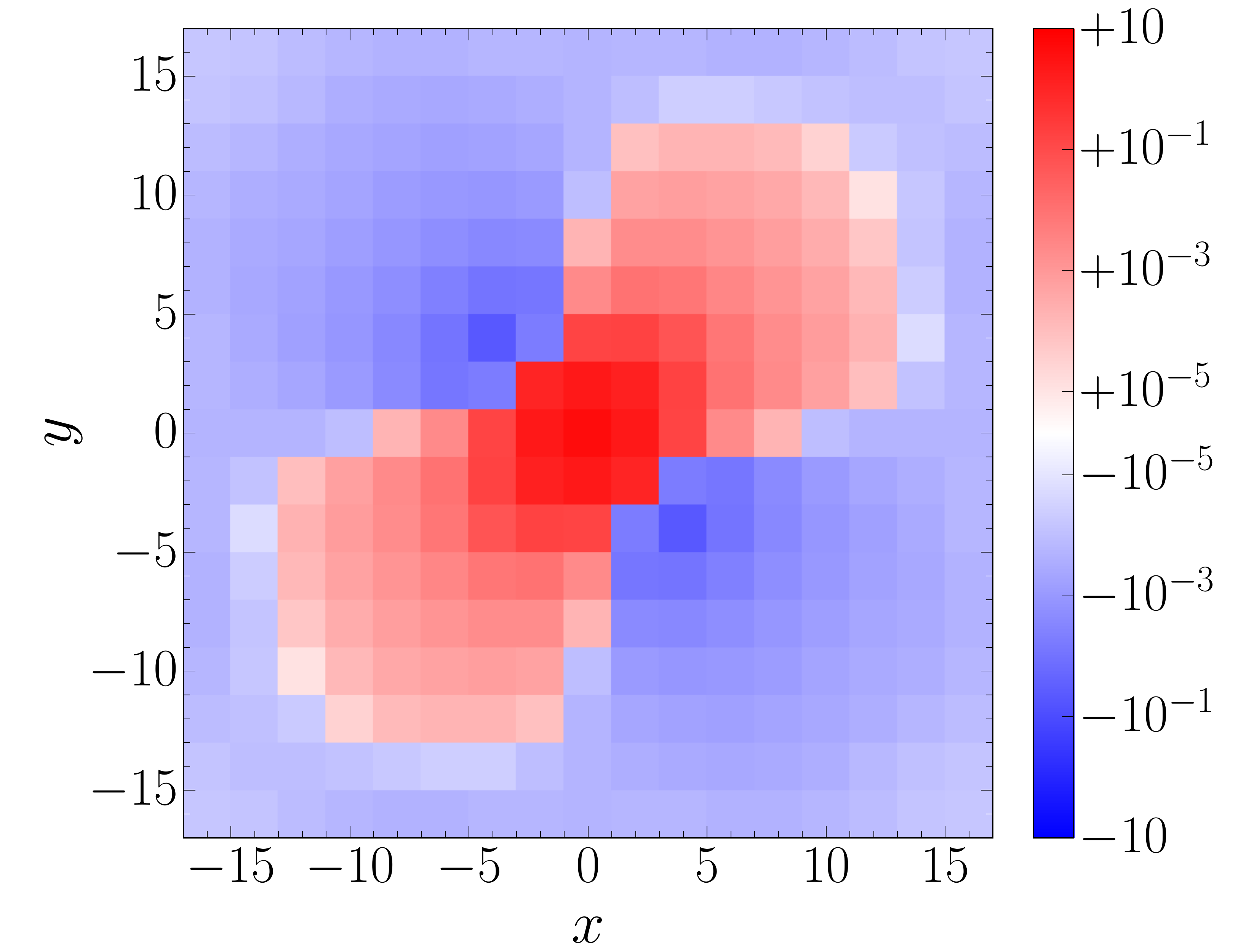}
\caption{Correlations of the staircase cell operator defined on $5\times 5$ spins (illustrated in Fig.~\ref{fig:model}) in the full space of distances
$x,y \in \{-L/2,\ldots,L/2\}$ (even values only) for the critical $J$-$Q_2$ model with $L=32$. Shades of red and blue correspond to positive and negative values,
respectively, as indicated by the color bar.}
\label{fig:corr2D}
\end{figure}

Let us now look at the $5\times 5$ cell correlations in the full space of distances $\mathbf{r} = (x,y)$. In Fig.~\ref{fig:scaledim} results were shown at the
extremal distances $\mathbf{r} = (L/2,0)$ and $\mathbf{r}=(L/2,-L/2)$, the latter being equivalent to $\mathbf{r}=(L/2,L/2)$ on the periodic lattices. The
correlation function is negative at all these distances. There is, however, a strong directional dependence, including sign changes, of the correlations away
from the longest distances in the respective directions. Fig.~\ref{fig:corr2D} shows results at all separations $(x,y)$ with even $x$ and $y$ for the $L=32$
system. Here we can see that the correlations along the $(1,1)$ direction are positive at short distances but change signs as the longest distance is approached. The
absolute values being very small and dominated by statistical errors at long distances for large systems, we do not have data of sufficient quality to examine the
sign changes systematically, but the behavior shown in Fig.~\ref{fig:corr2D} is qualitatively the same for all the systems we have studied.

We also point out here that the cell-cell correlation functions should formally be considered only at the distances corresponding to the tiling of the lattice
as in Eq.~(\ref{hhdef}), which in this case means that $x$ and $y$ should both be multiples of $4$. However, we find that the correlations at all $\mathbf{r}=(x,y)$
with even values of $x$ and $y$ form a common smooth function, and we therefore use all these values in Fig.~\ref{fig:corr2D}. For odd $x$ and $y$ the signs switch
relative to their neighboring points with even $x,y$ and we have not included these distances in our analysis.

Given the two large regions of different signs of the correlation function in Fig.~\ref{fig:corr2D}, it is important to study also an integrated correlation function,
corresponding explicitly to the scaling dimension of the entire operator sum $W$ perturbing the critical Hamiltonian in Eq.~(\ref{hhdef}). From
Fig.~\ref{fig:corr2D} it appears clear that the negative weights will dominate at long distances, and, therefore, the correlations studied so far should
capture the behavior also of the integrated correlation function. For completeness, we nevertheless also analyze two kinds of integrated correlators.

\begin{figure}[t]
\includegraphics[width=70mm, clip]{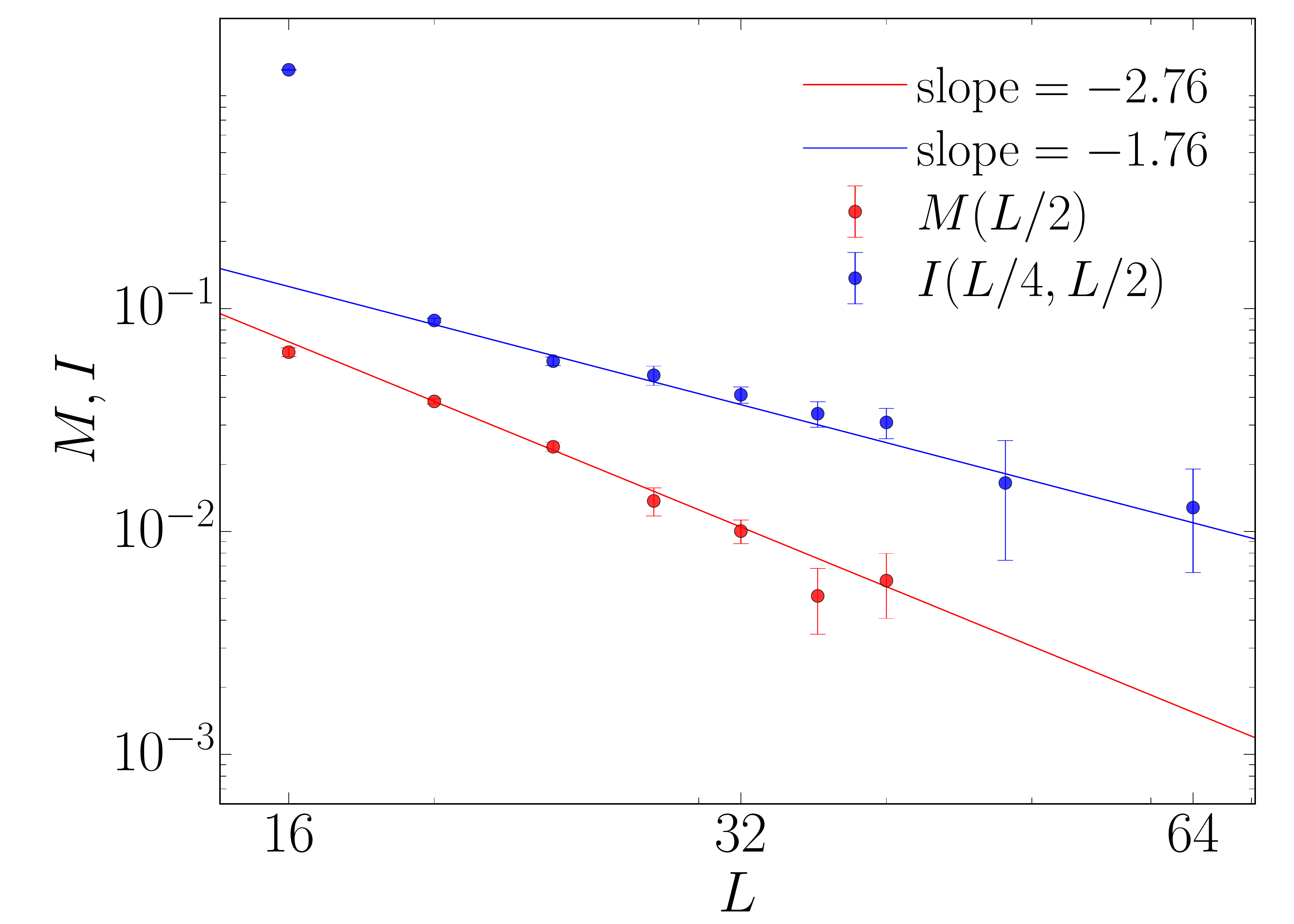}
\caption{The critical $5\times 5$ staircase correlation function integrated over the largest square frame, $M(R)$ defined in Eq.~(\ref{mrdef}), with
$R=L/2$, and the integrated correlation function over all distances starting from $r=L/4$, $I(L/4,L/2)$ defined in Eq.~(\ref{ildef}). The lines have the expected
slope (with $\Delta_W = 1.90$ determined in Fig.~\ref{fig:scaledim}) $1-2\Delta_W$ (red points) and $2-2\Delta_W$ (blue points).}.
\label{fig:MI}
\end{figure}

First, consider the distance space $x,y \in \{-L/2,\ldots,L/2\}$ of the correlations as in Fig.~\ref{fig:corr2D} and define a square ``frame'' of width
one, located at distance $R$ from the center (this  distance being defined in the $x$ or $y$ direction). Denoting the lattice sites belonging to such a frame
by ${\rm F}_R$, we define $M(R)$ as the sum of all correlations in the frame:
\begin{equation}
M(R)=\sum_{\mathbf{r} \in {\rm F}_R} C_W(\mathbf{r}).
\label{mrdef}
\end{equation}
In Fig.~\ref{fig:MI} we show results at the maximum distance $R=L/2$ (where equivalent distances are only counted once). Since the edge grows
as $L$, we expect $M(L/2) \propto L^{1-2\Delta_W}$, which indeed is satisfied with the value of $\Delta_W$ determined before.

The fully integrated correlator is difficult to analyze, because it is dominated by short-distance contributions, as seen clearly as the dark red region
at the center of Fig.~(\ref{fig:corr2D}). To circumvent this problem, we sum the frame quantities $M(R)$ only from $R=L/4$ to $L/2$, defining the quantity
\begin{equation}
I(L/4,L/2)=\sum_{R=L/4}^{L/2} M(R),
\label{ildef}
\end{equation}
for which we expect $I(L/4,L/2) \propto L^{2-2\Delta_W}$ asymptotically. As shown in Fig.~\ref{fig:MI}, this behavior is indeed realized, and, because
of the larger number of terms, the relative statistical errors are much smaller than in $M(L/2)$.

\subsection{5. Visualization of HVB order}

Here we discuss the color scheme we have used in illustrations such as Fig.~\ref{fig:colorful} of the local
VBS angle and amplitude. We also give some other examples, including a case with two spinons and their confining string.

\subsubsection{5-A. Color coding scheme for local VBS order}

In order to properly visualize typical QMC configurations (more precisely averages over short simulation segments) in the HVB phase, as shown in the color bar
in Fig. \ref{fig:colorful}, we associate the color hue with the angle $\phi \in [0, 2\pi)$. To define the angle, we have to perform coarse graining over some small
region of the lattice. We use the local VBS order parameter defined in Eq.~(\ref{txydef}), for each lattice link combining two overlapping neighboring $T_x$ terms
and two $T_y$ terms placed within a $4 \times 3$ site cell (if the targeted central link is $x$-oriented) or a $3 \times 4$ cell (for an $y$-oriented central link). 
The values $(\bar T_x,\bar T_y)$ are then used to extract the local VBS angle for the link in question. To represent these angles in the RGB scheme, distinct colors
red, blue, green, and yellow are assigned to the angles $\phi=0, \pi/2, \pi,$ and $3\pi/2$ corresponding to the four columnar VBS patterns. Angles in between
those values (which correspond to plaquette VBS order) were also made to have relatively easily distinguishable colors; purple, orange, yellow-green,
and cyan.  This design of the map can be clearly seen in the color bar of Fig.~\ref{fig:colorful}. Technically, the color code is a function
(the details of which are unimportant) $c : [0, 2\pi)$ $\rightarrow$ $\{0, 1, 2, \ldots, 255\}^3$ from the angle $\phi$ to an RGB value. 

We furthermore adjust the brightness of the colors for all the links to reflect how strongly correlated the two connected spins are, i.e., we use the value
of $-\langle S^z_iS^z_j\rangle$, which provides a better microscopic representation of the bond pattern than the coarse-grained amplitude extracted
from $(\bar T_x,\bar T_y)$. To do this systematically, we first collect the strongest and weakest bond correlation values and then rescale them so that their relative
strength $x$ varies from $0$ (weakest) to $1$ (strongest).  Now, we introduce a brightness function $f(x): [0,1]\rightarrow[0,1]$ and use this as rescaling factor
for the RGB values of each link. Thus, the final color code can be described as $f(x)c(\phi)$ (rounded up to integer RGB values).

\begin{figure}[t]
\includegraphics[width=80mm]{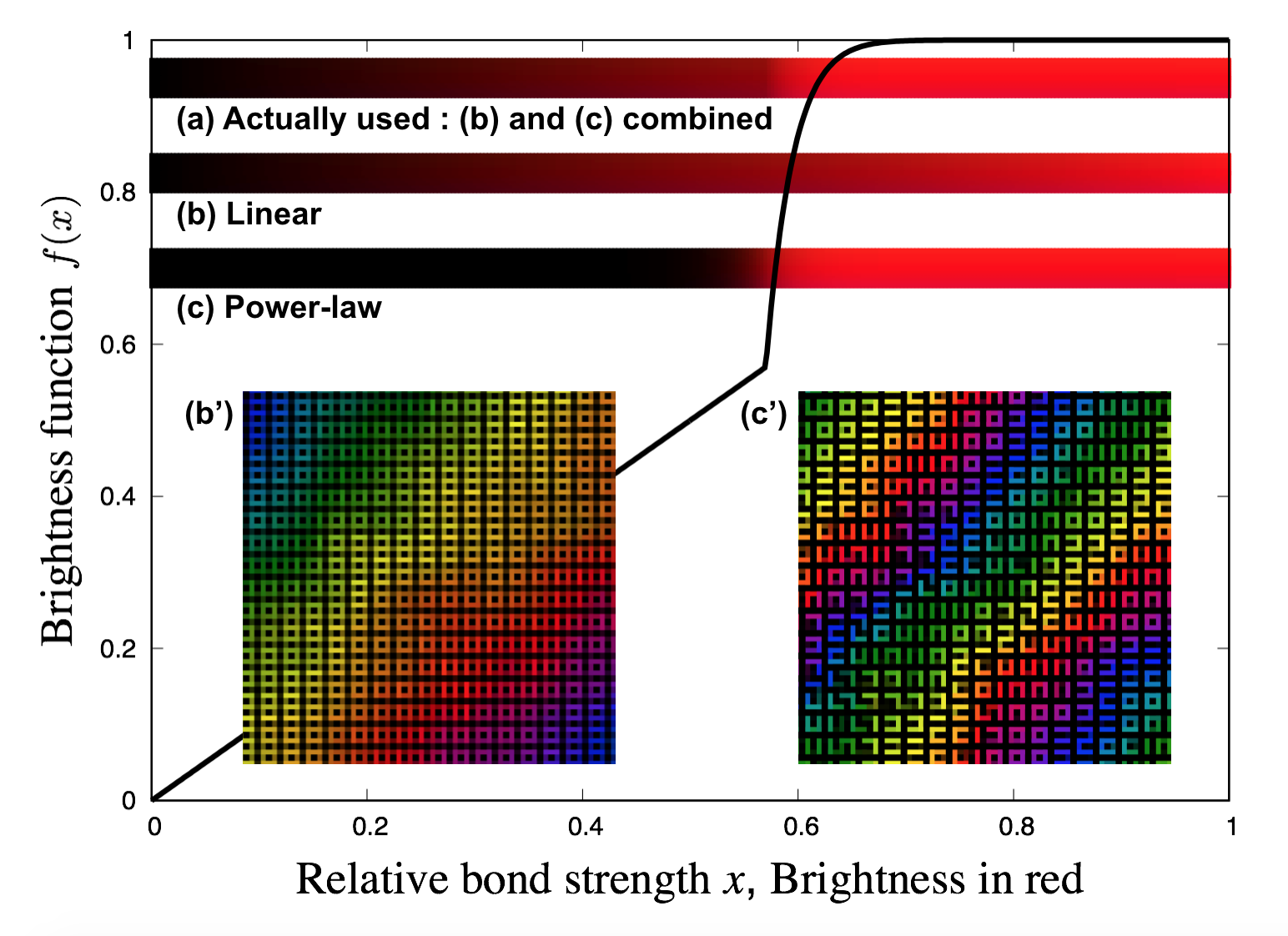}
\caption{Illustration of the brightness function $f$ used in Fig.~\ref{fig:colorful} along with two other options.
The red color is used to show (a) the nonlinear brightness scale in Fig.~\ref{fig:colorful}(a) as a function of relative
strength $x$, (b) the linear form $f(x)=x$, and (c) the power-law cut-off function (\ref{eq:cutoff}) with $x_c=0.57$ and $p=17$. The visualized configurations
marked (b') and (c') were constructed with $f(x)$ as in (b) and (c), respectively, and are from the same data as in Fig.~\ref{fig:colorful}.}
\label{fig:Brightness}
\end{figure}

The simplest brightness function is the linear identity function $f(x)=x$, but as we show in Fig.~\ref{fig:Brightness}(b') for the same QMC
configuration as in Fig.~\ref{fig:colorful}(a), this results in a rather pale picture  where the VBS patterns are somewhat blurred out.  Another extreme would
be to have a cut-off value $x_c$ and then set
\begin{eqnarray}\label{eq:cutoff}
f(x)=
\left\{
    \begin{array}{l}
      \frac{x^p}{x_c^{p-1}}           ~~~\hspace{1.5cm}(x\leq x_c)\\
        1-\frac{(1-x)^p}{(1-x_c)^{p-1}}       \hspace{0.54cm}(x\geq x_c).
    \end{array}
  \right. 
\end{eqnarray}
This is a step function at $x_c$ in the $p\rightarrow\infty$ limit and a power-law relaxation of a hard cut-off for finite $p$. In
Fig.~\ref{fig:Brightness}(c'), we show the same HVB configuration as in Fig.~\ref{fig:colorful}(b) scaled according to Eq.~(\ref{eq:cutoff}) with $p=17$
and $x_c=0.57$. While the picture emphasizes the pattern very clearly, it also extinguishes many bonds, sometimes resulting in apparent ``holes"
in the figure. 

For Fig.~\ref{fig:colorful}, we combined the above two extremes so that the strong bonds are brightly shown while weak bonds remain to some extent. 
More precisely, we have replaced the first half of Eq.~(\ref{eq:cutoff}) with the identity $f(x)=x$, or equivalently, used $p=1$
and $p=17$, for $x\leq x_c$ and $x\geq x_c$, respectively, in Eq.~(\ref{eq:cutoff}), as illustrated with the curve drawn in Fig.~\ref{fig:Brightness}.

\begin{figure}[t]
\includegraphics[width=65mm]{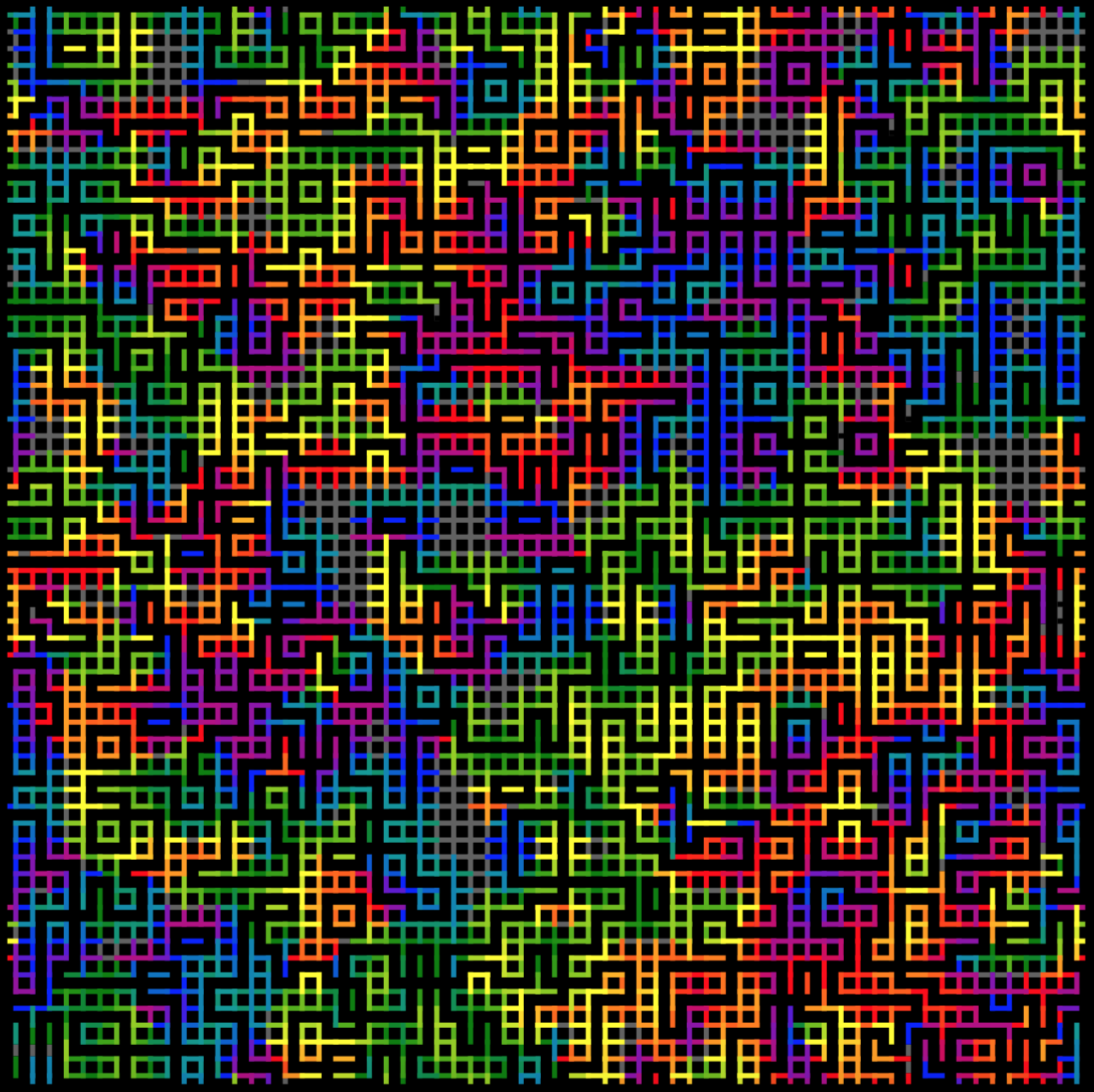}
\caption{Visualization of a single $z$-basis measurement of the HVB phase at $h=1$, $g=0.25$ for an $L=64$ system. The probability of having exactly
zero amplitude of the $T$ order parameter is relatively large when no averaging is performed. In such cases the local angle $\phi$ is ill-defined,
and we associate gray bonds to those.}
\label{fig:singleshot}
\end{figure}

The visualizations we present are constructed from the values of $\langle S^z_i S^z_j \rangle$ averaged over several thousands of updating sweeps in
SSE simulations---$2\times 10^4$ in the case of Fig.~\ref{fig:colorful}. There is of course a strong dependence here on exactly how much averaging is performed. 
A true snapshot of a $z$-basis configuration corresponds to $(\bar T_x,\bar T_y)$ evaluated in a single spin state (at one ``time slice''), but such a picture
looks too noisy to bring out the helical structure clearly (though some hints of the striped structure can often still be discerned). We show an example of such a
snapshot in Fig.~\ref{fig:singleshot}. In the other extreme, if we sample a very large number of statistically independent configurations (as in precise calculations
of expectation values), the resulting picture would be uniform (with the staircase pattern visible when $h>0$ in our model) due to the translational
invariance of the lattice. 

The fact that several thousands of QMC sweeps yields a reasonably sharp picture of helical order reflects the diverging QMC relaxation time of the HVB
phase as $L$ grows. Such averaged configurations should not be interpreted as something directly physically observable by a single projective measurement,
but rather atheoretical/numerical object for our intuitive and visual understanding of the HVB phase. 

\subsubsection{5-B. Additional visualizations}

        \begin{figure}[t]
                \centering
                \includegraphics[width=60mm]{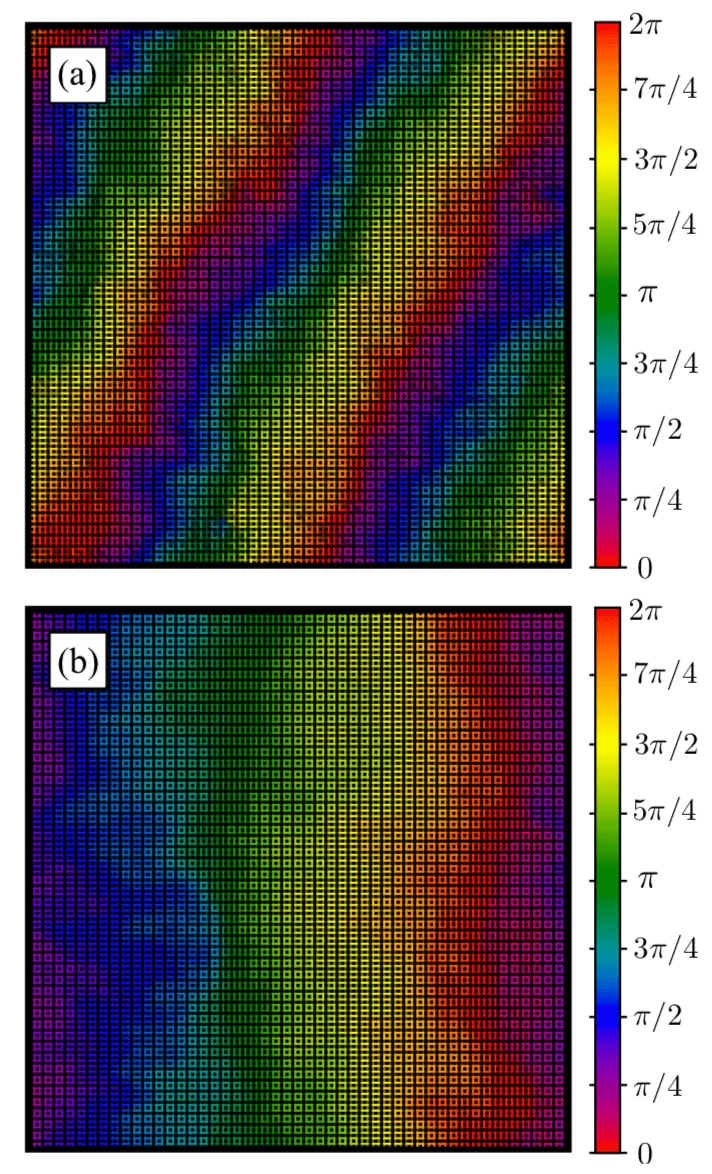}
                \caption{Visualization of HVB configurations with winding $w_x \not= w_y$.
                (a) Winding $(1,0)$ at $h=0.5$ and $g=0.35$, and (b) (1,2) at $h=0.25$ and $g=0.25$.}
                \label{fig:non45}
        \end{figure}

Here we present several interesting configurations to further illustrate the nature of the HVB phase. For the visualizations shown here
we averaged over $5\times 10^3$ SSE sweeps and used the coding of the dimer amplitude in Fig.~\ref{fig:Brightness}(a).

As we discussed briefly in the main paper, the HVB phase mostly comprises states with winding $w_x=w_y$, but $w_x \not = w_y$ states appear in the regions where
the diagonal winding numbers change due to avoided level crossings. In Fig.~\ref{fig:non45} we show examples of configurations with winding $(1,0)$
and $(2,1)$. The winding number can be regarded as conserved (i.e., the life time of the winding number is exceedingly long) in systems sufficiently inside
a segment of the HVB phase with $w_x=w_y$ (see also Ref.~\onlinecite{Shao15} for a demonstration of this for excited domain-wall states in the standard $J$-$Q$
model with $h=0$). At the boundary between two $w_x=w_y$ winding sectors, the winding numbers are not conserved, and the system can tunnel between different
sectors (though the characteristic tunneling time can still be very long).

\begin{figure}[b]
\includegraphics[width=60mm]{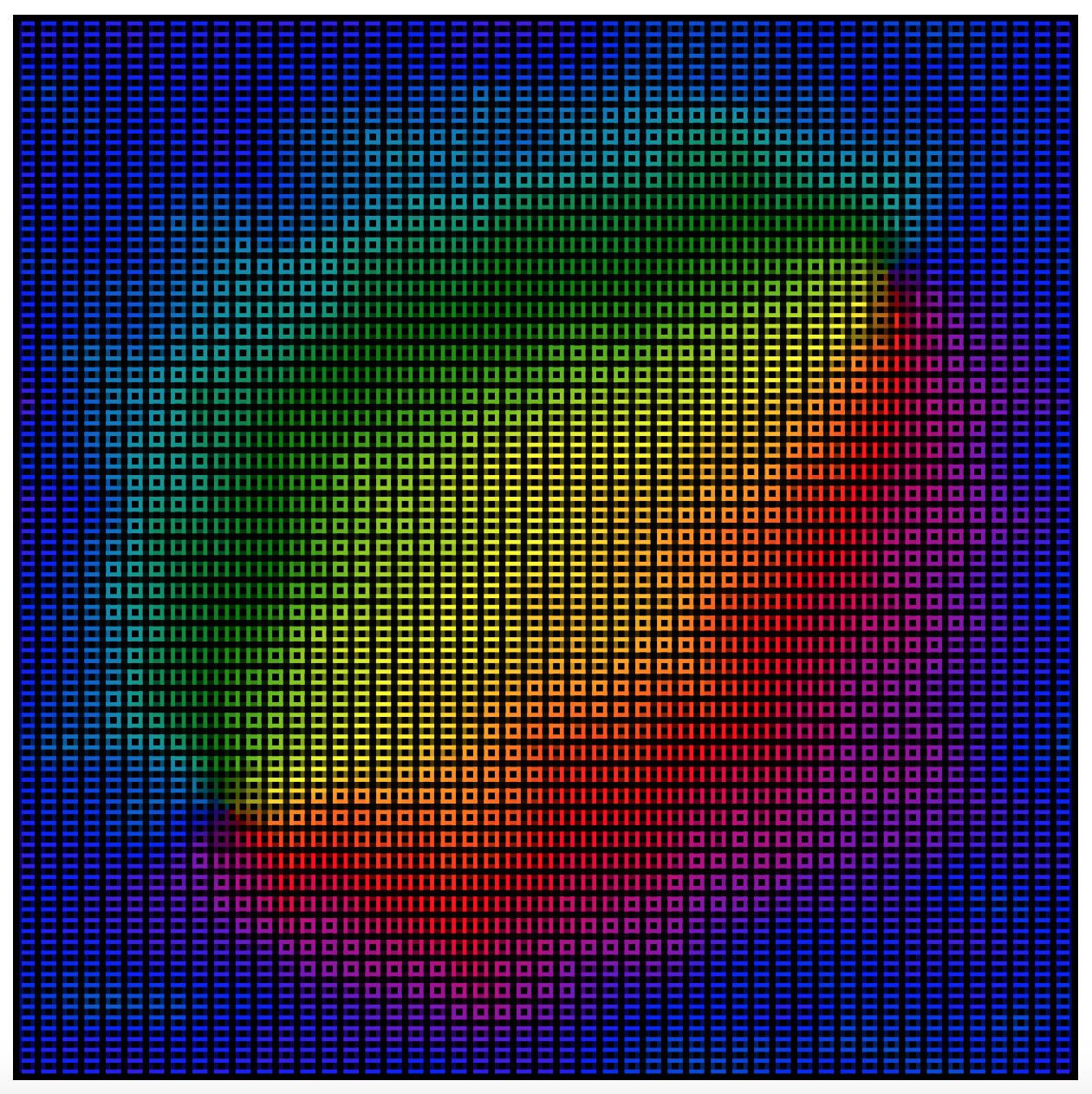}
\caption{An $L=96$ two-spinon configuration in the HVB phase at $h=1$ and $g=0.095$, in the transitional region between the VBS and the $(1,1)$ HVB.
The color scale is the same as in Fig.~\ref{fig:non45}.}
\label{fig:2spinons}
\end{figure}

\begin{figure*}[t]
\includegraphics[width=180mm, clip]{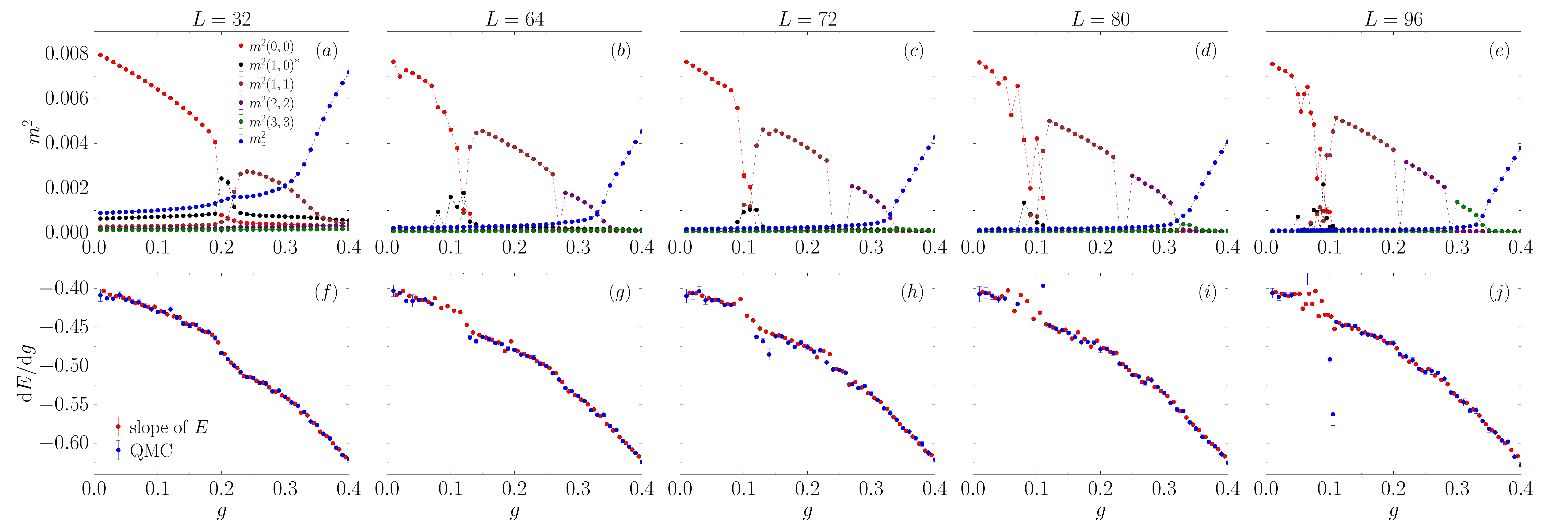}
\caption{(a)-(e) Squared order parameters for several system sizes versus $g$. (f)-(j) Energy derivatives computed in two different ways---by a direct QMC
estimator and by numerical derivatives (local slope computed with two adjacent $E$ values) of the computed energies on the discrete $g$ grid---for the same set
of system sizes. The large fluctuations in some cases are due to metastability and long tunneling times between winding sectors in regions of the avoided level
crossings.}
\label{fig:m2L}
\end{figure*}

The tunneling mechanism involves a ``vacuum fluctuation'', where two defects are created and separate from each other [see Fig.~\ref{fig:2spinons}].
At such defects, the four different
VBS patterns meet---in practice in a rather continuous way where the VBS angle changes by $\pm 2\pi$ along a path encircling the defect. The defect pair, a vortex
and an anti-vortex, in effect experiences a binding potential in the form of a string carrying winding in the region between the vortex and anti-vortex. In
the DQCP scenario, the vortex cores are associated with unpaired $S=1/2$ spin degrees of freedom. Such a spin along with its dimer vortex is referred
to as a spinon. Spinons have previously been visualized in disordered $J$-$Q$ models, where random couplings induce nucleation of pairs of localized
vortices \cite{Liu18}. In the system considered here, the spinons are mobile and able to wrap around the periodic boundaries and annihilate each other, leaving
behind the winding of the string and, thus, changing the winding number.

In the standard $h=0$ $J$-$Q$ model without disorder, it is difficult to observe the spinons directly in the ground state, because the bound pair is typically
small deep inside the VBS phase and when approaching the DQCP the spinons themselves grow and become difficult to distinguish from the fluctuating VBS background.
Some information has been gleaned from just studying the spin degrees of freedom in excited $S=1$ states \cite{Tang13}. In the HVB phase, in the regions where
several winding numbers coexist and fluctuate among each other, we have found configurations with clearly visible spinon pairs. An example is shown in
Fig.~\ref{fig:2spinons}. The two spinons are here separated approximately at the farthest distance on the torus, making the configuration very stable, surviving
for more than $10^5$ SSE updating sweeps. It appears that the ground state of this system may even be dominated by configurations with two spinons, which can
be regarded as transitional states between winding number $(0,0)$ and either $(1,0)$, $(0,1)$, or $(1,1)$ (or, in principle, even higher winding).

\subsection{6. Size dependent order parameters and phase boundaries}

Here we present additional results for the order parameters computed with different lattice sizes and $h$ values, complementing the results in Fig.~\ref{fig:OrderPara}
in the main paper. We also discuss how we have located the phase boundaries, and some aspects of finite-size scaling governed by the dynamic exponent.

\subsubsection{6-A. Order parameters}

In the main paper, in Fig.~\ref{fig:OrderPara} we showed results for several HVB winding order parameters as well as the AFM order parameter for lattice
size $L=96$. To show the dependence of these quantities on the system size, in Figs.~\ref{fig:m2L}(a)-(e) we show results also for several smaller
systems, and for $L=96$ we have included more points in the neighborhood of the VBS--HVB transition. Here we can see clearly how increasing $L$ can
accomodate gradually higher winding numbers. We also see consistently that the degenerate $(1,0)$ and $(0,1)$ winding states follow directly after the
conventional VBS order with $(0,0)$ winding, before the $(1,1)$ states. All other order parameters shown in Fig.~\ref{fig:m2L} are for $w_x=w_y$, but, as we
discussed in the preceding section, winding with $w_x \not= w_y$ appears in the transitional states between the different $w_x=w_y$ sectors in
the narrow regions where all the order parameters graphed in Figs.~\ref{fig:m2L}(a)-(e) are very close to zero. Such transitional winding states,
where the winding is not fully conserved, are similar to those with $(1,0)$ and $(0,1)$ winding between the $(0,0)$ and $(1,1)$ states.

\begin{figure}[t]
\includegraphics[width=70mm, clip]{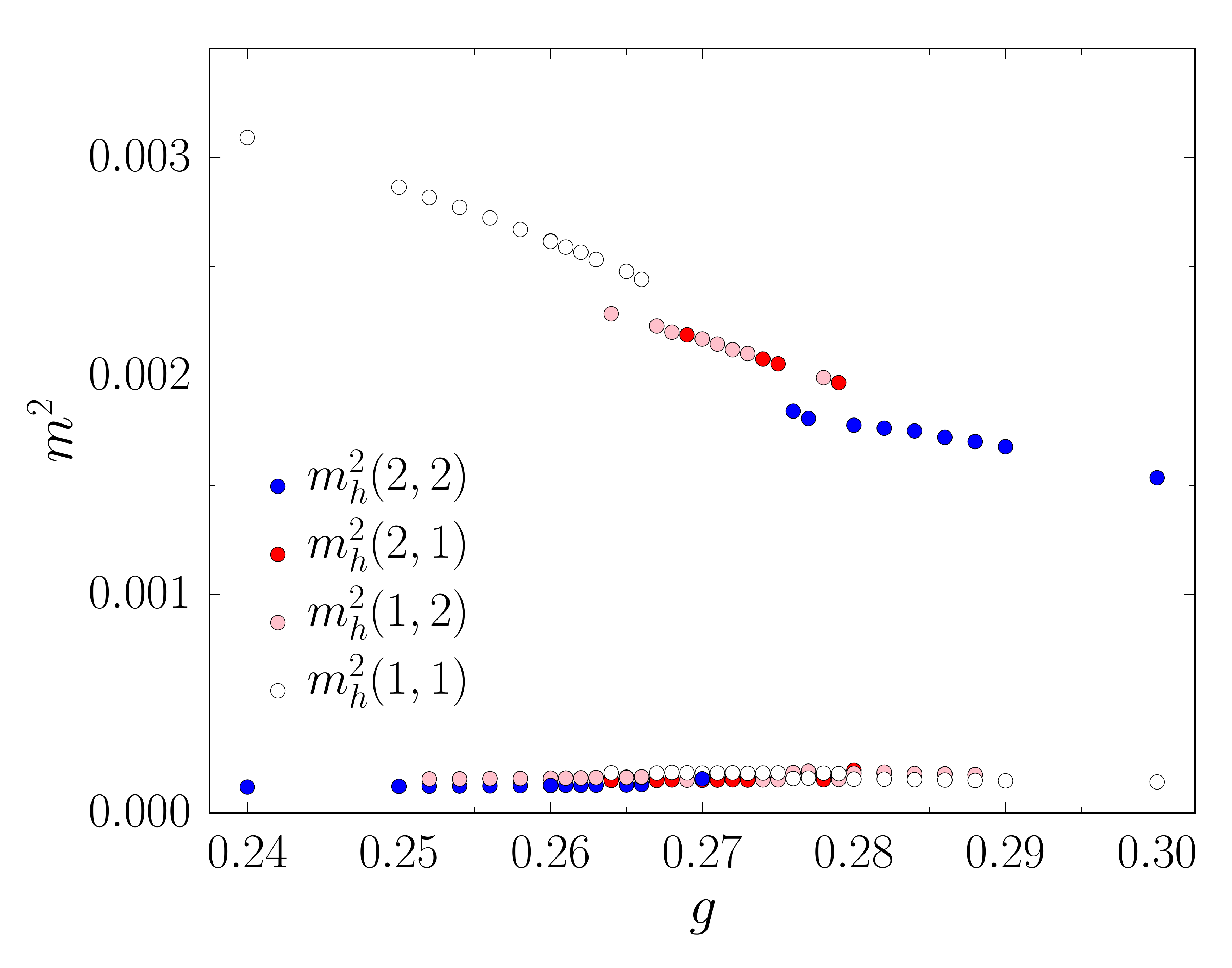}
\vskip-1mm
\caption{HVB order parameters for an $L=64$ system at $h=1$ [as in Fig.~\ref{fig:m2L}(b)] in the region where the winding number changes from $(1,1)$ at the smaller
$g$ values to $(2,2)$ for the larger $g$ values. The change of winding is mediated by spinon pairs (illustrated in Fig.~\ref{fig:2spinons}), whose creation and
subsequent destruction after wrapping around the periodic boundaries can change one of the winding numbers by $\pm 1$. This leads to a range of $g$ values for which
the stable winding sectors are $(2,1)$ and $(1,2)$ in this case. The different points in this graph were obtained in independent simulations, and the long-lived
metastability in transition regions between different winding sectors is reflected by simulations between $g \approx 0.26$ and $g \approx 0.28$ some times
being trapped in the wrong sector.}
\label{fig:meta}
\end{figure}

In Figs.~\ref{fig:m2L}(a)-(e) we can also observe how the overall smooth curves for $L=32$ become increasingly jagged close to the points between
different winding sectors, reflecting the first-order nature (avoided level crossings) of these ``micro transitions'' within the HVB phase. These level crossings
are associated with difficulties of the simulations to sample to ground state with co-existing winding numbers. The winding numbers are very stable in the simulations,
and simulations of reasonable length may be trapped in a metastable sector.

As mentioned, ``non-diagonal'' winding numbers $(n,n+1)$ and $(n+1,n)$ always appear in the transition regions between winding sectors $(n,n)$. and $(n+1,n+1)$.
Examples of metastability and non-diagonal winding are observed in Fig.~\ref{fig:meta}, which is based on a large number of independent simulations in the region
where the winding number of an $L=64$ system changes from $(1,1)$ to $(2,2)$. Sufficiently far from the transition region between $g\approx 0.26$ and
$g \approx 0.28$, all simulations consistenly converge to their stable $(1,1)$ or $(2,2)$ sectors, and at the center of the region around $g \approx 0.27$ the
stable sector is always $(1,2)$ or $(2,1)$. However, in the transition regions we observe some instances of metastability, and it is very difficult to determine
exactly the extent of the different winding sectors. This behavior is typical for systems with avoided level crossings between ground states that have the same
exact quantum numbers; here (by general theorems for bipartite spin Hamiltonians) the lattice momentum $k=0$, spin $S=0$, and all those related to lattice
reflection and rotation are $+1$ (i.e., fully symmetric states). The winding number is likely an emergent quantum number, becoming strictly conserved in the limit
of infinite size, as was confirmed in the case of winding number sectors inside the VBS state in Ref.~\cite{Shao15}. In principle, the winding number may not be
fully conserved if there are sectors of the HVB phase where the helical order is ``floating'' and not completely long-ranged (as discussed at the end of
the main paper).

We should also point out here that running a single simulation for each value of $g$, as done in Fig.~\ref{fig:meta}, is the way that most clearly shows the
presence of one true ground state and one (in practice reachable) metastable state. All other results reported in this paper were generated by running 
typically tens of independent simulations for each parameter value and averaging computed quantities of over all those runs. The metastability is then still 
reflected in very large fluctuations, but not just between two different values corresponding to the two quasi-degenerate states, but from averaging over 
a distribution over those two values. The distribution corresponds to how often a simulationsreaches the ground state versus staying trapped in the metastable 
state. Clear examples of such fluctuating mixed averages between two sectors are seen in Fig.~\ref{fig:OrderPara} in the main paper in the region 
$g \approx 0.05$-$0.12$ between the ${\bf w}= (0,0)$ and ${\bf w}= (1,1)$ sectors.

\begin{figure}[t]
\includegraphics[width=60mm, clip]{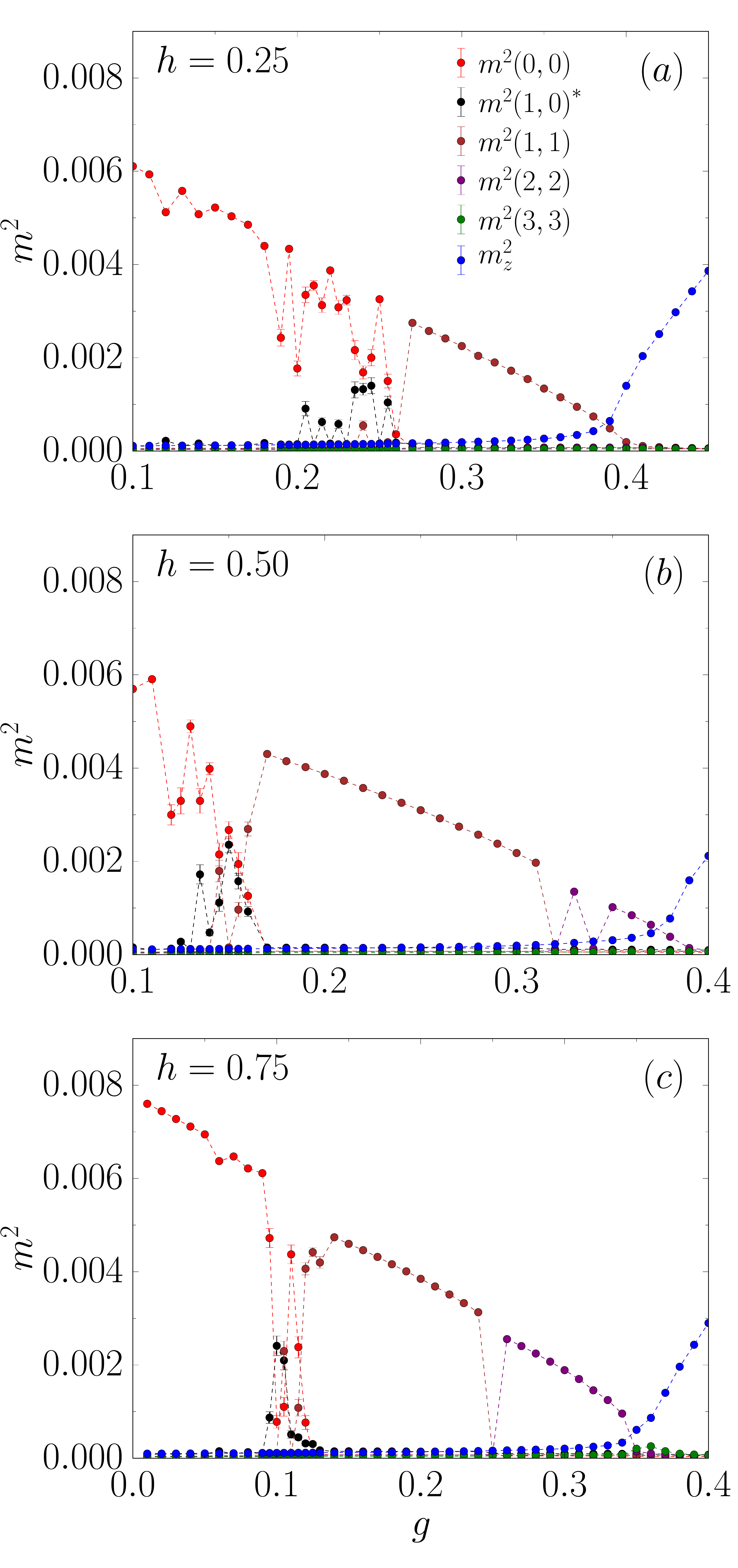}
\caption{Squared order parameters vs $g$ for systems with $h=0.25$, $0.5$, and $0.75$. The system size if $L=96$, and the corresponding results for $h=1$ are
shown in Fig.~\ref{fig:m2L}(e).}
\label{fig:m296}
\end{figure}

The avoided level crossings can be further confirmed by examining the derivative $dE/dg$ of the ground state energy, which should exhibit finite-size rounded
jumps. Figs.~\ref{fig:m2L}(f)-(j) show derivatives for the same system sizes as in the order parameter graphs Figs.~\ref{fig:m2L}(a)-(e). Rapid changes in
the derivatives can be observed in the region where the $(0,0)$ winding VBS states undergo transitions into the HVB phase, and some of
the other micro transitions within the HVB phase also have noticeable changes in the derivatives. For the larger systems the data are too noisy to detect unambiguous
shifts, however, which again relates to difficulties in sampling the system close to avoided level crossings.

Next, in Fig.~\ref{fig:m296} we present results for different values of the staircase strength $h$. We only show results for $L=96$, as the smaller
sizes exhibit systematic changes very similar to the $h=1$ results in Fig.~\ref{fig:m2L}. Here the results follow the expectation that larger $h$
induces higher winding. While at $h=1$ the highest winding realized in the $L=96$ system is $w_x=w_y=3$, at $h=0.75$ and $0.5$ the maximum value is $2$,
and at $h=0.25$ it is $1$. For all values of $h$, we observe $(1,0)$ and $(0,1)$ winding between $(0,0)$ and $(1,1)$. HVB order with non-diagonal winding
is also observed at between all other $w_x=w_y$ sectors, but those order parameters are not shown in Fig.~\ref{fig:m296}.

\begin{figure}[t]
\includegraphics[width=80mm, clip]{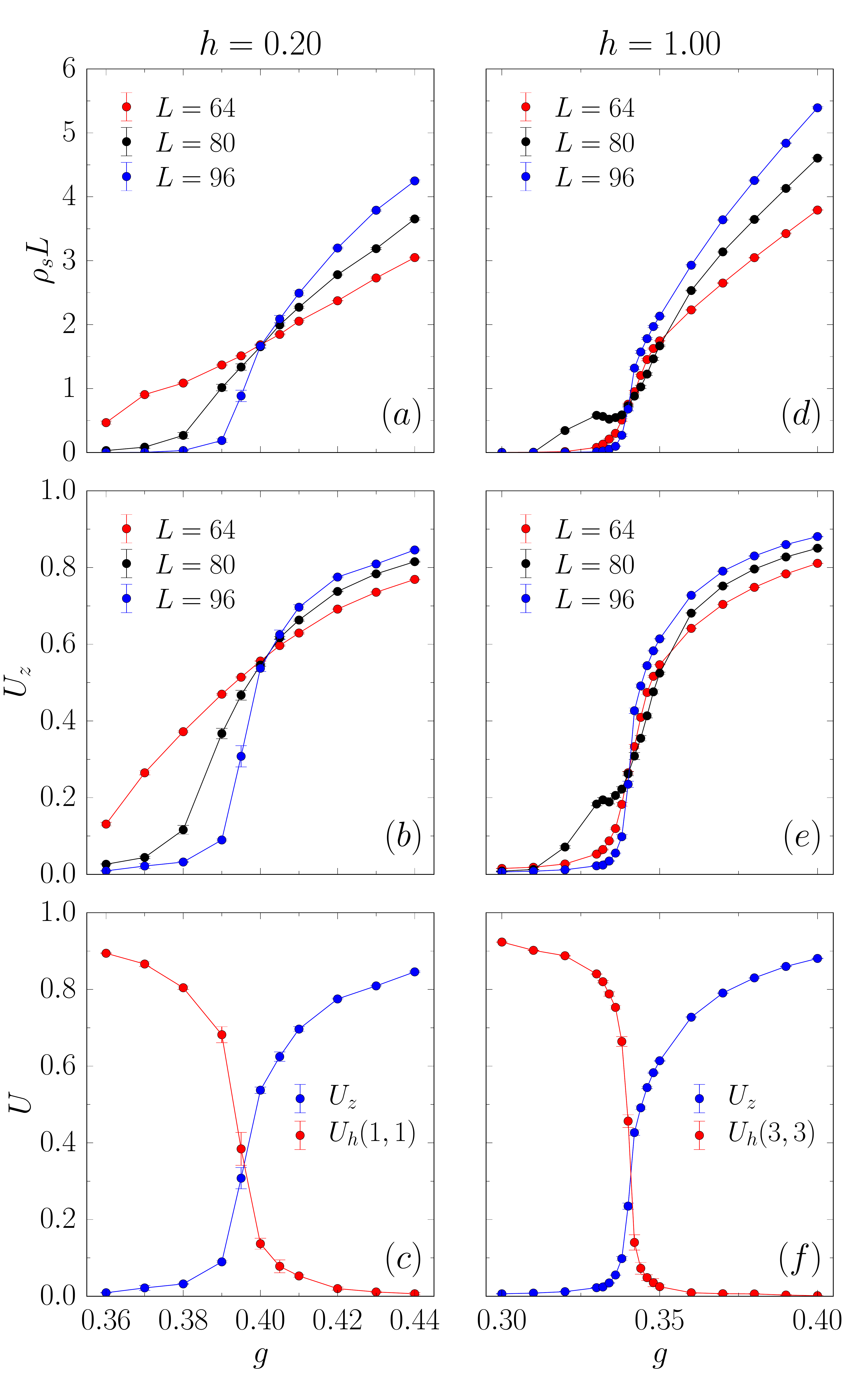}
\vskip-1mm
\caption{Quantities used to analyze the HVB--AFM transition by curve-crossing techniques versus $g$ exemplified by data for $h=0.2$ (left column)
 and $h=1$ (right column). (a) The spin stiffness multiplied by $L$ for system sizes $L=64,80,96$. (b) The AFM Binder cumulant for the same
 systems. (c) The AFM Binder cumulant and the cumulant of the HVB order parameter at winding number $(3,3)$ for $L=96$. The dashed lines show
 the estimated critical point based on the curve crossings.}
\label{fig:afmcross}
\end{figure}

\subsubsection{6-B. Phase diagram}

To construct the phase diagram of the model, we first need proper finite-size estimates of the VBS--HVB and HVB--AFM transition points for different $h$ values.
In Fig.~\ref{fig:afmcross} we show examples of quantities often used to locate AFM transitions. We show results for both the smallest and largest values of $h$
considered; $h=0.2$ in panels (a-c) and $h=1$ in (d-f).

The spin stiffness $\rho_s$ is shown for three different system sizes in in Figs.~\ref{fig:afmcross}(a,d). In the spatially anisotropic staircase system, there
are two independent stiffness constants, and different combinations of them can be defined. We here present results for the stiffness corresponding to a phase
twist along the diagonal $(x,x)$ direction. The spin stiffness is expected to scale as $L^{-z}$ at a quantum critical point, where $z$ is the dynamic exponent.
At a DQCP we should have $z=1$, and in Figs.~\ref{fig:afmcross}(a,d) we have therefore multiplied the QMC results for $\rho_s$ by $L$ in order for the curves for
different $L$ to asymptotically cross each other at the putative critical point. We indeed observe a well defined crossing point for all three system sizes both at
$h=0.2$ and $h=1$.

In the case of the $L=80$ data in Fig.~\ref{fig:afmcross}(d), a shoulder feature is visible that can be traced to a change in the dominant
winding number at $g$ only slightly below the transition point, as can be seen in Fig.~\ref{fig:m2L}(d) as a small (barely visible) maximum of the $m^2(3,3)$
HVB order parameter. In contrast, for $L=64$ and $L=96$ the order parameter of the winding sector at the phase transition is more substantially developed.
We find these non-monotonic features in $\rho_s$ for some system sizes for all the $h$ values studied, but in most cases their presence does not affect the
common crossing behavior of the $\rho_s L$ curves at the HVB--AFM transition. In some cases we do observe outlier points caused by
this winding effect.

The Binder cumulant $U_z$ (of the $z$ component of the sublattice magnetization) is shown in Figs.~\ref{fig:afmcross}(b,e). The cumulant is already
dimensionless and does not require any further scaling. We observe crossing points at the same value of $g$ as in the case of $\rho_s L$, and also
in this case the $L=80$ curve exhibits a small bump related to the winding number change.

\begin{figure}[t]
\includegraphics[width=65mm, clip]{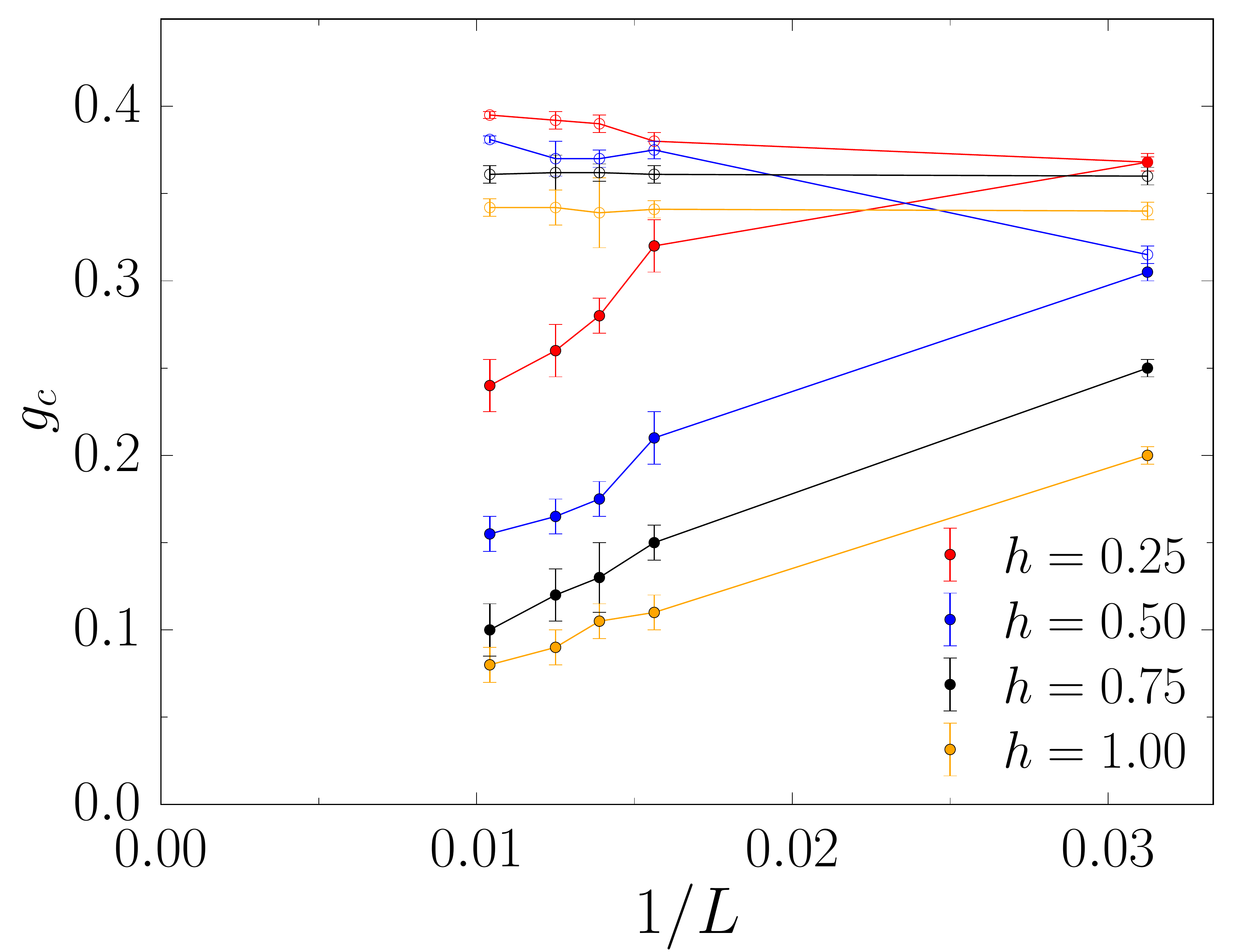}
\caption{Estimated VBS--HVB and HVB--AFM transition points vs the inverse system size for all the values of $h$ studied. The results are estimated based on data
such as those in Figs.~\ref{fig:m2L} and \ref{fig:m296}, as explained in more detail in the text.}
\label{fig:boundaries}
\end{figure}

In Figs.~\ref{fig:afmcross}(c,f) we show two Binder cumulants for the $L=96$ systems; in addition to the AFM cumulant $U_z$ we also consider the cumulant
of the HVB order parameter in the highest winding sector that is realized at the transition into the AFM state. The crossing point between the VBS and AFM
cumulants has previously been shown to perform well as an $L$ dependent quantity to locate the DQCP \cite{Zhao20},
and we use this method here with the HVB and AFM cumulants to define finite-size
HVB--AFM transition points. For $h=1$, Fig.~\ref{fig:afmcross}(f), the crossing point agrees very well with the $\rho_sL$ and $U_z$ crossings in
Figs.~\ref{fig:afmcross}(d,e), while for $h=0.2$ in Fig.~\ref{fig:afmcross}(c) the latter crossings take place at marginally larger $g$ values.
We also in general find somewhat more size dependence in the crossings of the same-size $(U_z,U_h)$ crossings than in the other crossing points discussed
here, defined, e.g., using system-size pairs $(L/2,L)$. The convergence of the single-size crossing points is still also rapid. Though we do not have data
on a very dense grid of $g$ values and only perform linear interpolations, we can still extract the crossing points to sufficient precision for our purposes
here. We show results for crossing points versus the inverse system size in Fig.~\ref{fig:boundaries} for several $h$ values.

In the case of the VBS--HVB transition, the avoided level crossings and quantum superposition of different winding numbers in the transition region cause problems
with metastability in the simulations, as seen in the large fluctuations in some of the data in Figs.~\ref{fig:m2L} and \ref{fig:m296}. As we have pointed out,
the $(1,0)$ and $(0,1)$ winding states appear at the transition between the $(0,0)$ and $(1,1)$ sectors. We therefore use the window where this order
parameter is peaked as a valid finite-size definition of the location of the VBS transition. Results are shown versus $1/L$ in Fig.~\ref{fig:boundaries}.

Ideally we would like to extrapolate the finite-size VBS--HVB and HVB--AFM transition points to the thermodynamic limit. However, while the points defining
the HVB--AFM transitions are rather well-behaved and converge to the thermodynamic with only small finite-size corrections, the VBS--HVB point show more
substantial $L$ dependence and also have much larger error bars. Though it appears clear that the data sets for all $h \le 1$ will converge to $g_c(h) > 0$,
it is difficult to extrapolate the values in a fully quantitative manner. Therefore, we chose to present the phase diagram in Fig.~\ref{fig:PhaseDiagram}
only based on the data for the largest system size, $L=96$. It is clear from Fig.~\ref{fig:boundaries} that the HVB--AFM boundary for $h \in [0.25,1]$ is
already close to the thermodynamic limit, while the VBS--HVB boundary will still shift to somewhat smaller $g$ values as $L \to \infty$. Here we also
note that studies of the HVB phase in the range $h \in (0,0.25)$ are difficult, because of the large system sizes required to realize winding in this
regime, as the maximum winding wavevector $\mathbf{k}^*_{\rm max}$ (at the HVB--AFM boundary) decreases with decreasing $h$.

\begin{figure}[t]
\includegraphics[width=65mm, clip]{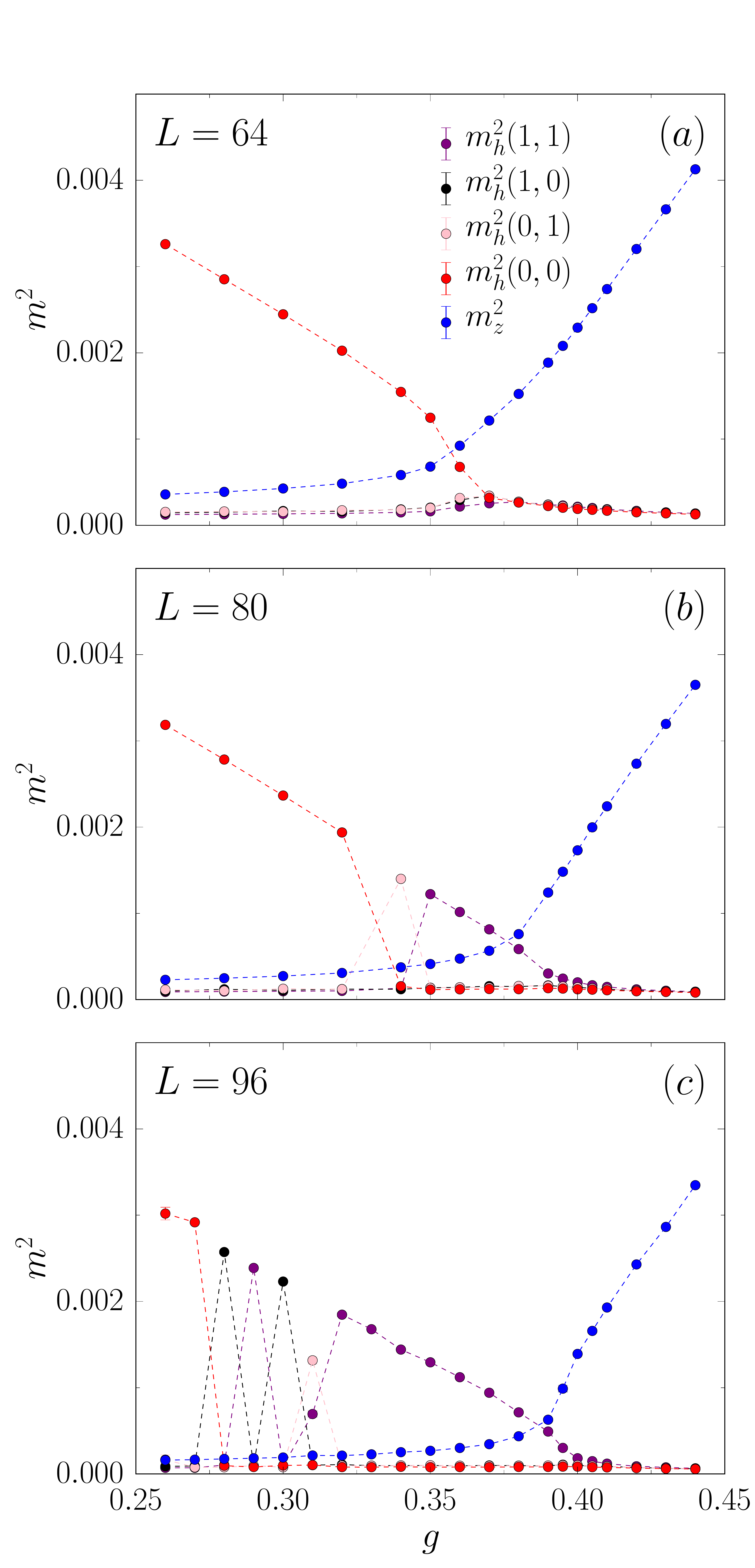}
\caption{Squared order parameters at $h=0.2$ for system sizes $L=64$ (a), $L=80$ (b), and $L=96$ (c). The strongly fluctuating order parameters
between $g=0.26$ and $0.31$ reflect metastability, as described in Sec.~6-A and further illustrated in Fig.~\ref{fig:meta}.}
\label{fig:h02}
\end{figure}

The smallest $h$ value for which we have been able to extract both the VBS--HVB and HVB--AFM transition points is $h=0.2$. As shown in
Fig.~\ref{fig:h02}(a), at $h=0.2$ an $L=64$ system still does not have a well-developed non-trivial winding sector. There are only very broad, low
maximums of the HVB order parameter at winding numbers $(0,1)$ and $(1,1)$ in the region where the system evolves from the VBS to the AFM. In the $L=80$ system,
Fig.~\ref{fig:h02}(b), the ${\bf w}=(1,1)$ sector has fully developed and is more prominent for $L=96$, Fig.~\ref{fig:h02}(c). While the transition into
the AFM state is not difficult to extract, as demonstrated above in Fig.~\ref{fig:afmcross}, the VBS--HVB boundary moves significantly to smaller $g$ between
$L=80$ and $L=96$. The transition point for $L=96$ shown in the phase diagram Fig.~\ref{fig:PhaseDiagram} is still likely somewhat too high (see also
the size dependence for $h=0.25$ in Fig.~\ref{fig:boundaries}). Currently we can not extract any meaningful VBS--HVB transition points for $h$
significantly less than $0.2$, as there is no non-trivial winding for the largest systems we can simulate within reasonable time.

\subsubsection{6-C. Dynamic critical exponent}

One important aspect of the DQCP is the dynamic exponent $z=1$ (emergent Lorentz invariance) \cite{Senthil04a}. In Figs.~\ref{fig:afmcross}(a) and
\ref{fig:afmcross}(d), $L\rho_S$ for both a large and a small value of $h$ appears
to be size independent at the crossing point, in support of a continuous $z=1$ transition. It should be noted, however, that $L\rho_s$ exhibits anomalous scaling in
the $J$-$Q$ model at $h=0$, decaying as a power-law, $\rho_s \propto L^{-a}$ with $a$ significantly less than the expected value $a=z=1$.
It has been proposed that this behavior reflects an ultimately weakly first-order transition \cite{Jiang08,Chen13,Wang17,Nahum19,Ma19},
that there are logarithmic scaling corrections at the DQCP \cite{Sandvik10a}, or that that the transition violates standard expectations for the finite-size
scaling of the order parameters \cite{Shao16,Sandvik20}. Even though some results can also be interpreted as scaling with $z \not =1$, the overall behaviors
cannot be explained by just a different value of $z$, as some quantities would demand $z<1$ and some $z>1$. Interestingly, our results presented in 
Figz.~\ref{fig:afmcross}(a) and \ref{fig:afmcross}(d) do not show any apparently anomalous behaviors of $L\rho_s$, but we still do not have enough data for
large system sizes to carry out a careful analysis and draw firm conclusions in this regard. In light of our finding here that the $h=0$ DQCP is a special
multi-critical point, it is possible that the HVB-AFM transitions constitute a line of generalized DQCPs without anomalous scaling. We plan to address
this interesting possibility in future work.

Here we also note that the spacing of
different winding sectors within the HVB phase should be $\Delta_g \propto L^{-1}$ if the sectors occupy roughly equal $g$ windows. Even if the windows
shrink slightly with increasing winding, as appears to be the case from Figs.~\ref{fig:m2L}(a-e), the spacing should still be safely larger than the
size of the critical window, $\delta \propto L^{-1/\nu}$. Though we have not yet calculated $\nu$ for the HVB--AFM transition, we note that $1/\nu \approx 2.20$
at $h=0$ \cite{Sandvik20}. Thus, a continuous transition between the HVB in its last winding number sector and the AFM should be amenable to standard
finite-size scaling methods, even though the change in winding numbers very close to the transition point some times causes anomalous features also
in the $g$ dependent spin quantities, as exemplified by the $L=80$ data in Figs.~\ref{fig:afmcross}(d) and \ref{fig:afmcross}(e).

\subsection{7. Correlation functions}

Here we show additional correlation function results for the $h=1$ system both inside the HVB phase and at the HVB--AFM transition. Critical correlation
functions often show different behaviors for even and odd distances, due to the presence of uniform and oscillating contributions governed by different
deay exponents, as we have also discussed above in Sec.~2. In the main paper, in Fig.~\ref{fig:Corr} we showed results for spin and dimer correlations at
distances $\mathbf{r}=(x,\pm x)$ for even $x$ inside the HVB phase. Here we will also consider the odd-$x$ branch, which behaves differently close
to the HVB--AFM transition.

\begin{figure}[t]
\includegraphics[width=60mm]{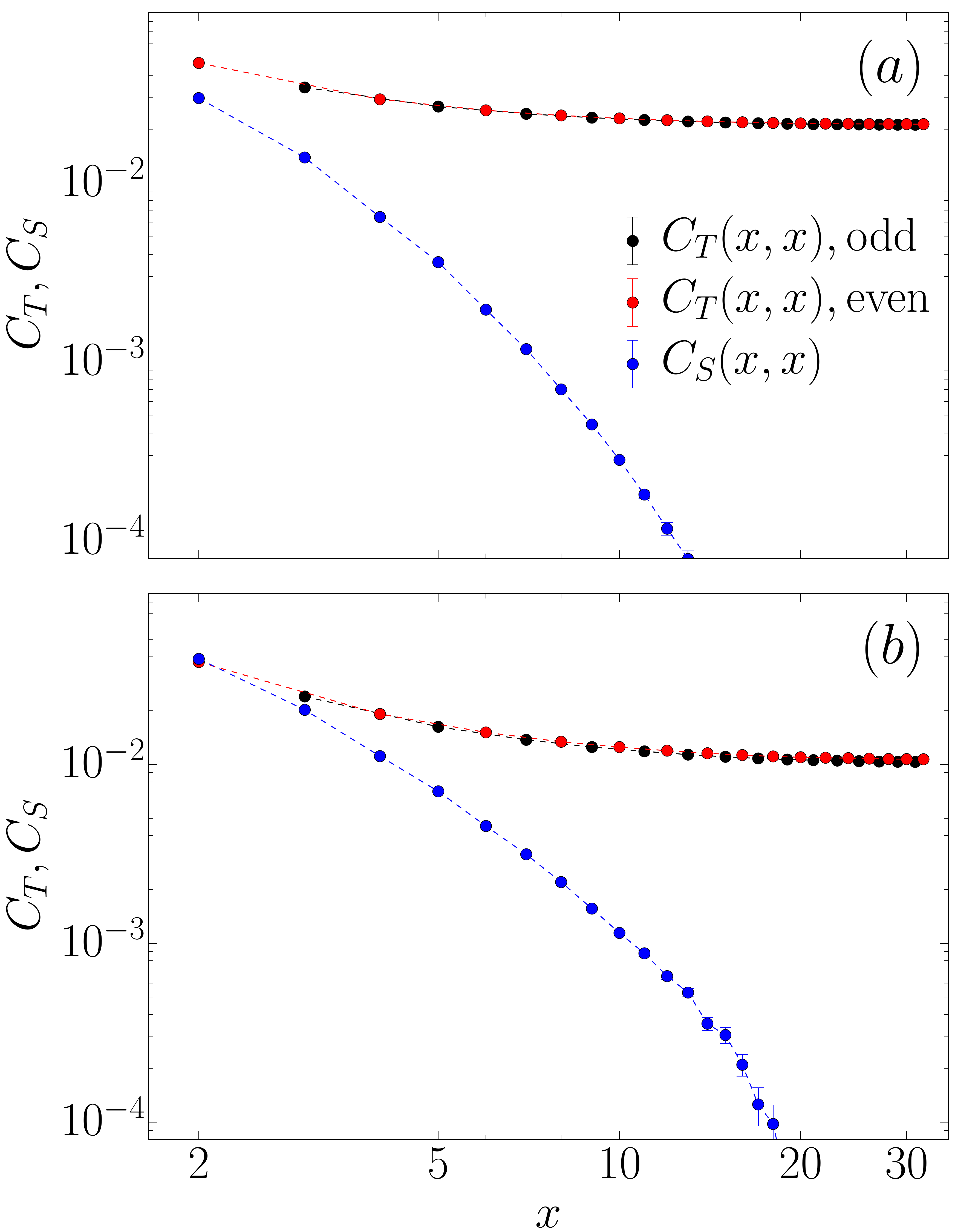}
\caption{Spin ($C_S$) and dimer ($C_T$) correlation functions in $L=64$ systems at $h=1$ with $g=0.25$ in (a) and $0.30$ in (b).
The dimer correlations are defined with with the $T_x$ operator in Eq.~(\ref{txydef}).}
\label{fig:corr64}
\end{figure}

First, we consider two points inside the HVB phase at $h=0$, with $g=0.25$ and $0.30$. In Fig.~\ref{fig:Corr} we showed results inside the HVB phase
at $g=0.25$ for $L=128$, while here we use $L=64$ and focus on the correlations in the $(x,x)$ direction. In Fig.~\ref{fig:corr64}(a), the results for even
$x$ are very similar to those for the larger system, and we note that there are no discernible even-odd oscillations. Some differences between even and
odd $x$ become visible when moving to $g=0.30$, shown in Fig.~\ref{fig:corr64}(b).

\begin{figure}[t]
\includegraphics[width=60mm]{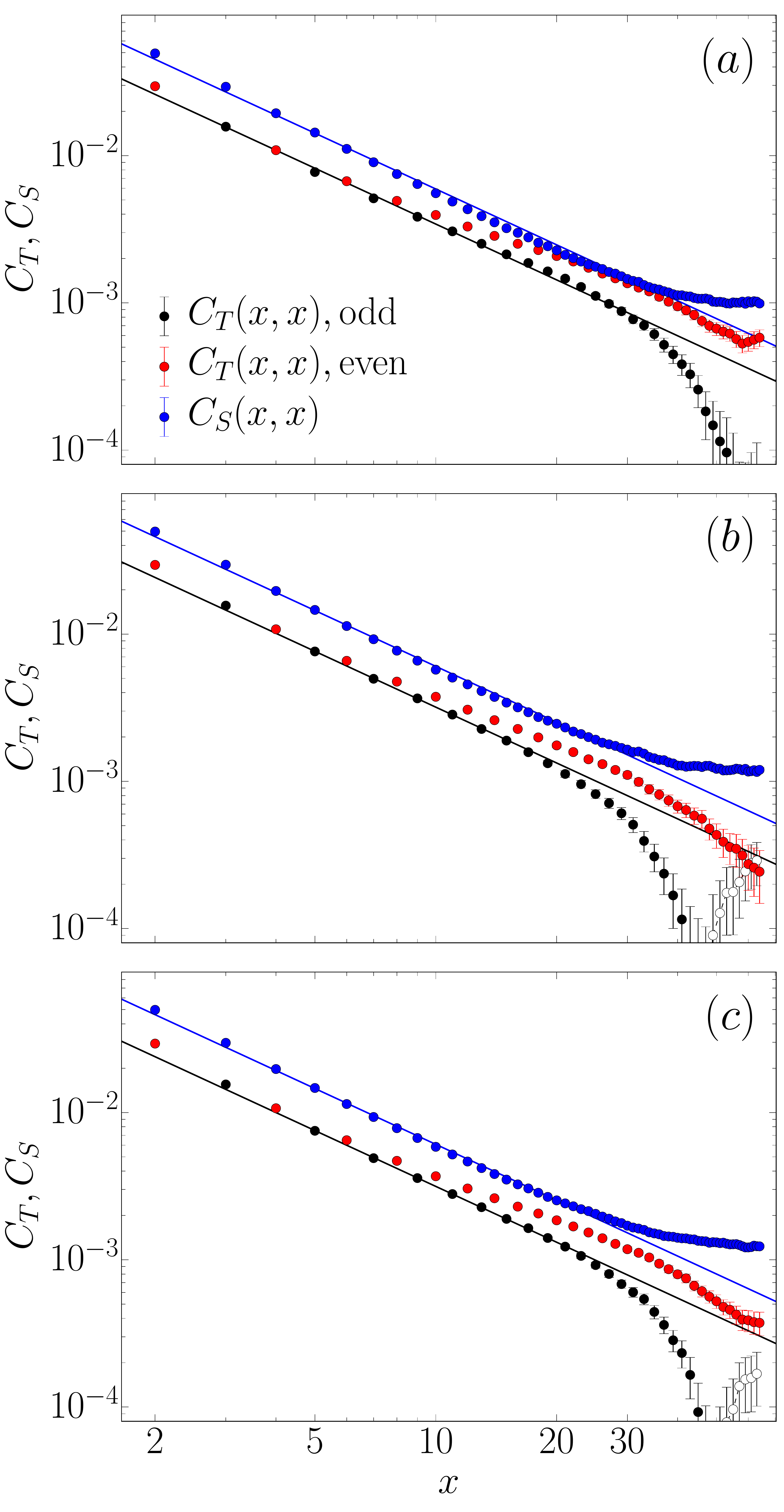}
\caption{The same correlation functions as in Fig.~\ref{fig:corr64} for $L=128$ systems with $h=1$ and $g=0.3410$ in (a), $g=0.3415$ in (b), and $g=0.3420$ in (c).
The case in (b) should be the closest to the critical point $g_c$. The solid lines all have slope $-1.26$, corresponding to the (approximately) known values
\cite{Nahum15b} of the scaling dimensions of the order parameters at the conventional DQCP. The open black circles show $-1$ times the negative values of the
odd-$x$ $C_T$ branch.}
\label{fig:corr128}
\end{figure}

In Fig.~\ref{fig:corr128} we show $L=128$ correlation functions for three values of $g$ close to the HVB--AFM transition. Here again we do not include the results
along the $(x,-x)$ direction, which have no visible even-odd effects. In the diretion $(x,x)$, a separation of the dimer correlations into even and odd branches
is very apparent, however, while the spin correlations show no such effects. From the overall changes in both the spin and dimer correlations as $g$ is increased
in the neighborhood of $g=0.34$, it appears that the HVB--AFM transition point should be $g_c(h=1) \approx 0.3415$, Fig.~\ref{fig:corr128}(b). This critical
value also agrees well with the estimates based on curve crossings in the preceding section. The spin correlations and the odd-$x$ dimer correlations exhibit
reasonably good power-law decays, with decay exponents in good agreement with the best estimates at the conventional DQCP; $-2\Delta{\rm AFM}$ $\approx$
$-2\Delta_{\rm VBS} \approx 1.26$ \cite{Nahum15b}---see also Sec.~2 above, where we presented plaquette-operator correlation functions exhibiting the
same decay in Fig.~\ref{fig:bondpair}. The fact that we observe the same decay of the correlations as at the DQCP at $h=0$ clearly suggests that the
line of HVB--AFM transitions at $h>0$ also should be described by the same field theory.

\begin{figure*}[t]
\includegraphics[width=155mm]{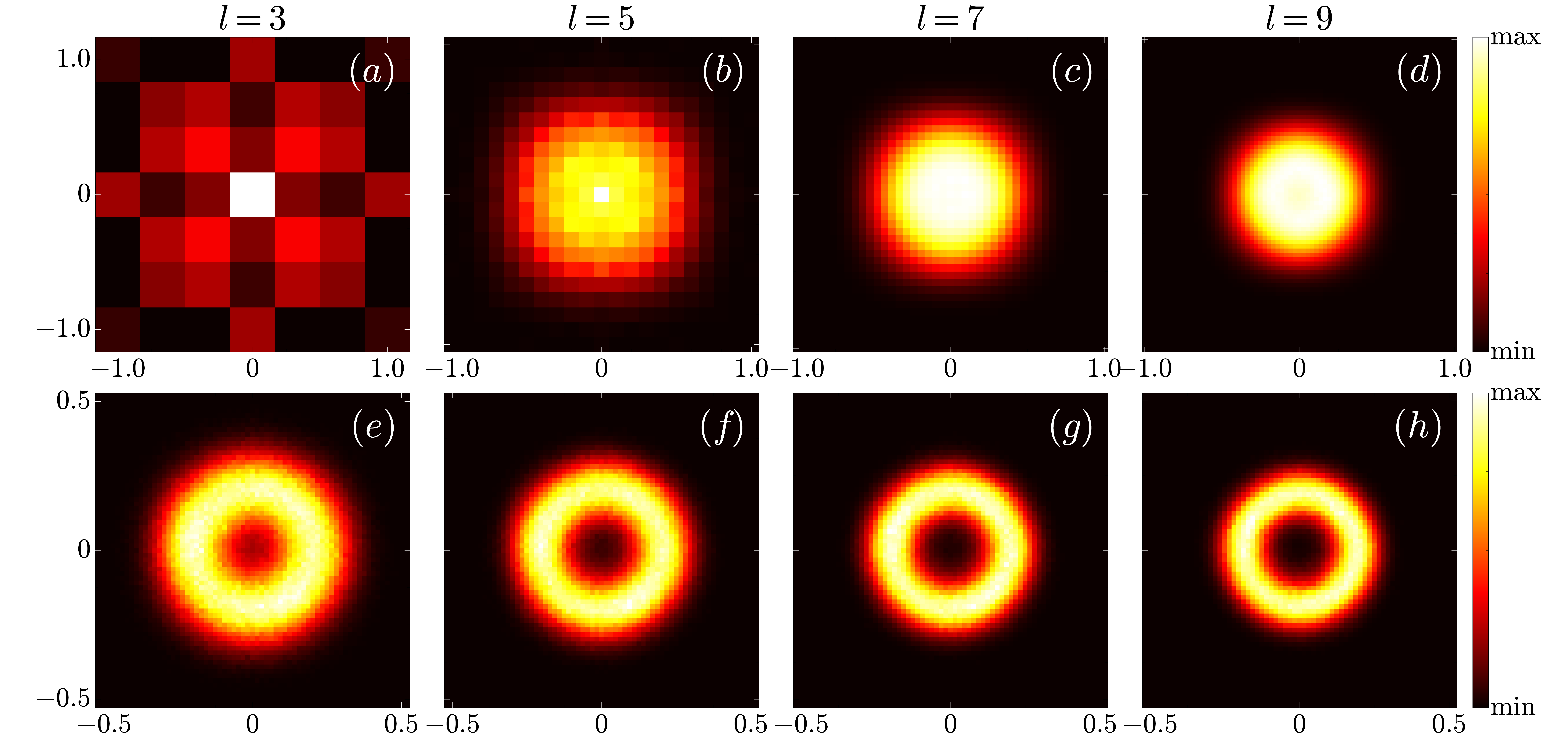}
\caption{Order parameter distributions $P_l(D_x,D_y)$
for an $L=64$ system at $h=1$, $g=0.25$, where the system is in the HVB phase with winding number $(1,1)$. Square cells are used in (a)-(d) and strip
cells in (e)-(h), with $l=3,5,7,9$ in both cases as indicated above the four columns. The scale used for the horizontal and vertical axes are relative to
the maximum value of $D_x$ and $D_y$ in an ideal VBS cell.}
\label{fig:u1histos}
\end{figure*}

As already also discussed in Sec.~2, power-law decays of the correlation functions of the order parameters can only be observed within a rather narrow
window of $1 \ll x \ll L/2$. In the present case, anomalous behaviors can be observed where the even- and odd-$x$ branches of $C_T$ diverge from each other
even though we do not expect different critical exponents, with the results for even $x$ turning slighly up and
those for odd $x$ beoming very small and eventually changing signs for the longest distances. We currently do not have any explanation for these intriguing
differences between the even and odd branches, but they should be related to the periodic boundary conditions, which normally enhance the correlations at the
longest distances, as in the case of $C_S$ in Fig.~\ref{fig:corr128}, but may in some cases have the opposite effect. The behaviors are also most
likely related to the fact that different winding numbers start to mix significantly in the neighborhood of the phase transition (where the conservation
of the winding numbers also should break down \cite{Shao15}), and this
includes winding sectors with $w_x \not= w_y$. Thus, there are some components of the wave function with long-wavelength helicity that can account
for modulation of the  correlation functions, though the reason for the stark differences for even and odd $x$ are presently unclear.

\subsection{8. Emergent U($1$) symmetry in the HVB phase}

In the standard $J$-$Q$ model, emergent U($1$) symmetry has been detected in the near-critical VBS phase and at the DQCP, using order parameter
distributions $P(D_x,D_y)$ such as those discussed in Sec.~1 above \cite{Sandvik07,Jiang08,Lou09,Sandvik12}. The distributions are essentially circular-symmetric
for system sizes up to a length scale $\xi' \gg \xi$ (with $\xi$ being the standard correlation length) governed by a dangerously irrelevant clock
perturbation \cite{Oshikawa00,Leonard15,Okubo15,Shao20} that in the field theory languages corresponds to the presence of quadrupled monopoles \cite{Senthil04a}.
Here we will investigate order parameter distributions in the HVB phase.

In the conventional VBS phase, the distribution of the $(D_x,D_y)$ order parameter evaluated on the full $L \times L$ lattice can be used to test the
symmetry. This method cannot be used with the HVB parameter defined in Eq.~(\ref{mkdef}) (or similar definitions), however, because spatially translating
a given QMC configuration will result in a phase change. Therefore, an ergodic simulation running long enough to restore translational symmetry in the HVB
phase will artificially produce a histogram with a large number of clock angles. When also broadened by fluctuations, for a large system the histogram
should then always be ring-shaped, even in cases where the local dimer order favors certain directions (e.g., if the stripes have order predominantly at
the four clock angles).

We here study distributions of local VBS order parameters defined without phases. We simply use a definition of $(D_x,D_y)$ as in Eq.~(\ref{dxdydef}), but sum
over only a lattice cell much smaller than the system size. In order to observe a putative emergent symmetry, the cell size $l$ must be much smaller than the system
size $L$, and ideally $l$ should also be much smaller than the wavelength of the HVB order. Then, if both $l$ and $L$ are taken to infinity at fixed winding
number while keeping $l \ll L$, the distribution should correctly reflect the symmetry of the local VBS order.

We here consider two types of cells: i) of size $l\times l$ lattice sites, i.e., $(l-1)^2$ plaquettes, with $l$ odd, and ii) in the form of a strip along the
$(1,1)$ diagonal direction, with $l$ (odd) defining the thicknes of the strip in the $(1,0)$ direction [i.e., the total number of plaquettes on the periodic lattice
is $L(l-1)$]. We investigate distributions $P_l(D_x,D_y)$ for several cases of $l$ less than the system size, using $L=64$. Results for both types of cells are
shown in Fig.~\ref{fig:u1histos}.

For small square cells, the number of possible values of $D_x$ and $D_y$ is small, and this is reflected in the large pixels in Figs.~\ref{fig:u1histos}(a,b), where
$D_x=D_y=0$ is also seen to be the most frequently occurring value. Starting from $l=7$, Fig.~\ref{fig:u1histos}(c), the histogram develops a circular shape, and with
$l=9$, Fig.~\ref{fig:u1histos}(d), a ring shape with a shallow local minimum at $(0,0)$ has formed (at about $94\%$ of the maximum probability density, barely
visible on the color scale). In an ordered dimer phase with $U(1)$ symmetry we would indeed expect the ring shape, regardless of the wavelength of the HVB order
as long as $l$ is less than (or at least not much larger) than the wavelength. The radius of the circle corresponds to the amplitude of the local VBS order.

The ring shape is more clearly visible for the strip cells, Figs.~\ref{fig:u1histos}(e-h), since the total size of the strip for given $l$ is much larger than
the $l^2$ cell. Note that the strip is defined in the direction along the stripes in the HVB phase, so that the VBS angle has a finite mean value along the entire
strip. For all four strip widths, the ring shape is very pronounced, and it should be noted that the ``doughnut hole'' in the center is growing with $l$ while
the thickness of the ring is decreasing. These are distinct signatures that are difficult to explain unless the HVB phase truly develops U($1$) symmetry of
the coarse-grained order parameter.

In summary, the results presented here are strongly suggestive of emergent U($1$) symmetry of the helical order, not only close to the transition into
the AFM phase but in the entire HVB phase. The test discussed above was for a system with winding number $(1,1)$, while the transition into the AFM phase
for the $L=64$ system at $h=1$ takes place from the $(2,2)$ winding sector, as seen in Fig.~\ref{fig:m2L}(b). Moreover, as seen in the same figure, the
magnitude of the HVB order parameter here is rather large, not much smaller than that of the VBS order parameter before the transition into the $(1,1)$
winding state. In the VBS phase, we observed in Sec.~1-A that there is no emergent U($1$) symmetry, exactly because the system has strong VBS order
and is far from the DQCP in the neighborhood of which the emergent U($1$) symmetry forms. To definitely confirm emergent U($1$) symmetry in the HVB phase,
it would still be useful to study larger system sizes, so that the requirement $l \ll L$ can be satisfied more stringently than was possible here, especially
with the square cells.

\subsection{9. Microscopic origin of the winding}

As we describe in the main text, the HVB phase does not have (and therefore does not break) chiral $\mathbb{Z}_2$ symmetry of the winding. This may appear counterintuitive,
since the Hamiltonian has reflection symmetry with respect to the diagonal $(1,1)$ axis. However, the VBS order (a given pattern) is not invariant under such
a reflection, and the overall chirality of the winding is therefore fixed and reflected in the fact that the winding numbers $w_x$ and $w_y$ are only positive
with our definitions in the main paper. We can also understand the microscopic mechanisms causing finite winding in the HVB phase and the lack of chiral symmetry
breaking within a dimer-like picture, as we outline in this section. 

\begin{figure}[t]
\includegraphics[width=80mm]{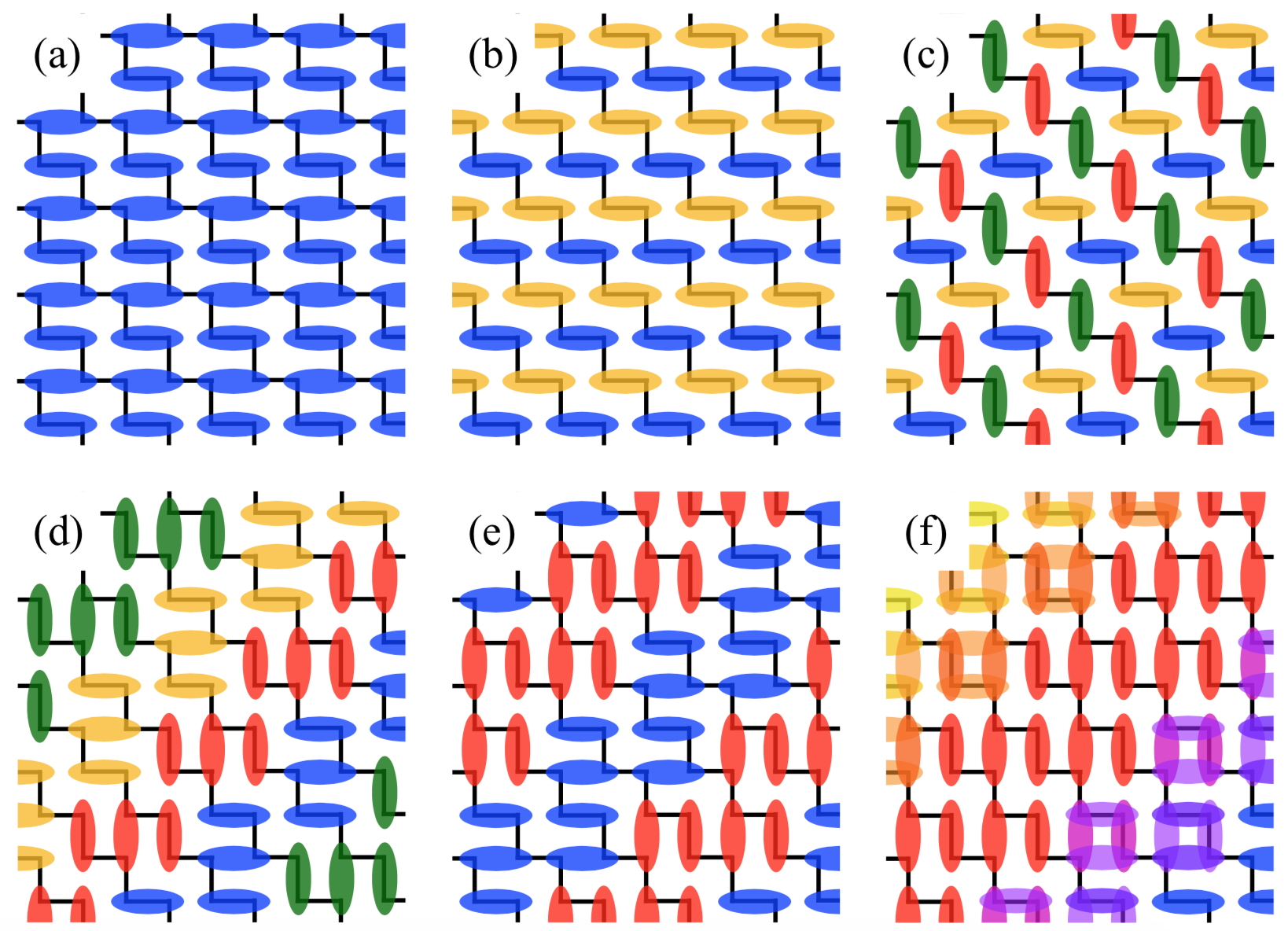}
\caption{Depictions of different valence-bond patterns, where the ovals represent valence bonds, and different colors correspond to
the four different columnar angles as in the actual visualizations (Sec.~3). (a) Columnar VBS pattern. (b) Staggered VBS pattern. 
(c) Maximally winding (tilted) HVB pattern. (d) HVB pattern with width-3 winding. (e) HVB pattern with width-4 winding. 
(f) HVB pattern with domain walls of the plaquette VBS type.}
\label{fig:Schematic}
\end{figure}

For simplicity we fix $h=1$. At $g=0$, the Hamiltonian Eq.~(\ref{hhdef}) then has no staircase pattern because $g=J/(J+Q)$ and $J$ is the overall amplitude
of the exchange terms. We also know that this system has VBS order, since it is just the standard $Q_3$ model \cite{Lou09,Sandvik12} and the $Q_3$ term is
naturally a valence-bond inducing interaction that favors the columnar dimer pattern illustrated schematically (without fluctuations) in Fig. \ref{fig:Schematic}(a). 

The dimer-inducing effect of the $Q_3$ term must remain even when the stair-case modulation of $J$ is turned on, because the $\mathbb{Z}_4$ symmetry is initially
preserved, as discussed above in Sec.~3. With $h=1$, alternating staircase (zig-zag) shaped chains have internal couplings $2J$ and $0$. Consider the limit
when $J$ is large and for the moment neglect the fact that the system AFM orders. The system can be regarded as a set of weakly coupled zig-zag chains,
and to minimize the energy the dimers (singlets) should cover those chains as much as possible. This results in dimers forming on every other bond of the
zig-zag chains (as in the Heisenberg chains with frustrating interactions, which has a VBS ground state). There are two ways of arranging the dimers in this
way, which in the staircase geometry correspond to all horizontal (even) or all of vertical (odd) bonds. In the presence of the $Q_3$ terms, in general
adjacent zig-zag chains must be energetically favored to either have in-phase or out-of-phase dimer coverings. For the in-phase case, the resulting state
is depicted in Fig.~\ref{fig:Schematic}(b), which is conventionally called the staggered VBS. The resulting out-of-phase configuration is depicted
Fig.~\ref{fig:Schematic}(c). Note that Figs.~\ref{fig:Schematic}(b) and (c) are both degenerate ground states of the quantum dimer model in the staggered
limit when viewed as dimer configurations. 

If we carefully observe Fig.~\ref{fig:Schematic}(c) and focus on one color of the valence bonds, we can see that they actually form a very thin (and sparse)
strip in the diagonal direction, and this can be regarded as a {\it maximally winding} HVB state. This kind of maximal winding is very well known in the context
of quantum dimer models, which when mapped to the height model \cite{Henley97} is the state of maximum tilt. In the work on Cantor deconfinement in quantum dimer
models by Fradkin {\it et al.} \cite{Fradkin04}, the maximally tilted state was taken as the staggered configuration with the same orientation of all dimers,
Fig.~\ref{fig:Schematic}(b), because in their
setting it was assumed that winding takes places in either the $x$ or $y$ direction only. The geometry of the square lattice then allows a maximal $\pi$ rotation
per single lattice spacing and this results in the conventional staggered configuration. In our case, the staircase perturbation induces diagonal winding (tilt)
but the square lattice does not have a well-fitting pattern of $\pi$ rotation in the diagonal direction, but it does accommodate the $\pi/2$ shift corresponding to
Fig.~\ref{fig:Schematic}(c).

The conventional staggered VBS pattern in Fig.~\ref{fig:Schematic}(b) has no columnar order at all to satisfy the $Q_3$ terms,
and if the regions of ``blue'' and ``yellow'' dimers are expanded to accommodate some of the ``columnar desires'' of the $Q_3$ terms, energetically expensive
$\pi$ domain walls must form. In contrast, the maximally winding diagonal HVB state in Fig.~\ref{fig:Schematic}(c) can be modified to host thicker stripes
with less expensive $\pi/2$ domain walls between the stripes, and must be favored over the pattern in Fig.~\ref{fig:Schematic}(b) for this reason. Examples
of widened strips of width three and four dimers are shown Fig.~\ref{fig:Schematic}(d,e).

There are several aspects of Fig.~\ref{fig:Schematic}(d) that deserves careful consideration as an approximation to the ground state of our model.
First, let us make some simplistic but useful energy arguments that will explain the origin of winding.
The fraction of the $2J$ bonds that are satisfied by having a valence bond is $1/4$ for (a), $1/2$ for (c), and $1/3$ for (d). The fraction
of $Q_3$ terms that are completely satisfied in each state is $1/4$ for (a), $0$ for (c), and $1/6$ for (d). At least qualitatively, we can say that (a)
satisfies the $Q_3$ terms the most while ignoring the $J$ terms, (c) does the complete opposite of satisfying the $J$ terms as much as possible while ignoring
$Q_3$, and (d) can be seen as an energetically favored compromise between them.  This schematic argument shows that it is natural that, as we increase $g$ in the
Hamiltonian (i.e., increasing $J$ or decreasing $Q$), the ground state changes from the VBS phase (a) to the HVB phase with suvvessively higher winding.
In principle, it should be possible to reconstruct the
phase diagram of different winding numbers with quantum or classical dimer models, but we will leave such a calculation for future work. In the actual Hamiltonian,
the winding never reaches the extreme case depicted in Fig.~\ref{fig:Schematic}(c), but undergoes the transition into the non-dimerized AFM phase at some maximal
winding vector $\mathbf{k}_{\max}$ as we have discussed in the main paper.

Looking a bit more carefully at details of the dimer patterns, we note that simplified depictions of widened stripes such as those in
Fig.~\ref{fig:Schematic}(c,d,e) should have odd widths of the stripe, as the width-one case in (c) and width-three in (d), in order to be realistic. For even
width, as depicted in Fig.~\ref{fig:Schematic}(e), the two adjacent VBS states on either sides of a single stripe (say, blue) become the same one (red),
making a stripe pattern consisting of two colors rather than four. This would result in an instability of the state; whenever two red VBS patterns touch
each other due to quantum fluctuation those stripes will merge and eventually the blue region will be eliminated. Moreover, we can also see that such even-width
states are energetically unfavorable. If we change the stripe width from 4 to 5, not only does that result in satisfying more $Q_3$ terms, but also more $J$
terms. This is because when the width is odd, a single layer of the stripe can match the two edge bonds with the $J$ preferred bonds. This could be seen as another
reason why the chirality is always fixed from the outset: if we try to select adjacent VBS states to match the strong $J$ bonds, it will always result in the
same chirality (and odd widths).

Going further toward the actual HVB state observed in our simulations, we should keep in mind that the true state also has a domain wall between the 
different columnar-ordered VBS stripes. These domain walls are essentially plaquette VBS ordered, as shown in Fig.~\ref{fig:Schematic}(f). The plaquette stripes
have even width in both directions and can be inserted without changing the above argument about the fixed chirality. There are four different plaquette VBS
states, corresponding to the four different domain walls, which can be seen as a compromise of both (columnar) VBS orderings. In the next section, we will show
evidence that the plaquette domain walls actually smoothly connect to the columnar stripes in continuous rotation of the VBS angle, in analogy with the emergent
U($1$) symmetry associated with fluctuation-thickening of domain walls within the DQCP scenario \cite{Levin04}.

We should note that the discussion here has been framed for commensurate HVB order. It is possible that the observed HVB phase in the staircase
$J$-$Q$ model also hosts incommensurate helical order, which also would be induced by the same winding (or tilt) mechanism discussed above. It is beyond the
scope of this paper to determine whether indeed incommensurate order appears, but in the future we plan address this question by also studying systems with
other, non-periodic boundary conditions.

As a conclusion for this section, by considering the dimer configurations maximally satisfying the $Q_3$ terms (columnar VBS) on the one hand and on the other
hand a different dimer configuration maximally satisfying the $J$ terms (maximum winding HVB state), we have argued that there is a natural interpolation
between these cases that gives a qualitatively understanding of why the winding increases with $g$ in the HVB phase. In the simplest dimer model description
with analogous $Q_3$ and $J$ terms, only a direct VBS to maximal-winding HVB transition takes place, with no intermediate windings. This difference with
respect to the staircase $J$-$Q$ model is not surprising, considering the fact that we were only using static, ideal configurations. In reality, there are
of course fluctuations, including longer valence bonds, and it may be possible to take these aspects of the HVB phase into account in a tilted quantum dimer
model similar to that of Fradkin et al.~\cite{Fradkin04}.
However, it should be noted that he AFM aspect of the problem is neglected in that approach, and a complete description of
the HVB phase and its quantum phase transition into the AFM phase is beyond that description. A generalization of the DQCP approach \cite{Senthil04a} will be
required to develop a full understanding of the problem, and our results presented here will serve as guides to formulating the appropriate field-theory with
spinons and gauge fluctuations originating from the winding HVB phase. In that regard, the applicable symmetries of the HVB phase are important and
in the next section we discuss a possible emergent U($1$) symmetry.

\end{document}